\documentclass[11pt,a4paper]{article}
\pdfoutput=1

\usepackage{jheppub}
\usepackage{color}
\usepackage[dvipsnames]{xcolor}
\usepackage{graphicx}
\usepackage{wrapfig}
\usepackage{verbatim}
\usepackage{amsmath}
\usepackage{amssymb}
\usepackage{url}
\usepackage{bbold}
\usepackage{xspace}
\usepackage{slashed}
\usepackage{multirow}
\usepackage{threeparttable}
\usepackage{paralist}
\usepackage{subcaption}
\usepackage{floatrow}
\usepackage{multirow}
\usepackage{soul}
\usepackage[export]{adjustbox}
\usepackage[normalem]{ulem}
\usepackage{braket}
\usepackage{feynmp-auto}
\usepackage{hyperref}
\usepackage{aas_macros}
\usepackage{fontawesome5}

\usepackage{placeins}

\usepackage{dsfont}

\usepackage{relsize}

\usepackage{nccmath}

\newcommand{\MeV}{\,\mathrm{MeV}}

\newcommand{\hc}{\mathrm{h.c.}}

\newcommand{\beq}{\begin{equation}}
\newcommand{\eeq}{\end{equation}}
\newcommand{\bea}{\begin{eqnarray}}
\newcommand{\eea}{\end{eqnarray}}

\newcommand{\Tr}{\mathrm{Tr}}
\newcommand{\Lag}{\mathcal{L}}

\DeclareRobustCommand{\Sec}[1]{Sec.~\ref{#1}}

\DeclareRobustCommand{\App}[1]{App.~\ref{#1}}
\DeclareRobustCommand{\Tab}[1]{Table~\ref{#1}}

\DeclareRobustCommand{\Fig}[1]{Fig.~\ref{#1}}
\DeclareRobustCommand{\Figs}[2]{Figs.~\ref{#1} and \ref{#2}}

\DeclareRobustCommand{\Eq}[1]{Eq.~(\ref{#1})}
\DeclareRobustCommand{\Eqs}[2]{Eqs.~(\ref{#1}), (\ref{#2})}

\DeclareRobustCommand{\Refcite}[1]{Ref.~\cite{#1}}
\DeclareRobustCommand{\Refscite}[1]{Refs.~\cite{#1}}

\newcommand{\cm}{c_{u-d}}
\newcommand{\cp}{c_{u+d}}
\newcommand{\ca}{\tilde{c}_D}
\newcommand{\ci}[1]{c_{#1}}
\newcommand{\ai}[1]{a_{#1}}
\newcommand{\ch}[1]{\hat{c}_{#1}}

\newcommand{\as}{a^{s}}
\newcommand{\Nv}{N_{v}}

\newcommand{\Nvbar}{\bar{N}_{v}}
\newcommand{\Hv}{H_{v}}
\newcommand{\LOneRel}{\mathcal{L}^{\left(1\right)}_{\pi N}}
\newcommand{\LOneHB}{\hat{\mathcal{L}}^{\left(1\right)}_{\pi N}}
\newcommand{\LTwoRel}{\mathcal{L}^{\left(2\right)}_{\pi N}}
\newcommand{\ga}{g_{A}}
\newcommand{\go}{g_0}
\newcommand{\mn}{m_{N}}
\newcommand{\Pp}{P_{v+}}
\newcommand{\Pm}{P_{v-}}
\newcommand{\Ppm}{P_{v\pm}}

\DeclareRobustCommand{\LagHBn}[1]{\hat{\mathcal{L}}^{\left(#1\right)}_{\pi N}}
\DeclareRobustCommand{\LagHBnm}[1]{\hat{\mathcal{L}}^{\left(#1\right)}_{\pi N,1/m_{N}}}
\DeclareRobustCommand{\LagHBnc}[1]{\hat{\mathcal{L}}^{\left(#1\right)}_{\pi N,c_i}}

\newcommand\figref[2]{(\hyperref[#1]{#2})}

\newcommand{\ie}{\emph{i.e.}~}
\newcommand{\eg}{\emph{e.g.}~}

\newcommand{\github}[1]{%
   \href{#1}{\faGithubSquare}%
}

\newcommand{\orcid}[1]{\begingroup
  \hypersetup{hidelinks}\href{https://orcid.org/#1}{\includegraphics[width=10pt]{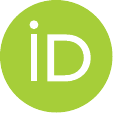}} \endgroup}

\begin{document}

\title{From Supernovae to Neutron Stars: A Systematic Approach to Axion Production at Finite Density}

\author[a,b]{Konstantin Springmann,}
\author[a,c]{Michael Stadlbauer,}
\author[d]{Stefan Stelzl,}
\author[a]{Andreas Weiler}

\affiliation[a]{Technical University of Munich, TUM School of Natural Sciences, Physics Department, James-Franck-Str. 1, 85748 Garching, Germany}
\affiliation[b]{Department of Particle Physics and Astrophysics,
Weizmann Institute of Science, Rehovot, Israel 7610001}
\affiliation[c]{Max Planck Institute for Physics, Boltzmannstr. 8, 85748 Garching, Germany}
\affiliation[d]{Institute of Physics, Theoretical Particle Physics Laboratory, \'Ecole Polytechnique F\'ed\'erale de Lausanne, CH-1015 Lausanne, Switzerland}

\emailAdd{konstantin.springmann@weizmann.ac.il}
\emailAdd{michael.stadlbauer@tum.de}
\emailAdd{stefan.stelzl@epfl.ch}
\emailAdd{andreas.weiler@tum.de}

\date{\today}

\abstract{
We present a systematic study of QCD axion production in environments with finite baryon density and temperature, implying significant changes to axion phenomenology.
Within heavy baryon chiral perturbation theory, we derive the effective Lagrangian describing axion interactions with nucleons and mesons up to next-to-leading-order in the chiral expansion.
We focus on corrections to the axion-nucleon couplings from higher orders and finite density. These couplings are modified by up to an order of magnitude near nuclear saturation density, significantly impacting axion production in supernovae and neutron stars.
Density-dependent corrections enhance the axion luminosity in supernovae by an order of magnitude, strengthening current best bounds by a factor of three.
We stress the importance of including all axion production channels up to a given chiral order for a consistent luminosity calculation and classify the missing contributions up to the third chiral order.
The modified axion-nucleon couplings also affect neutron star cooling rates via axion emission. 
A re-evaluation of existing neutron star cooling bounds, constrained to regions where perturbative control is reliable, weakens these bounds by a factor of four. 
Lastly, our results have implications for terrestrial axion searches that rely on precise knowledge of axion-nucleon couplings. \href{https://github.com/michael-stadlbauer/Axion-Couplings.git}{\faGithub}
}

\preprint{TUM-HEP-1528/24}
\maketitle

\section{Introduction}

In recent years, there has been a resurgence of interest in the QCD axion, both as a solution to the strong CP problem \cite{PhysRevLett.40.223,Wilczek:1977pj,Peccei:1977hh,PhysRevLett.53.535} and as a compelling dark matter candidate \cite{Preskill:1982cy,Abbott:1982af,Dine:1982ah}.
This attention is driven by a robust experimental program, accompanied by the emergence of novel theoretical concepts and refinements of astrophysical constraints.

Astrophysical environments, such as supernovae (SNe) and neutron stars (NSs) are excellent laboratories to test the physics of weakly coupled (light) scalar fields such as the QCD axion \cite{Raffelt:1987yt,Raffelt:1990yz,Raffelt:2006cw,Iwamoto:1984ir,Brinkmann:1988vi,Burrows:1988ah,Burrows:1990pk,Turner:1987by}, for recent reviews see \eg \cite{DiLuzio:2020wdo,Caputo:2024oqc}.
They are particularly relevant because of the large temperatures or large number density of SM particles, which can compensate for the weak coupling of the QCD axion to these particles. 
The constraints from the observation of the neutrino signal from SN 1987A and the observation of NS cooling rates set the most stringent constraints on the QCD axion, see \eg \cite{Turner:1988bt,Carena:1988kr,Iwamoto:1984ir,Brinkmann:1988vi,Keil:1996ju,Ericson:1988wr,Buschmann:2021juv,Carenza:2019pxu,Chang:2018rso}.

In this work, we revisit the theory prediction of axion production in dense and hot baryonic matter, focussing on the regime where the axion can freely escape the SN or the NS.
While previous works include density and temperature effects on production rates in a phenomenological manner or ignore them altogether, a systematic treatment of the interactions of axions with nucleons in a thermal bath and finite density background is still missing \cite{Chang:2018rso,Carenza:2019pxu}.
Concretely, for the first time, we systematically calculate axion production rates within the framework of Chiral Perturbation Theory (ChPT) expansion, where reliable theory error estimates are possible. 
We focus on the dominant axion production mechanism in NS and SN matter environments, $NN\to NNa$, where $N=(p,n)^{\mathrm{T}}$ is the nucleon field.
In this work, we take the first step towards a systematic evaluation of this rate. 
In particular, we include higher order and finite density corrections to the axion-nucleon coupling.
This leads to significant changes in the axion luminosity during SN 1987A and from NSs and, hence, affects resulting bounds on the QCD axion decay constant $f_a$.

Our main tools for this systematic calculation are various effective field theories (EFTs) for the QCD axion.
Starting at energies above the QCD scale, we match to the ChPT Lagrangian of mesons and nucleons at two flavors $N_f=2$, which is valid at energies below the cut-off $\Lambda_\chi$, which is around the QCD scale where baryons are non-relativistic.
We then integrate out the pions and arrive at an effective theory of baryons and the axion only \cite{GrillidiCortona:2015jxo}. 
Note that at densities within SNe or NSs, pion dynamics are crucial.
Keeping the pions as part of our effective theory, we investigate the couplings of the QCD axion to nucleons within the controlled expansion of heavy baryon ChPT.

We find that the couplings of the QCD axion to protons and neutrons depend on axion energy $\omega_a$, the nucleon chemical potential as well as the temperature. 
We estimate the uncertainty due to the truncation of the chiral expansion and due to the uncertainty in low energy constants (see e.g.~\Fig{fig:axion_nucleon_coupling_KSVZ_vac} for the KSVZ axion).
Furthermore, a background of nucleons can change the potential of the QCD axion as well as other light particles. 
The phenomenology of these effects has largely been worked out in \cite{Hook:2017psm,Balkin:2020dsr,Balkin:2021zfd,Balkin:2021wea,Balkin:2022qer,Balkin:2023xtr,Gomez-Banon:2024oux}. 

Here, we calculate the modification of the axion production rate, focusing on the QCD axion coupling to nucleons due to higher-order corrections and interactions with the density background. 
The impact of finite density on the QCD axion-nucleon coupling was first explored in \cite{Balkin:2020dsr}, where an initial estimate of the coupling magnitude was provided.
Notably, a change of up to $\mathcal{O}(10)$ in the coupling of a KSVZ axion to the neutron has been estimated.
Our findings indicate that at nuclear saturation density, $n_0\simeq 0.16~ \text{fm}^{-3}$, the KSVZ axion couplings undergo significant changes.
Specifically, the proton coupling exhibits a relative change of $\mathcal{O}(1)$, while the neutron coupling shows a change of  $\mathcal{O}(10)$, consistent with the previous estimate (see also \eg~\Fig{fig:axion_nucleon_coupling_KSVZ_sym_matter}).

The substantial magnitude of these effects can be attributed to the proximity of the theory's cutoff. 
In particular, the dominant effects are naively $(k/ \Lambda_\chi)^3$ suppressed, where $k$ is the typical momentum and $\Lambda_\chi$ is the UV-cutoff of the theory. 
However, the naive power counting is mitigated by the existence of a low-lying $\Delta(1232)$-resonance, which amplifies the effects.
Thus, certain contributions are only suppressed by $(k/\Lambda_\chi)^2 (k/\delta m_{N-\Delta})$, \ie by the mass difference $\delta m_{N-\Delta}$ between the nucleons and the $\Delta(1232)$-resonance \cite{Bernard:1995dp}. 
Note that while for the form factors in vacuum, we generically find $k = k_0=\{\omega_p,\omega_a\}$, \ie an expansion in energy, in the case of a density background, we instead find an expansion in $k_f/\Lambda_\chi$, where $k_f$ is the Fermi momentum of the background nucleons. 
This shows that in the limit $\omega_a  \to 0$, density effects dominate since $\omega_p \simeq k_f^2/2m_N \ll k_f $, where $m_N$ is the nucleon mass. The systematic expansion allows us to estimate an error in our predictions due to missing higher-order terms.
In a supplementary material, we provide numerical tables of the density and temperature dependence of the axion-nucleon couplings for use in dedicated supernova simulations.
These tables can be accessed at our GitHub repository~\footnote{\hbox{\url{https://github.com/michael-stadlbauer/Axion-Couplings.git}}}~\href{https://github.com/michael-stadlbauer/Axion-Couplings.git}{\faGithub}.

The effective couplings are of significant importance both for astrophysical constraints on the QCD axion as well as for terrestrial experiments.
On the astrophysical side, we explore the modifications of the supernova bound as well as neutron star cooling bounds.
Within supernovae, axions are dominantly produced by nucleon axion Bremsstrahlung process, $NN\to NNa$ \cite{Brinkmann:1988vi}.
If the axion luminosity during the supernova exceeds the neutrino luminosity, a shorter neutrino signal would have been observed from SN 1987A. 
We find that by including density dependent axion-nucleon couplings, the supernova axion bound on $f_a$ is strengthened by a factor of a few compared to previous works.
In particular, we find 
\begin{equation}
    f_a \gtrsim  1.0^{+0.5}_{-0.2} \times 10^9 \, \mathrm{GeV},\quad m_a \lesssim 5.9^{+1.8}_{-2.0} \, \mathrm{meV},\quad\text{(KSVZ)}
\end{equation}
for the KSVZ axion.

Density-dependent axion-nucleon couplings are, however, not the only contributions to the $NN\to NNa$ process.
We hence classify all the relevant topologies that contribute to the process up to order $(k/\Lambda_\chi)^3$. 
The evaluation of these extends the scope of this work and is left for a separate publication, which replaces phenomenological approaches and paves the way to a consistent and systematic axion production in SNe and NS environments.

The breakdown of the effective theory at high densities, as found in NSs, challenges the common interpretation of NS bounds \cite{Buschmann:2021juv}. 

While a significant portion of the NS's luminosity originates from regions with densities around $(0.5-2)\,n_0$, the exact contribution depends on the NS's maximum mass and the chosen equation of state. 
However, an order-one fraction of the luminosity comes from the deep core, where the ChPT expansion is not valid.

This limitation presents two potential approaches for addressing the high-density region. 
The first approach sets the axion cooling bound by considering only the parts of the NS where controlled calculations are feasible—specifically, regions with densities below nuclear saturation density. 
This approach neglects a substantial portion of the NS, typically resulting in a weakening of the cooling bound by roughly a factor of four. 
Within this framework, the emissivity can be systematically calculated using finite-density couplings. 
For the KSVZ axion, this method reduces the bound by an additional factor of two due to the density corrections to the couplings.

The second, less rigorous approach involves estimating axion production in the high-density region by approximating the axion couplings using naive dimensional analysis (NDA). 
In this scenario, there is generally no clear theoretical prediction that differentiates between various axion models. 
Instead, all models predict roughly the same rate in the high-density region and have an uncertainty of $\mathcal{O}(1)$. 
This suggests that vacuum couplings have minimal influence on NS cooling via axions.

Finally, our findings have significant implications for axion dark matter experiments that target the axion-nucleon coupling~\cite{Brandenstein:2022eif,Jiang:2021dby,JEDI:2022hxa,Abel:2017rtm,Gao:2022nuq,Lee:2022vvb,Bloch:2019lcy,2009PhRvL.103z1801V,JacksonKimball:2017elr,Wu:2019exd,Garcon:2019inh,Wei:2023rzs,Xu:2023vfn,Chigusa:2023hmz,Bloch:2021vnn,Bloch:2022kjm,Graham:2020kai,Mostepanenko:2020lqe,Adelberger:2006dh,Bhusal:2020bvx}; the bulk of a nucleus effectively serves as a background density for the nucleon interacting with the axion. 
This effect is reminiscent of the quenching of the axial coupling $g_A$ as measured in beta decay in large nuclei~\cite{Menendez:2011qq,Gysbers:2019uyb}. 
Understanding the QCD axion couplings to matter is particularly crucial in spin precession experiments, especially if axion detection occurs, as it could provide insights into possible ultraviolet (UV) completions.
Similarly, non-derivative axion-nucleon couplings undergo corrections in a comparable manner, although these corrections have yet to be fully calculated. This is relevant for both current and future searches for new physics using nuclear clocks,  including those involving the recent measurement of the low-lying transition in the $^{229}$Th nucleus \cite{EPeik_2003,Flambaum2012,Caputo:2024doz,Kim:2022ype,Fuchs:2024edo}.

The paper is organized as follows. 
In \Sec{sec:HBCHPTfromUV}, we review the EFTs describing the interactions of the axion at different energies, with a particular emphasis on the EFT that will be used later on, heavy baryon ChPT (HBChPT). 
In \Sec{sec:vac_couplings}, we calculate the axion couplings within this framework, including two subleading orders. 
In \Sec{sec:finite_density_couplings}, we extend this analysis to calculate the effective axion-nucleon coupling in the presence of a nucleon density background.
Finally, in \Sec{sec:implications}, we explore phenomenological implications of the above findings. 
In \Sec{sec:supernova} we re-evaluate the axion luminosity during SN explosions, in \Sec{sec:neutron_star} we study the impact on neutron star cooling and in \Sec{sec:terrestrial} we comment on implications for terrestrial axion searches. 
We conclude in \Sec{sec:conclusion}. 
A comprehensive derivation of the next-to-leading order terms of the heavy baryon Lagrangian is provided in \App{app:construction}. 
In \App{app:ratedensity}, we review calculations of particle production rates at finite temperature and chemical potential, focusing specifically on the case of axion-nucleon couplings. 
In \App{app:explicit_calc}, we show additional details of the explicit loop calculation, as well as some further results for the DFSZ axion. 
Finally, in \App{app:temp}, we investigate the relative importance of temperature compared to chemical potential.

\section{Axion Heavy Baryon Chiral Perturbation Theory}\label{sec:HBCHPTfromUV}

In this section, we review chiral perturbation theory in the heavy baryon limit, including the QCD axion. 
We start with the QCD Lagrangian extended by the axion in the UV and construct the effective theory of non-relativistic baryons, pions, and the axion for two light quark flavors. 
Finally, at momenta below the pion mass, we integrate out the pions and match our EFT with the one found in \cite{GrillidiCortona:2015jxo}.

\subsection[From the Axion QCD Lagrangian ...]{The Axion QCD Lagrangian}\label{sec:From_IR_to_UV}

Below the scale of electroweak symmetry breaking the effective Lagrangian of QCD, including the QCD axion field $a(x)$, with decay constant $f_a$ is given by
\begin{equation} \label{eq:QCD_axion_Lagrangian}
\mathcal{L}_{\mathrm{QCD}}=\mathcal{L}_{\mathrm{QCD}, 0}-\bar{q} M_{q} q + \frac{1}{2}\left(\partial a\right)^{2} + \frac{a}{f_{a}} \frac{g^{2}}{32 \pi^{2}} G^{\mu \nu} \tilde{G}_{\mu \nu}+\frac{\partial_{\mu} a}{2 f_{a}} J^{\mu}_{\mathrm{PQ,0}},
\end{equation}
where
\begin{subequations}
\begin{align}
\mathcal{L}_{\mathrm{QCD}, 0} = &-\frac{1}{4} G^{\mu \nu} G_{\mu \nu}+i \bar{q} \slashed{D} q,\label{eq:LQCD0} \\ 
J_{\mathrm{PQ},0}^{\mu}=&\sum_{q} c_{q}^{0} \bar{q} \gamma^{\mu} \gamma_{5} q. \label{eq:PQ_current_UV}
\end{align}
\end{subequations}
In the chiral limit, the theory is classically invariant under the symmetry group 
\begin{equation}
    SU(3)_c\times SU(N_f)_L \times SU(N_f)_R \times U(1)_V \times U(1)_A.
\end{equation}
For simplicity we neglect the effects of the strange quark and take the number of light quark flavors to be $N_f=2$, such that $M_q=\operatorname{Diag}\left(m_u, m_d\right)$.
$J^{\mu}_{\mathrm{PQ,0}}$ is the model-dependent axial quark current associated with the spontaneously broken Peccei-Quinn (PQ) symmetry $U(1)_{PQ}$, with the axion as Nambu-Goldstone Boson (NGB).
The axion-gluon coupling can be removed by performing a chiral rotation on the quark fields
\begin{equation}
    q \rightarrow e^{\frac{i a}{2 f_a} \gamma_5 Q_a} q,\quad q = (u,d)^T,
\end{equation}
with $Q_a$ an arbitrary matrix in flavor space, as long as $\Tr[Q_a]=1$. It is defined as 
\begin{equation}
Q_a = \frac{\operatorname{Diag}[1, z]}{1+z}+ \text{corrections}, \quad z \equiv \frac{m_u}{m_d},
\end{equation}
to remove the leading order tree level mixing of the axion with the pion \cite{GrillidiCortona:2015jxo}, resulting from \Eq{equ:lag_LpiN_2}.
At next-to-leading order additional mixing terms arise \Eq{eq:l7NLOPionLag}, which modify $Q_a$, see \Eq{eq:QaNLO}.
In this basis, the Lagrangian above the QCD confinement scale reads
\begin{equation} \label{eq:QCD_axion_Lagrangian_after_chiral_rotation}
\mathcal{L}_{\mathrm{QCD}}=\mathcal{L}_{\mathrm{QCD}, 0} - \left(\bar{q}_{L} M_{a} q_{R}+\text {h.c.}\right) + \frac{1}{2}\left(\partial a\right)^{2} + \frac{\partial_{\mu} a}{2 f_{a}} J^{\mu}_{\mathrm{PQ}},
\end{equation}
where $M_a$ is the axion dressed quark matrix and $J^{\mu}_{\mathrm{PQ}}$ the shifted PQ current 
\begin{subequations}
\begin{align}
M_{a} & \equiv e^{\frac{i a(x) }{2 f_{a}}Q_{a}} M_q e^{\frac{i a(x) }{2 f_{a}}Q_{a}} , \\
J_{\mathrm{PQ}}^{\mu} & =\sum_{q=u,d} c_{q}^{\mathrm{UV}} \bar{q} \gamma^{\mu} \gamma_{5} q,\quad c_{q}^{\mathrm{UV}} \equiv c_{q}^{0}-\left[Q_{a}\right]_{q}.
\end{align}
\end{subequations}
The couplings $c_{q}^{\mathrm{UV}}$ are defined at energies around $f_a$. At the scales of interest, \ie around the QCD scale, the couplings are given by \cite{GrillidiCortona:2015jxo}
\begin{equation} \label{eq:IR_constants_quarks}
   c_q \equiv c_q^{\mathrm{IR}}\simeq C_{q q^{\prime}}c_{q^{\prime}}^{\mathrm{UV}}, \quad C_{q q^{\prime}}=\left\{\begin{array}{ll}
    0.975 \quad & q=q^{\prime} \\
    -0.024 \quad & q \neq q^{\prime}
    \end{array} \, ,\right.
\end{equation}
which follow from RG evolution. Decomposing the PQ current into isoscalar and isovector, we find
\begin{equation}
    \label{eq:JPQ}
    J^{\mu}_{\text{PQ}}=\bar{q}\gamma^{\mu}\gamma_5 \left(c_{u+d}\frac{\partial_{\mu} a}{2 f_a}\mathbb{1}+c_{u-d}\delta_{b3}\frac{\partial_{\mu} a}{2 f_a}\tau^b\right)q,
\end{equation}
where 
\begin{equation}c_{u\pm d}=(c_u\pm c_d)/2
\end{equation}
contain the corresponding contribution from $Q_a$. 

\subsection[... to Axion Chiral Perturbation Theory ...]{Axion Chiral Perturbation Theory}

At low energies QCD confines and forms a chiral condensate $|\left\langle\bar{q}_L q_R+\hc\right\rangle|\equiv B f_\pi^2$, spontaneously breaking the approximate chiral symmetry
\begin{equation}
    SU(2)_L \times SU(2)_R \times U(1)_V \rightarrow SU(2)_{L+R} \times U(1)_V.
\end{equation}
The low energy degrees of freedom are described as fluctuations around the chiral condensate by the unitary $2\times 2$ matrix in flavor space
\begin{equation}
\label{eq:Upion}
  U=e^{i\frac{\Pi}{f_{\pi}}},\quad \Pi=\pi^a \tau^a=
  \begin{pmatrix}
  \pi^0 & \sqrt{2}\pi^{+}\\
  \sqrt{2}\pi^{-} & -\pi^0
  \end{pmatrix},
\end{equation}
with the pion decay constant $f_{\pi}\simeq93\,\text{MeV}$. 
This leads to the EFT of pions, a systematic expansion in small pion momenta $p/(4\pi f_\pi)$, called chiral perturbation theory (ChPT) \cite{Weinberg:1978kz,Gasser:1983yg, Gasser:1984gg}. Note that a generalization to three flavors is straightforward but will not alter the results below. 
The pion Lagrangian is
\begin{equation}
\Lag_{\pi\pi}=\Lag_{\pi\pi}^{(2)}+\Lag_{\pi\pi}^{(4)}+\dots,
\end{equation}
where $\Lag_{\pi\pi}^{(2)}$ is the leading pion axion Lagrangian given by
\begin{equation}
\Lag_{\pi\pi}^{(2)}=\frac{f_{\pi}^2}{4}\Tr\Big[\nabla_{\mu}U(\nabla^{\mu}U)^{\dagger}+\left(\chi U^{\dagger}+\chi^{\dagger}U\right)\Big], \label{equ:lag_LpiN_2}
\end{equation}
with\footnote{
Note that, in accordance with the standard convention in ChPT, the strong coupling constant and the Wilson coefficient, which relate the pion mass and the product of $f_\pi$ and the quark masses, are included in $\chi$. Consequently, the coefficients for operators involving higher powers of $\chi$ are small, as evidenced e.g. by the coefficient $\ell_7$.}
$\chi=2 B M_{a}^\dagger$, and where the covariant derivative $\nabla_{\mu}$ is given by 
\begin{equation}
\label{eq:covlin}
\nabla_{\mu} U=\partial_{\mu} U-i\left\lbrace a_{\mu}, U \right\rbrace - 2 i a_{\mu}^{s} U.
\end{equation}
Here the isovector and isoscalar axion currents are 
\begin{equation}
    a_{\mu}=\cm\frac{\partial_{\mu} a}{2 f_a}\tau^{3},\quad \text{and} \quad a_{\mu}^s=\cp\frac{\partial_{\mu} a}{2 f_a}\mathbb{1},
\end{equation}
as can be seen from \Eq{eq:JPQ}. 
For details about the construction, see \App{app:construction}. 
Note that \Eq{equ:lag_LpiN_2} gives the leading order QCD axion mass 
\cite{GrillidiCortona:2015jxo},
\begin{equation}
  m_a^2 = \frac{z}{(1+z)^2} \frac{m_{\pi}^2f_{\pi}^2}{f_a^2}.
\end{equation}
We now want to eliminate all the axion-pion mixing at next-to-leading order. The mixing comes from loops from the leading order Lagrangian $\Lag_{\pi\pi}^{(2)}$ as well as tree level contributions from the next-to-leading order Lagrangian $\Lag_{\pi\pi}^{(4)}$. The loop level mixing between $a$ and $\pi^0$ generated from $\Lag_{\pi\pi}^{(2)}$ is however suppressed by $m_a^2$ and thus negligible. 
There is a single term in $\Lag_{\pi\pi}^{(4)}$ that gives rise to tree level mixing \cite{Gasser:1983yg}
\begin{equation} \label{eq:l7NLOPionLag}
    \Lag_{\pi\pi}^{(4)}\supset-\frac{\ell_7}{16}\Big(\Tr\left[\chi U^\dagger-\chi^\dagger U\right]\Big)^2.
\end{equation}
We can remove this additional contribution to the mixing by choosing
\begin{equation}\label{eq:QaNLO}
    Q_a=\frac{\text{Diag}\left[1,z\right]}{1+z}-4\ell_7 \frac{m_\pi^2}{f_\pi^2}\frac{(1-z)z}{(1+z)^3}\tau^3.
\end{equation}
We see that it contributes to the isovector coupling $c_{u-d}$ but leaves the isoscalar coupling $c_{u+d}$ unchanged, see \Eq{eq:JPQ}.

Next, we include baryons in our theory.
Since their mass is at the ChPT cut-off, no consistent power counting scheme can be found for relativistic nucleons and antinucleons. 
Treating the nucleons as heavy and non-relativistic, a systematic expansion in $p/\Lambda_{\chi}$ is available, see e.g.~\cite{Weinberg:1978kz, Jenkins:1990jv, Weinberg:1990rz, Weinberg:1991um, Weinberg:1992yk}. Here $\Lambda_{\chi}\sim 700\MeV$ is the cutoff chosen such that most Wilson coefficients are $O(1)$. 
The leading order chiral pion-nucleon Lagrangian \cite{Gasser:1983yg} can be extended to include the axion \cite{GrillidiCortona:2015jxo}
\begin{equation} \label{eq:baryon_lag_ChPT}
\mathcal{L}_{\pi N}^{(1)} = \bar{\Psi}\left(i \slashed{D}-m_N+\frac{{g}_{A}}{2}   \slashed{u}\gamma_{5} + \frac{\go}{2} \slashed{\hat{u}}  \gamma_{5} \right) \Psi,
\end{equation}
with $\Psi=(p, n)^{\mathrm{T}}$ being the nucleon isospin doublet. 
The usual building blocks of chiral perturbation theory are given by
\begin{subequations}
\begin{align} 
&D_{\mu} \Psi=\partial_{\mu} \Psi+\Gamma_{\mu} \Psi,\\
&\Gamma_{\mu} = \frac{1}{2} [u^{\dagger}, \partial_{\mu} u]-\frac{i}{2} u^{\dagger}a_{\mu} u + \frac{i}{2} u a_{\mu} u^{\dagger},\\
&u_{\mu} = i \left\{ u^\dagger, \partial_\mu u \right\} + u^{\dagger} a_{\mu} u + u a_{\mu} u^{\dagger}, \quad \hat{u}_{\mu}  = u^{\dagger} a^s_{\mu} u + u a^s_{\mu} u^{\dagger}, 
\end{align}
\end{subequations}
with $u=\sqrt{U}$. Note that $u_\mu$ and $\hat{u}_\mu$ are isovector and isoscalar quantities, respectively.
See \App{app:construction} for a detailed derivation of the NLO Lagrangian $\LTwoRel$.

\subsection[... to Axion Heavy Baryon Chiral Perturbation Theory...]{Axion Heavy Baryon Chiral Perturbation Theory...}\label{sec:HBChPT}
The Lagrangian in \Eq{eq:baryon_lag_ChPT} describes the leading order interactions of non-relativistic nucleons with pions and the axion. To simplify, we decompose the nucleon momentum into a piece proportional to its mass and a small residual momentum
\begin{equation} \label{eq:splitng_momenta}
p^\mu = m_N v^\mu + k^\mu,
\end{equation} 
where we choose $v^\mu = (1,\vec{0})^{T}$, which leads to a double expansion in $k/m_N$ and $k/\Lambda_{\chi}$, called heavy baryon chiral perturbation theory (HBChPT) \cite{Jenkins:1990jv,Bernard:1992qa}, and see \eg for reviews \cite{Bernard:1995dp,Scherer:2002tk,Epelbaum:2008ga}.
We split $\Psi$ into velocity-dependent heavy and light components and integrate out the heavy component (see \App{app:construction} for details).
This is analogous to the construction of heavy quark effective theory (HQET) \cite{Isgur:1989vq, Eichten:1989zv,Georgi:1990um,Manohar:2000dt}.
The Lagrangian \Eq{eq:baryon_lag_ChPT} at leading order in HBChPT is
\begin{equation} \label{eq:leading_order_HBChPT_lag}
\hat{\mathcal{L}}_{\pi N}^{(1)} = \bar{N} \bigg( i v \cdot D+g_{A} S \cdot u+\go S \cdot \hat{u} \bigg) N,
\end{equation}
with $S^{\mu}=\frac{i}{2} \gamma_{5} \sigma^{\mu \nu} v_{\nu}=(0,\Vec{\sigma}/2)$. 
Here ${N} \equiv e^{i m_N v \cdot x} P_{+} \Psi$ is the light component of the field $\Psi$, with $P_{+} = (1 + \slashed{v})/2$ the corresponding projection operator. 
The leading order nucleon contact interactions are
\begin{equation}\label{eq:nuclcontlag}
\hat{\mathcal{L}}_{N N}^{(1)} = -\frac{1}{2} C_{S}(\bar{N} N)(\bar{N} N)+2 C_{T}(\bar{N} S N) \cdot(\bar{N} S N).
\end{equation}
In this context, leading order refers to a systematic diagrammatic power counting in HBChPT, developed in \cite{Weinberg:1978kz, Jenkins:1990jv, Weinberg:1990rz, Weinberg:1991um, Weinberg:1992yk}, where we treat $k/\Lambda_{\chi}$ on the same footing as $k/m_N$.
We define the chiral order $\nu$ of the expansion as $\left( k/\Lambda_\chi\right)^\nu$, where $k$ can denote both $|\vec{k}|$ or $k_0$. For a Feynman diagram with two external nucleons $\nu$ is given by
\begin{equation} \label{eq:pc}
\nu=2L+\sum V_{i} \Delta_{i} , \quad \Delta_{i}=d_{i}+\frac{1}{2} n_{i}-2,
\end{equation}
where $L$ is the number of loops and $\Delta_i$ is the order of each vertex, appearing $V_i$ times in the diagram. 
Here $d_i$ is the number of derivatives or $m_{\pi}$ insertions, and $n_i$ denotes the number of nucleon lines associated with the vertex $i$. 
Note, that for the vertices of the leading order Lagrangians \Eqs{eq:leading_order_HBChPT_lag}{equ:lag_LpiN_2}, and \Eq{eq:nuclcontlag} one finds $\Delta_i=0$.

On the other hand, for very low momenta, the existence of bound states interferes with our power counting. In the environments we are interested in, typical nucleon energies are large compared to typical bound state energies, allowing us to ignore their effects. Implicitly, this coincides with only keeping the leading term in an expansion in $(m_N E_\text{B}/p^2) \ll 1$, where $E_\text{B}\sim\mathcal{O}(\text{MeV})$ is the binding energy of bound states such as deuteron.

In the non-relativistic limit, the nucleon propagator only has a single pole in the complex plane. 
Consequently, pure nucleon loops (with more than one nucleon propagator) do not contribute. In other words, anti-nucleons are not included in our EFT.

The next-to-leading order pion-nucleon and contact interaction interactions, which include the QCD axion, are constructed in \App{app:construction}, see \Eq{eq:NLOLagApp} and \Eqs{eq:NLOLagApp2}{eq:NLOLagApp3}. We find
\begin{subequations}
\begin{align}
\begin{split}
         \LagHBn{2}=&\,\,\bar{N}\left[-\frac{1}{2\mn}\left(D^2-\left(v\cdot D\right)^2+i\ga\left\lbrace S\cdot D,v\cdot u\right\rbrace+i\go\left\lbrace S\cdot D,v\cdot\hat{u}\right\rbrace\right)\right.\\
        &+\ch1\braket{\chi_{+}}+\frac{\ch2}{2}\left(v\cdot u\right)^2+\ch3\left(u\cdot u\right)+\frac{\ch4}{2} i\epsilon^{\mu\nu\rho\sigma}\left[u_{\mu},u_{\nu}\right]v_{\rho}S_{\sigma}\\
        &+\left.\ch5\widetilde{\chi}_{+}+\frac{\ch8}{4}\left(v\cdot u\right)\left(v\cdot\hat{u}\right)+\ch9\left(u\cdot\hat{u}\right)\right]N,\label{eq:NLOLagpiN}
\end{split}\\
    \mathcal{L}_{\pi N N}^{(2)}=&\,\, \frac{c_{D}}{2 f_{\pi}^{2} \Lambda_{\chi}}(\bar{N} N)\left(\bar{N} \,S\cdot u \,N\right) + \frac{\ca}{2 f_{\pi}^{2} \Lambda_{\chi}}(\bar{N} N)\left(\bar{N} \,S\cdot \hat{u}\, N\right),\label{eq:ContactIntLag}
\end{align}
\end{subequations}
where the terms $\ch8$, $\ch9$ and $\ca$ have not been considered before.
Here
\begin{equation}
\chi_{\pm}=u^{\dagger} \chi u^{\dagger} \pm u \chi^{\dagger} u, \quad \widetilde{\chi}_\pm={\chi}_{\pm} -\frac{1}{2}\langle {\chi}_\pm\rangle,
\end{equation}
and $\langle \cdot \rangle$ denotes the flavor trace. 
Note that hatted constants are divided by one power of the cut-off, \ie $\hat{c}_i=c_i/\Lambda_\chi$. Without any additional suppression, one naively expects $c_i \sim \mathcal{O}(1)$. 
Without a pseudo-scalar source, the coefficient $\hat{c}_5$ has only isospin-breaking effects. 
Its size is smaller than the naive estimate, as it is only generated by isospin-breaking physics in the UV. 

However, other coefficients can be larger due to the low-lying $\Delta(1232)$-resonance, which is only $\sim 300~\text{MeV}$ above the nucleon mass \cite{Bernard:1995dp,Jenkins:1991es}.
The $\Delta(1232)$ resonance has isospin $3/2$, and thus its presence can enhance the Wilson coefficients of the nucleon coupling to two isovectors, \ie the coefficients $\hat{c}_2, \hat{c}_3$ and $\hat{c}_4$. 
This enhancement has important implications, as it interferes with naive power counting. 
As we will see, contributions that are naively suppressed can, in fact, be dominant due to this enhancement. 
This is the reason why we keep certain $\nu=3$ terms in our calculation while including all terms up to $\nu=2$ in our expansion\footnote{In a separate publication \cite{Springmann:2024ret} we investigate the model-independent bound on the QCD axion, \ie assuming that all derivative couplings of the axion are negligible, as \eg in the astrophobic axion \cite{DiLuzio:2017ogq}. 
In this case, the enhanced terms vanish, and we keep the non-enhanced terms at this order.
We show that these non-derivative operators give rise to novel axion production channels, which we evaluate in supernova environments.
The constraint we find is model-independent and exceeds similar current best bounds \cite{Lucente:2022vuo} by two orders of magnitude.
}.
Since the axion decay constant $f_a$ is orders of magnitude larger than any other scale, terms with more than one axion are strongly suppressed.
All relevant Feynman rules can then be calculated by expanding the fundamental building blocks, and keeping only the leading terms we find
\begin{subequations}
\begin{align}
\hat{u}_{\mu} = & \,\,\cp \left(\frac{\partial_{\mu} a}{f_{a}}\right) \mathbb{1}+\ldots,\\
u_{\mu} = & - \left(\frac{\partial_{\mu} \pi^{a}}{f_{\pi}}\right) \tau^{a} + \cm \left(\frac{\partial_{\mu} a}{f_{a}}\right) \tau^{3}+\cm \left(\frac{ \pi^a \pi^b\partial_\mu a}{2 f_a f_\pi^2} \right) \left( \tau^b \delta^{3 a} - \tau^3 \delta^{a b} \right) + \ldots, \\ 
D_{\mu} =&\,\,\partial_{\mu}+\ldots , 
\label{eq:covdevexp}  \\
\hspace{-1cm} \langle \chi_{+} \rangle = & \,\,
4m_\pi^2
\left[1-\frac{\pi^a\pi^b}{2f_{\pi}^2}\delta^{ab}\right]+\ldots , \\
\widetilde{\chi}_{+}=& \,\,
-2m_\pi^2\frac{1-z}{1+z}\tau^3
-\, m_\pi^2\frac{4z}{(1+z)^2}
\left( \frac{\pi^a a}{ f_\pi f_a} \right) \tau^a + \ldots , \\
\hspace{-1cm}  \langle \chi_{-} \rangle = & \,\,  i m_\pi^2 \frac{8z}{(1+z)^2} \left( \frac{a}{f_a} \right) +  \dots,\\
\hspace{-1cm}  \widetilde{\chi}_{-}  = & \,\,  -2im_\pi^2 \left(\frac{\pi_0}{f_\pi}\right)\tau^3 +  \dots
.
\end{align}
\end{subequations}
Contributions to $D_{\mu}$ including pion or axion fields, are kinematically suppressed by $v\cdot k \sim \vec{k}^2/m_N$, and we do not include them here. 
For completeness, we show $\chi_-$ terms, which do not appear at NLO since no C and P invariant terms can be constructed to this order.
They are however relevant for higher-order tree-level interactions. 

In astrophysical environments such as neutron stars and supernovae, typical momenta can be high, and pion dynamics cannot be integrated out. 
While terrestrial experiments are generally performed at much lower energies, the nucleons can still have non-negligible (Fermi-)momenta \cite{Menendez:2011qq}, such that the pions are still important.

\subsection{... to Axion Heavy Baryon Theory}\label{sec:VillaEFT}

At even lower nucleon momenta $p\ll m_\pi$ one can integrate out the pion, see \cite{GrillidiCortona:2015jxo}.
At these energies, the EFT at leading order takes the form
\begin{equation}
\begin{aligned}
    \mathcal{L}_{a N} =& \bar{N}\bigg[i  v \cdot \partial  + \frac{S\cdot\partial a}{f_a}\left(G_A c_{u-d}\tau^3  + G_0c_{u+d}\mathbb{1}\right) \\
    &+ \sigma\left\langle \operatorname{Re}\left(M_a\right)\right\rangle +b \operatorname{Re}\left( M_a-\frac{1}{2}\left\langle M_a\right\rangle\right) +\ldots\bigg]N.
\end{aligned}
\end{equation}
The isovector axial coupling $G_A = \Delta u - \Delta d =1.2754(13)$ is known to high precision from measurements of nuclear $\beta$-decay \cite{Workman:2022ynf} and the isoscalar axial combination $G_0=\Delta u + \Delta d = 0.440(60)$ is inferred from the lattice \cite{FlavourLatticeAveragingGroupFLAG:2021npn}, where we linearly added the statistical and systematical errors. From now on, we assume that the combined error is Gaussian.

Note that while the leading order isospin breaking effect is taken into account, \ie $\propto \cm$, higher order isospin breaking terms, including the subleading axion pion mixing, are neglected here. 
The corrections from neglecting these terms are small and within the uncertainty of the choice of $G_A$, see \cite{GrillidiCortona:2015jxo}.
Also, note that we slightly redefined the operators proportional to $\sigma$ and $b$ in comparison to \cite{GrillidiCortona:2015jxo} in order to separate them into isospin symmetric and isospin breaking contributions, respectively.

This Lagrangian gives rise to an axion nucleon vertex
\begin{equation}
    \label{eq:VertexAHBT}\includegraphics[scale=1,valign=c]{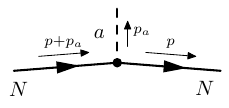} \quad = \quad - \frac{1}{f_a}\left(G_A\cm \tau^3+G_0\cp\mathbb{1}\right) S \cdot p_a,
\end{equation}
with model-dependent couplings 
\begin{equation}\label{eq:axioncouplingswithoutpions}
\begin{aligned}
     C_{ap}=  G_0 \cp+ G_A\cm  = & -0.439(48) + 0.847(29) c^0_u - 0.427(29) c^0_d, \\  
     C_{an}= G_0 \cp - G_A\cm  = & \; 0.019(48) - 0.427(29) c^0_u + 0.847(29) c^0_d.
\end{aligned}
\end{equation}
Note that for the KSVZ axion ($c_q^0=0$), there is a cancellation between two a priori unrelated terms in $C_{an}$, leading to a value compatible with zero.

\section{Axion Couplings in Heavy Baryon Chiral Perturbation Theory}\label{sec:AxioncouplingsHBCHPT}

In this section, we examine the couplings of the QCD axion to nucleons both in vacuum and at finite density. 
By including pions in our theory, we first derive loop corrections to the axion-nucleon coupling identified in Axion Heavy Baryon Theory (see \Eq{eq:axioncouplingswithoutpions}) in \Sec{sec:vac_couplings}. 
To evaluate these corrections, we match the Lagrangian constants to their experimentally determined values at the corresponding order in the chiral expansion. 
This approach reveals the non-trivial momentum dependence from higher-order corrections and reproduces the couplings of \cite{GrillidiCortona:2015jxo} in the small momentum limit. In the second part of this chapter, \Sec{sec:finite_density_couplings}, we calculate the corrections to the couplings that arise in a finite-density background.

\subsection{Axion-nucleon couplings in vacuum} \label{sec:vac_couplings}

We now discuss the couplings of the QCD axion to nucleons in vacuum up to chiral order $\nu=3$. At this order, we focus on contributions enhanced by large Wilson coefficients from integrating the low-lying $\Delta(1232)$-resonance, as they are numerically as significant as the $\nu=2$ terms. These same diagrams will also lead to the dominant density modifications of the axion coupling in the subsequent section.

The leading order $\nu=0$ axion-nucleon Lagrangian, according to \Eq{eq:leading_order_HBChPT_lag}, reads
\begin{equation}
    \hat{\mathcal{L}}_{\pi N}^{(1)}  \supset  \left(\frac{\partial_\mu a}{f_a}\right)  \bar{N}c_N S^{\mu} N,\quad N = \left(p,n\right)^{T},
\end{equation}
with the coupling given by 
\begin{equation} \label{eq:tree_axion_couplings}
  c_N=\operatorname{Diag}\left(c_p,c_n\right)=g_A\cm \tau^3+\go\cp\mathbb{1}.
\end{equation}
This leads to the $\Delta=0$ (see \Eq{eq:pc}) Feynman rule of the tree-level axion-nucleon vertex within the HBChPT approximation, 
\begin{equation}
    \label{eq:LOVertex}\includegraphics[scale=1,valign=c]{figures/LOVertex2.pdf} \quad = \quad - \frac{1}{f_a} \mathcal{A}^{(\nu=0)} S \cdot p_a=- \frac{1}{f_a} c_N S \cdot p_a,
\end{equation}
where we introduced the form factor $\mathcal{A}^{(\nu=0)}=c_N$.
We denote the vertices of this order ($\Delta=0$) in all diagrams with a filled circle.

Higher order corrections start with a single tree level $\nu=1$ diagram, containing only contributions from relativistic corrections to the leading order Lagrangian.
At this order, no new low-energy constants appear. Instead, we expect the contributions to be $\propto c_N\left(\frac{v\cdot p}{m_N}\right)$, where $p$ is a typical nucleon momentum. Indeed, we find 
\begin{equation}
\begin{aligned}
    \label{eq:NLOVertex}\includegraphics[scale=1,valign=c]{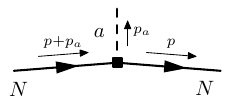} \quad &= -\frac{1}{f_a}\mathcal{A}^{(\nu=1)}S\cdot p_a-\frac{1}{f_a}\mathcal{B}^{(\nu=1)}S\cdot p\\ &= \frac{c_N}{2 f_a\mn}  \left(\omega-\omega_p\right)\left(S \cdot p_a + 2S\cdot p\right),
\end{aligned}
\end{equation}
where we defined
\begin{equation}
    \omega_a \equiv v \cdot p_a, \qquad \omega_p \equiv v \cdot p, \qquad \text{and} \qquad \omega \equiv \omega_a + \omega_p,
\end{equation}
and where we denote $\Delta=1$ vertices by a filled square.

\begin{figure}[t!]
    \centering
    \includegraphics[scale=1]{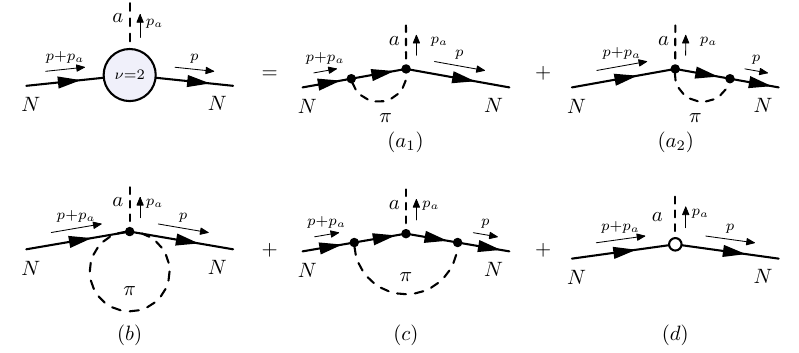}
    \caption{Corrections to the axion nucleon vertex at chiral order $\nu=2$. 
    Filled dots $(a_1),\dots,(c)$ denote $\Delta=0$ vertices according to \Eq{eq:pc}. The empty circle $(d)$ denotes the tree-level contributions from $\LagHBn{3}$, including terms needed to renormalize the previous diagrams and relativistic corrections, see \cite{Vonk:2020zfh}.}
    \label{fig:Nu2Vac}
\end{figure}

At chiral order $\nu=2$, one-loop corrections start contributing and are renormalized by corresponding tree-level terms from $\LagHBn{3}$.
All relevant, \ie non-zero $\nu=2$ diagrams are shown in \Fig{fig:Nu2Vac} and have been calculated in \cite{Vonk:2020zfh}. 
Note, however, that our result, which can be written as
\begin{equation}\label{eq:Nu2VacResult}
   \includegraphics[scale=1,valign=c]{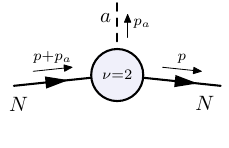} \quad=\quad - \frac{1}{f_a}\mathcal{A}^{(\nu=2)} S\cdot p_a - \frac{1}{f_a}\mathcal{B}^{(\nu=2)} S\cdot p,
\end{equation}
slightly differs from the result found in \cite{Vonk:2020zfh}.\footnote{Our result is also valid for momenta above the pion mass and differs in the evaluation of one of the diagrams, see \App{ap:vac_loops}.}
We find the form factors $\mathcal{A}$ and $\mathcal{B}$ are given by
\small
\begin{equation}
\begin{aligned}
\mathcal{A}^{(\nu=2)}  = &  -\frac{c_N}{4 m_N^{2}}\left(\omega_p^{2}-\omega^2 + \omega_p \omega - p^{2}\right) -\frac{3 {c}_N}{2}\left(\frac{g_{A} m_\pi}{4 \pi f_{\pi}}\right)^{2} \\
& -\frac{\hat{c}_N}{6}\left(\frac{g_{A}}{4 \pi f_{\pi}}\right)^{2}\bigg[m_{\pi}^{2} + \frac{1}{\omega_a} \left(\frac{-\omega_a^3}{3}+m_\pi^3 F\left(\frac{- \omega_a}{m_\pi} \right) - m_\pi^3 F\left(0\right) \right)\bigg] \\
& +4 m_{\pi}^{2}\left[\left(\bar{d}_{16} \tau^{3} 
{-d_{17}\frac{1-z}{1+z}}
\right) c_{u-d}+\bar{d}_{16}^{u+d} \cp -\left(d_{18}+2 d_{19}\right) 
{\frac{z}{(1+z)^2}}
\right],
\end{aligned}
\end{equation}
\normalsize
and 
\begin{equation}
    \mathcal{B}^{(\nu=2)} =  \frac{c_N}{4 m_N^2}\bigg[2\left(\omega^{2}-\omega_p^{2}\right)-(p+p_a)^{2}+p^{2}\bigg],
\end{equation}
where we defined $\hat{c}_N=\tau^a c_N \tau^b\delta_{ab} = -g_{A} \cm \tau^{3}+3 g_{0} \cp \mathbb{1}$ and 
\begin{equation}
    {F}(x)= - ({x}^2-1+i \varepsilon)^{3/2}\left[\log \left({x}-\sqrt{{x}^2-1+i \varepsilon}\right)-\log \left({x}+\sqrt{{x}^2-1+i \varepsilon}\right)\right],
\end{equation}
see also \App{ap:vac_loops}.

\begin{figure}[t!]
    \centering
    \includegraphics[scale=1]{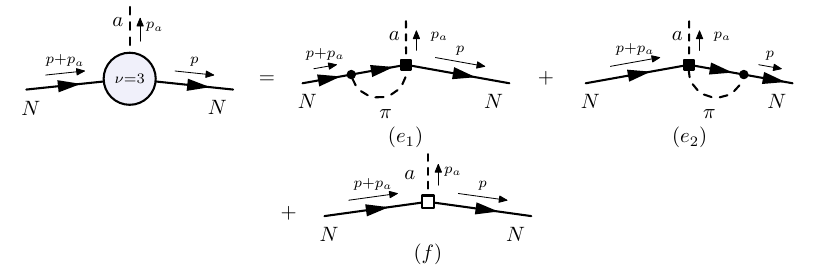}
    \caption{Corrections to the axion nucleon vertex at chiral order $\nu=3$. Filled squares $(e)$ denote $\Delta=1$, filled circles denote $\Delta=0$ vertices according to \Eq{eq:pc}. The empty square $(f)$ denotes tree-level contributions from $\LagHBn{4}$, see \Refcite{Meissner:1998rw}, needed to renormalize the previous diagrams.}
    \label{fig:Nu3Vac}
\end{figure}

While the next order $\nu=3$ is, of course, naively suppressed, it is well known that the low energy constants $\ch{3/4}$ from the NLO pion-nucleon Lagrangian, see \Eq{eq:NLOLagpiN}, are larger than the naive power counting expectation due to the low-lying $\Delta(1232)$-resonance \cite{Bernard:1995dp}.
Note that while the coefficient $\hat{c}_2$ is also enhanced, its effects are kinematically suppressed and thus neglected.
Including only the enhanced diagrams captures the dominant effects of the next order without calculating all diagrams at $\nu=3$. 
In \Fig{fig:Nu3Vac}, we show the enhanced diagrams and the corresponding tree-level diagram necessary for their renormalization. 
Note that the analogous isoscalar diagrams, generated by $\hat{c}_9$ are not enhanced by the $\Delta(1232)$-resonance and thus dropped.
The result can be written in terms of a single form factor
\begin{equation}
   \includegraphics[scale=1,valign=c]{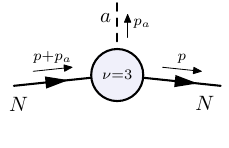} \quad=\quad - \frac{1}{f_a}\mathcal{A}^{(\nu=3)} S\cdot p_a ,
\end{equation}
with
\begin{equation} \label{eq:Anu3}
\mathcal{A}^{(\nu=3)} = \,\,\frac{4g_{A} \cm}{3\left(4\pi f_{\pi}\right)^{2}}\left( \hat{c}_{3}-2 \hat{c}_{4}\right) \tau^{3}\bigg[ m_{\pi}^{2} \omega_a - \omega_a^{3} - m_\pi^3 F\left(\frac{-\omega_a}{m_\pi} \right) -  m_\pi^3 F\left(0 \right) \bigg] .
\end{equation}
To arrive at this result, we subtracted the piece
\begin{equation}
    -\frac{2g_A}{3f_{\pi}^2}\left(\hat{c}_3-2\hat{c}_4\right)c_{u-d}\tau^3\Lambda(\lambda)\left(6m_{\pi}^2\left(\omega+\omega_p\right)-4\left(\omega^3+\omega_p^3\right)\right),
\end{equation}
where $\Lambda(\lambda)$ collects the scale-dependent and divergent pieces in dimensional regularization.
The isovector divergent parts are renormalized by the appropriate $\Delta=3$ operators found in \cite{Meissner:1998rw}, and no finite pieces are generated. 
For details, see \App{app:reno}. 

Summarizing our findings, the result up to the relevant $\nu=3$ contributions is given by
\begin{equation}\label{eq:vacuumcouplingfull}
\begin{aligned}
    & \includegraphics[scale=1,valign=c]{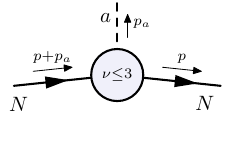} \quad=- \frac{1}{f_a}\mathcal{A}^{(\nu\le3)} S\cdot p_a - \frac{1}{f_a}\mathcal{B}^{(\nu\le3)} S\cdot p \\
   \hspace{-5mm} =&  - \Bigg\{ \frac{c_N}{\mn}  \left(\omega_p-\omega\right) + \frac{c_N}{4 m_N^2}\bigg[2\left(\omega^{2}-\omega_p^{2}\right)-\left((p+p_a)^{2}-p^{2}\right)\bigg] \Bigg\} \frac{S \cdot p}{f_a} \\
   & - \left\{ c_N + \frac{c_N}{2 \mn}  \left(\omega_p-\omega\right) -\frac{c_N}{4 m_N^{2}}\left(\omega_p^{2}-\omega^2 + \omega_p \omega - p^{2}\right) -\frac{3 {c}_N}{2}\left(\frac{g_{A} m_\pi}{4 \pi f_{\pi}}\right)^{2} \right. \\
   & -\frac{\hat{c}_N}{6}\left(\frac{g_{A}}{4 \pi f_{\pi}}\right)^{2}\Bigg[m_{\pi}^{2} + \frac{1}{\omega_a} \left(\frac{-\omega_a^3}{3}+ m_\pi^3 F\left(\frac{-\omega_a}{m_\pi} \right) - m_\pi^3 F\left(0\right)\right)\Bigg] \\
& + \frac{4g_{A} \cm}{3\left(4\pi f_{\pi}\right)^{2}}\left( \hat{c}_{3}-2 \hat{c}_{4}\right) \tau^{3}\bigg[ m_{\pi}^{2}\omega_a-\omega_a^3\bigg.  - \left. \left.  m_\pi^3 F\left(\frac{- \omega_a}{m_\pi} \right) -  m_\pi^3 F\left(0\right) \right] \right. \\
& + \left. 4 m_{\pi}^{2}\left[\left(\bar{d}_{16} \tau^{3}
{-d_{17}\frac{1-z}{1+z}}
\right) c_{u-d}+\bar{d}_{16}^{u+d} \cp -\left(d_{18}+2 d_{19}\right) 
{\frac{z}{(1+z)^2}}
\right] \right\} \frac{S \cdot p_a}{f_a}.
\end{aligned}
\end{equation}
The form factors are diagonal matrices in isospin space, and we define their components as
\begin{subequations} \label{eq:def_A_B}
    \begin{align}
        \mathcal{A}^{(\nu\le3)}&=\text{Diag}\left(\mathcal{A}_{ap},\mathcal{A}_{an}\right),\\ 
        \mathcal{B}^{(\nu\le3)}&=\text{Diag}\left(\mathcal{B}_{ap},\mathcal{B}_{an}\right).
    \end{align}
\end{subequations}

We want to stress at this point that it is important to make the correct choice of the (bare) Lagrangian constants such that our EFT predicts the measured coupling constants in the zero momentum limit. 
That is, as we go to chiral order $\nu=3$, the constants have to be matched consistently at this order to the experimentally measured values. 

\subsubsection{Consistent choice of Lagrangian parameters}
\label{sec:match}
We now determine the Lagrangian parameters consistently using experimental input.
To do so, we calculate the matrix elements $\epsilon_A^\mu \langle N(p^\prime) | A_\mu^a | N(p) \rangle$ as well as $\epsilon_A^{\mu} \langle N(p^\prime) | A_\mu^s | N(p) \rangle$ to order $\nu=3$, where $A_\mu^a$ and $A_\mu^s$ are isovector and isoscalar axial currents respectively while $\epsilon_A^\mu$ are the corresponding polarization vectors for the isovector and isoscalar.
The form factors $G_A(t)$ and $G_0(t)$ within HBChPT are related to the matrix elements by
\begin{subequations}
    \begin{align}
        \epsilon_A^\mu \langle N(p^\prime) | A_\mu^a | N(p) \rangle &\supset G_A(t) \epsilon_A^\mu \bar{u}(p^{\prime}) \gamma_\mu \gamma_5 ({\tau^a}/{2}) \Pp u(p),\label{eq:GA_form_factor}\\[10pt]
        \vspace{0.5pt}
        \epsilon_A^\mu \langle N(p^\prime) | A_\mu^s | N(p) \rangle &\supset  G_0(t) \epsilon_A^{\mu} \bar{u}(p^{\prime}) \gamma_\mu \gamma_5 ({\mathbb{1}}/{2})\Pp u(p),\label{eq:G0_form_factor}
    \end{align}
\end{subequations}
where $\Pp$ is the projector onto the light baryon component, see App. \ref{App:EffBaryonLag}.
In the limit $t = (p^\prime - p)^2 \rightarrow 0$ these are fixed experimentally, from beta decay $G_A(0) \equiv G_A =1.2754(13)$ \cite{Workman:2022ynf}, and the lattice $G_0(0) \equiv G_0=0.440(60)$ \cite{FlavourLatticeAveragingGroupFLAG:2021npn}. 

\begin{table}[b!]
\begin{tabular}{|c||c|c|c|c|c|}
 \hline
Parameter & $G_A$ \cite{Workman:2022ynf}  & $G_0$ \cite{FlavourLatticeAveragingGroupFLAG:2021npn} & $M_N$ \cite{Workman:2022ynf} & $F_\pi$ \cite{FlavourLatticeAveragingGroupFLAG:2021npn} & $M_\pi$ \cite{Workman:2022ynf}   \\ 
\hline
Value & $1.2754(13)$ & $0.440(60)$& $938.9(5) ~\mathrm{MeV}$& $92.07(57) ~\mathrm{MeV}$& $139.6(5) ~\mathrm{MeV}$  \\
\hline
\end{tabular}
\caption{Numerical values for the measured parameters. Note that we add the electromagnetic mass splitting as an error for the proton and pion mass and choose as the central value the average between charged and neutral masses for the nucleons, while for the pion, we choose the charged pion mass.} \label{tab:constantsinput}
\end{table}

\begin{figure}[t] 
  \centering
\includegraphics[width=1.\textwidth]{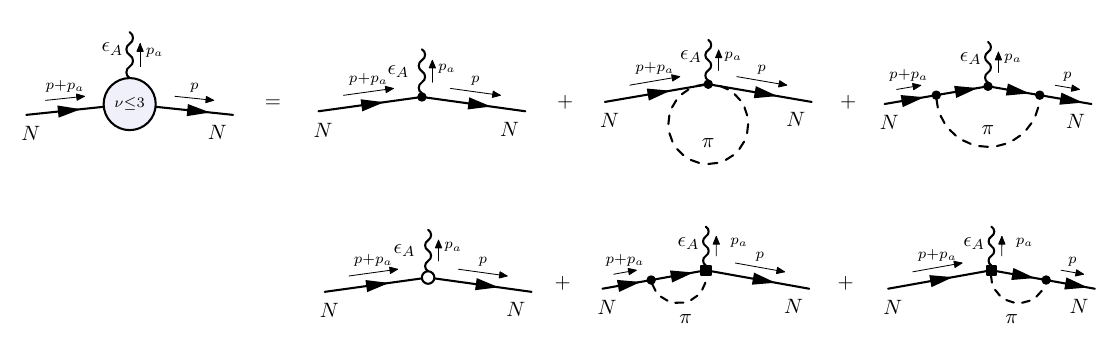}
 \caption[]{Diagrams contributing to the isovector form factor.} \label{fig:IsovectorRenorm}
\end{figure}

In order to calculate the matrix elements, we need to evaluate the diagrams shown in \Figs{fig:IsovectorRenorm}{fig:IsoscalarRenorm}, and multiply each by two factors of the nucleon wave function renormalization $\sqrt{\mathrm{Z}_N}$, one for each external leg, given by
\begin{equation}
  \label{eq:ZN}
    \mathrm{Z}_N=1+\Sigma^{\prime}(0)=1-\frac{3 g_A^2 m_\pi^2}{32 \pi^2 f_\pi^2},
\end{equation}
where at $\nu=3$, consistent with our calculation above, we dropped the terms that are not $\Delta(1232)$-enhanced.
\begin{figure}[b] 
  \centering
\includegraphics[width=1.\textwidth]{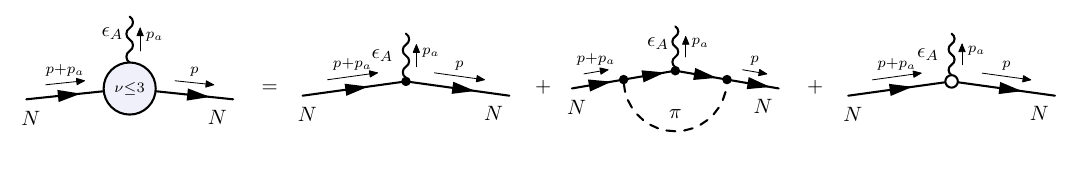}
 \caption[]{Diagrams contributing to the isoscalar form factor.} \label{fig:IsoscalarRenorm}
\end{figure}
The isovector axion form factor in the limit $t\to 0$ to this order is given by \cite{Kambor:1998pi,Bernard:2006te}
\begin{equation}
    G_A = g_A - \frac{g_A^3 m_\pi^2}{16 \pi^2 f_\pi^2} + 4 m_\pi^2 \bar{d}_{16} + \frac{g_A m_\pi^3}{6 \pi f_\pi^2} (2\ch4- \ch3 ),
\end{equation}
 while we find that the analogous result for $G_0$ at this order reads
\begin{equation}
   G_0 = g_0 - \frac{3 g_0 g_A^2 m_\pi^2}{16 \pi^2 f_\pi^2} + 4 m_\pi^2 \bar{d}_{16}^{u+d}.
\end{equation} 
A similar calculation at this order can be done for the pion mass and decay constant, as well as the nucleon mass, which up to small isospin-breaking corrections gives \cite{Colangelo:2001df,Bernard:1995dp}
\begin{align} 
  \label{eq:constants_renorm2}
    M_\pi^2 & = m_\pi^2 \left[ 1 - \frac{m_\pi^2}{32 \pi^2 f_\pi^2} \bar{\ell}_3 
    \right], \\
    \label{eq:constants_renorm1}
    F_\pi & =\: f_\pi \, \left[1+\frac{m_\pi^2}{16 \pi^2 f_\pi^2} \bar{\ell}_4\right], \\
    \label{eq:constants_renorm3}
    M_{N} & = m_N\left[1-4 \ch1 \frac{m_\pi^2}{m_N}-\frac{3 g_A^2 m_\pi^3}{32 \pi f_\pi^2 m_N}\right].   
\end{align}

\begin{table}[t]
\begin{tabular}{|c|c|}
\hline
\multicolumn{1}{|l|}{Lagrangian parameter} & Value  \\ \hline
\hline
$g_A$ & $1.23(11)$ \\
$g_0$ & $0.478(90)$ \\
$m_N$ & $870.8(36) ~\mathrm{MeV}$   \\
$f_\pi$ & $85.52(77) ~\mathrm{MeV}$  \\
$m_\pi$  & $141.7(75) ~\mathrm{MeV}$  \\
$z$    \cite{Workman:2022ynf}      & $0.474(65)$     \\
$\bar{\ell}_3$ \cite{FlavourLatticeAveragingGroupFLAG:2021npn} & 3.41(82) \\
$\bar{\ell}_4$ \cite{FlavourLatticeAveragingGroupFLAG:2021npn} & 4.40(28)  \\
${\ell}_7$ \cite{GrillidiCortona:2015jxo} & $7(4) \times 10^{-3}$ \\
$\bar{d}_{16}$ \cite{Siemens:2017opr}     & $-3.0(1.6) ~\mathrm{GeV}^{-2}$  \\
$d_{18}$ \cite{Hoferichter:2015tha}   & $-0.44(24) ~\mathrm{GeV}^{-2}$  \\
$\ch1$  \cite{Hoferichter:2015tha}  & $-1.07(2) ~\mathrm{GeV}^{-1}$   \\
$\ch3$ \cite{Hoferichter:2015tha}   & $-5.32(5) ~\mathrm{GeV}^{-1}$ \\
$\ch4$   \cite{Hoferichter:2015tha}  & $3.56(3) ~\mathrm{GeV}^{-1}$  \\
$C_{S}$   \cite{Epelbaum:2001fm}  & $ -119(7) ~\mathrm{GeV}^{-2}$  \\
\hline
\end{tabular}
\caption{Numerical values of the bare parameters. Note that the first ones are derived from the physical parameters in Tab. \ref{tab:constantsinput}. For simplicity, we assumed that all errors in the input parameters are Gaussian around the central value. Note, however, that the errors in the derived quantities are strongly correlated.} \label{tab:constants}
\end{table}
We denote physical parameters with capital letters, \ie $G_A, G_0, M_\pi, F_\pi$ and $M_N$ summarized in Tab.~\ref{tab:constantsinput}, while Lagrangian (bare) quantities are referred to by lowercase letters and are summarized in Tab.~\ref{tab:constants}. Clearly, they are different from the tree-level results shown in~\Sec{sec:VillaEFT}. 

Since the isospin breaking effects to the nucleon masses cancel in the average, we use $2M_N=m_p+m_n$ (see \Tab{tab:constantsinput}) to match. On the other hand, we identify $M_\pi$ with the mass of the charged pion, as the neutral pion gets an additional isospin-breaking contribution to the mass not included in \Eq{eq:constants_renorm2}. In this work, we neglect electromagnetic effects, which we account for by adding a $\pm 0.5~\text{MeV}$ uncertainty when identifying the above masses with their measured value. Note that up to isospin-breaking electromagnetic and tiny Yukawa corrections, the ratio of quark masses $z$ is scale and scheme-independent.
 
In analogy to \cite{Vonk:2020zfh}, we assume that the undetermined low-energy constants $\bar{d}_{16}^{u+d}$, $d_{17}$, and $d_{19}$ are drawn from a superposition of two normal distributions with $d_i=\pm 0.5(5)~ \text{GeV}^{-2}$. 
Note that while we assume a Gaussian error for all input parameters, the errors of derived quantities are strongly correlated. 
We can now discuss our predictions \Eq{eq:vacuumcouplingfull}.

\subsubsection{Results} \label{sec:vac_results}
Let us start with evaluating the axion coupling to protons and neutrons up to $\nu=3$ in the limit of all external momenta $p_i \to 0$. As evident from \Eq{eq:vacuumcouplingfull} the analytic result consists of two form factors $\mathcal{A}^{(\nu\leq3)}$ and $\mathcal{B}^{(\nu\leq3)}$, however in this limit, the only surviving contribution comes from $\mathcal{A}^{(\nu\leq3)}$.
Here, our results necessarily agree with~\cite{GrillidiCortona:2015jxo}, as both EFTs are equally valid.
All relevant Lagrangian parameters are summarized in \Tab{tab:constants}. 

The model-dependent result for the axion neutron and axion proton coupling is
\begin{equation}
\begin{aligned}
    C_{ap}\equiv\lim_{p_i\to0} \mathcal{A}^{(\nu\le3)}_{ap}= & -0.423(54) + 0.847(30) c^0_u - 0.427(30) c^0_d, \\  
    C_{an}\equiv\lim_{p_i\to0} \mathcal{A}^{(\nu\le3)}_{an}= & \; 0.020(54) - 0.427(30) c^0_u + 0.847(30) c^0_d,
\end{aligned}
\end{equation}
where we choose the corresponding entries of the isospin matrix $\mathcal{A}$.
This agrees with \Eq{eq:axioncouplingswithoutpions}. However, here we included isospin-breaking effects, including the change in $c_{u-d}$ due to NLO axion-pion mixing.
The remaining numerical differences compared to \cite{GrillidiCortona:2015jxo} come from an updated choice for the central values of physical input parameters, \ie mostly due to taking $\Delta u$ and $\Delta d$ from \cite{FlavourLatticeAveragingGroupFLAG:2021npn}.
Note further that we do not take the strange quark as well as heavier quarks into account.
Finally, for a KSVZ axion ($c_q^0 = 0$) we find 
\begin{equation} \label{eq:res_vac_KSVZ_constants}
    C_{ap}^{\mathrm{KSVZ}}=-0.423(54), \quad C_{an}^{\mathrm{KSVZ}}=0.020(54),
\end{equation}
while for a DFSZ axion ($c^0_u = \sin^2{\beta} / 3 $, $c^0_d = 1- c^0_u$), we obtain
\begin{equation}
\begin{aligned}
     C_{ap}^{\mathrm{DFSZ}}= & -0.565(50) + 0.425(7)\sin^2\beta, \\
     C_{an}^{\mathrm{DFSZ}}= & \; 0.300(50) -0.425(7)\sin^2\beta.
\end{aligned}
\end{equation}

While the calculation for the $\nu=2$ correction closely resembles \cite{Vonk:2020zfh}, the interpretation of our results differ; we determine the Lagrangian parameters from experiment at the order of our perturbative expansion, thereby ensuring we reproduce the results of~\cite{GrillidiCortona:2015jxo} in the limit of small momentum. 
This ensures that there is no additional uncertainty due to the truncation of the ChPT expansion at zero momentum.
More importantly, our main result is the non-trivial momentum dependence of the axion coupling in \Eq{eq:vacuumcouplingfull}.

\begin{figure}[t!] 
  \centering
  \begin{subfigure}{.49\textwidth}
  \centering
  % include first image
\includegraphics[width=1.\textwidth]{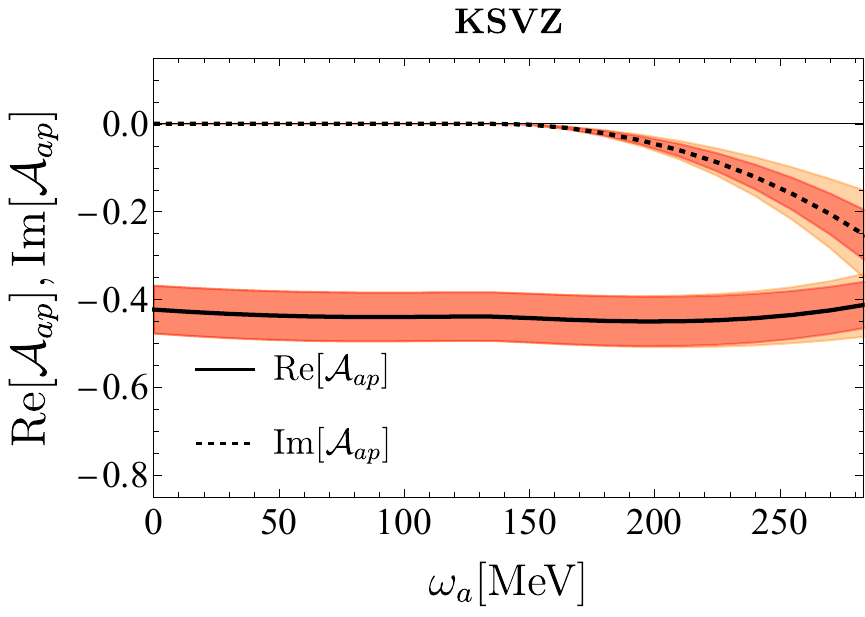}
\end{subfigure}
  \begin{subfigure}{.49\textwidth}
  \centering
  % include second image
\includegraphics[width=1.\textwidth]{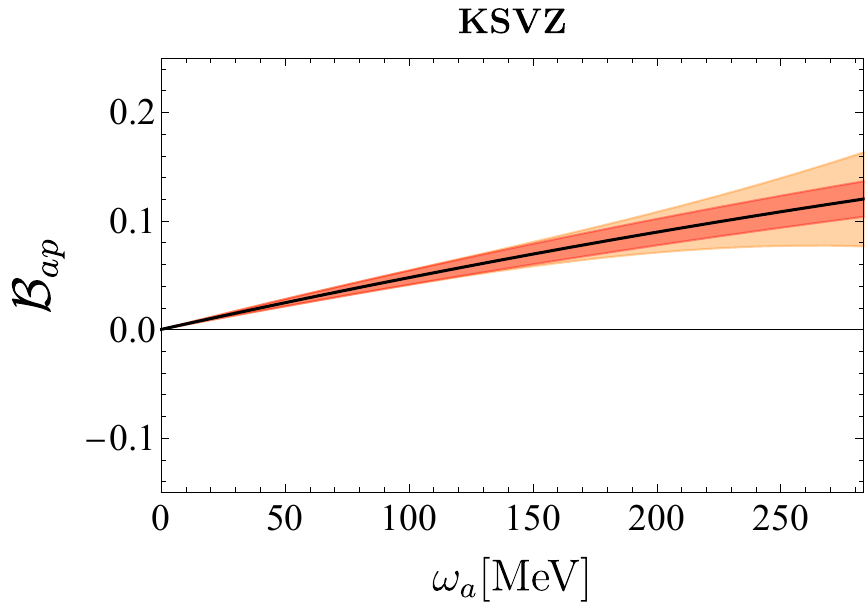}
\end{subfigure}
\begin{subfigure}{.49\textwidth}
  \centering
  % include first image
\includegraphics[width=1.\textwidth]{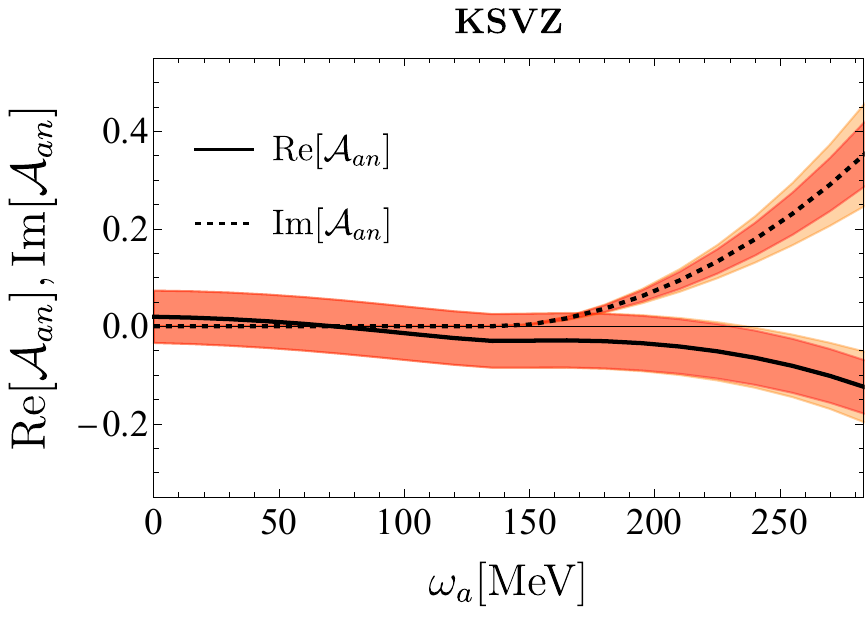}
\end{subfigure}
  \begin{subfigure}{.49\textwidth}
  \centering
  % include second image
\includegraphics[width=1.\textwidth]{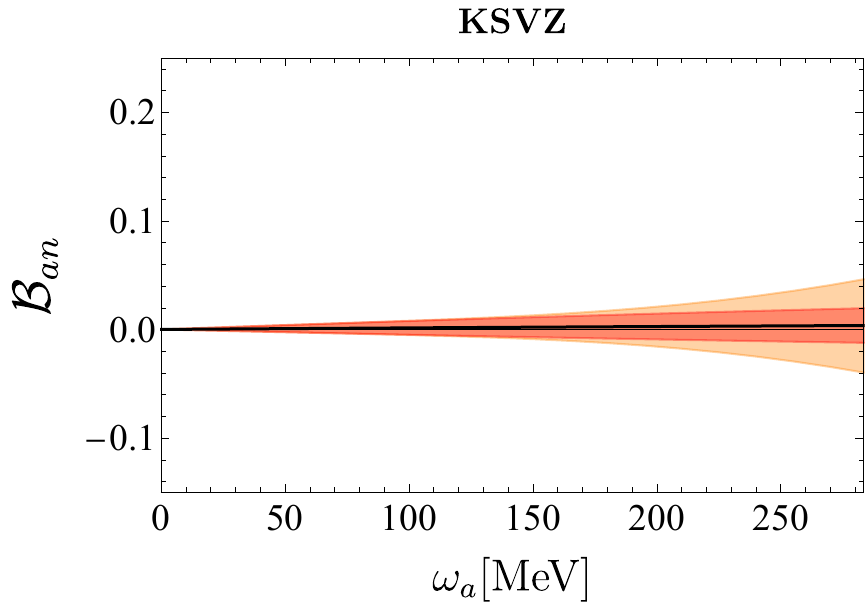}
\end{subfigure}
 \caption[]{Dependence of the KSVZ axion-nucleon couplings on the axion energy $\omega_a=v \cdot p_a$. The red-shaded region represents the error bars resulting from the uncertainties of the constants given in \Tab{tab:constants}, the orange-shaded region represents the estimated error emerging from the truncation of the chiral expansion. At $\omega_a\sim m_\pi$ one can see the pion production threshold, at which the coupling gets an imaginary part that becomes the dominant contribution at large energies. Note that since $\mathcal{B}$ does not include loop corrections, hence there is no imaginary part.}\label{fig:axion_nucleon_coupling_KSVZ_vac}
\end{figure}

For non-negligible axion momenta both form factors $\mathcal{A}^{(\nu \le 3)}$ and $\mathcal{B}^{(\nu \le 3)}$ contribute.
We show our full results in the rest frame of the outgoing nucleon, $p=0$, for a KSVZ axion in Fig. \ref{fig:axion_nucleon_coupling_KSVZ_vac}
 plotted against the axion energy $\omega_a$. Note that we keep the outgoing axion and outgoing nucleon momenta on-shell.

Due to the $\Delta(1232)$-resonance enhancement, the $\nu=3$ diagrams are numerically more important than the $\nu=2$ corrections
. 
These higher-order diagrams also lead to more substantial modifications than those from the strange quark, as detailed in~\cite{Vonk:2021sit}.

We address the uncertainties associated with truncating the chiral expansion by including an error in orange. 
We estimate the size of this neglected contribution by taking the $\nu =3$ result and multiplying it by an additional factor of $\omega_a/(4\pi f_\pi)$.

In addition, we can determine the momentum dependence of an ALP, \ie an axion without QCD anomaly, from \Eq{eq:vacuumcouplingfull}; one first needs to drop terms that are not $\propto \cp$ or $\propto \cm$, which amounts to dropping terms $\propto d_{18},d_{19}$.
Further, the contributions to $\cp$ and $\cm$ only arise in the UV, and thus $c_{u\pm d} = (c_u^0 \pm c_d^0)/2 $. 

In summary, our findings show a distinct momentum dependence of the axion-nucleon coupling at higher orders in the chiral expansion, which is relevant for phenomenological applications as discussed in \Sec{sec:implications}.

\subsection{Axion-nucleon couplings at finite density} \label{sec:finite_density_couplings}
In the following, we systematically calculate the leading order corrections to the axion-nucleon coupling induced by a background nucleon density. 
Within the real-time formalism of thermal field theory, the effects of finite nucleon densities are captured by the nucleon propagator, which is extended by a term depending on the nucleon chemical potential $\mu$ and the temperature $T$~\cite{Niemi:1983ea, Niemi:1983nf, Landsman:1986uw}. For a detailed discussion, we refer to \App{app:ratedensity} and the associated references. 
Note that an initial estimate for the effective finite density modification of the axion-nucleon coupling has already been performed in \cite{Balkin:2020dsr}. 
Recently, these results have been used to show that density corrections do not spoil suppressed axion-nucleon couplings in tuned models \cite{DiLuzio:2024vzg}.

In the non-relativistic limit, as $T \rightarrow 0$, the finite density nucleon propagator in HBChPT takes the following form:

\begin{equation} \label{eq:HBChPT_propagator_isopin_Asymetric_T0}
    i G(k) = \frac{i}{v \cdot k + i \epsilon} - 2 \pi \delta(v \cdot k) \left[ \frac{1+\tau^{3}}{2} \theta(k^p_{f}-|\vec{k}|)+\frac{1-\tau^{3}}{2} \theta(k^n_{f}-|\vec{k}|) \right],
\end{equation}
where $k^\mu$ is the residual momentum as defined in Eq.~\eqref{eq:splitng_momenta}.
Here
\begin{equation}\label{eq:fermimomenta}
    k^p_f \equiv ( 3 \pi^2 n_p )^{1/3} \qquad \text{and}\qquad k^n_f \equiv ( 3 \pi^2 n_n )^{1/3}
\end{equation} 
are the proton and neutron Fermi momenta and $n_n$ and $n_p$ denote their respective number densities.
Note that the propagator modification due to density dependence in the $T\to0$ limit can also be derived without using thermal field theory, see App. A in \Refcite{Ghosh:2022nzo}.

The axion nucleon form factor at finite density is calculated, similarly to before, by evaluating loop diagrams using the above propagator.
In fact, the first term of \Eq{eq:HBChPT_propagator_isopin_Asymetric_T0} is the zero density heavy baryon propagator, which, if chosen for every internal nucleon line, leads to the vacuum corrections of the coupling calculated in the previous section. 
The second term of the propagator accounts for the interaction of the nucleon with the background and is referred to as medium- or density-insertion.
Diagrammatically \Eq{eq:HBChPT_propagator_isopin_Asymetric_T0} reads
\begin{equation}
    i G(k) \quad = \quad \includegraphics[scale=.85,valign=c]{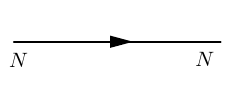} \quad + \quad \includegraphics[scale=.85,valign=c]{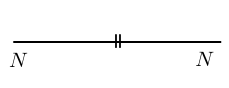},
\end{equation}
where follow the convention to denote the density insertion by the short double-stroke. 
In the isospin symmetric case, where $n/2 \equiv n_n = n_p$ with $k_f = ( 3 \pi^2 n/2 )^{1/3}$, \Eq{eq:HBChPT_propagator_isopin_Asymetric_T0} simplifies to
\begin{equation} \label{eq:HBChPT_propagator_isopin_Symetric_T0}
    i G(k) = \frac{i}{v\cdot k + i \epsilon} - 2 \pi \delta(v\cdot k) \theta(k_f - |\vec{k}|).
\end{equation}
In the following we discuss the vertex corrections to the axion-nucleon coupling that arise due to the medium insertion.

\begin{figure}
    \centering
    \includegraphics[scale=1]{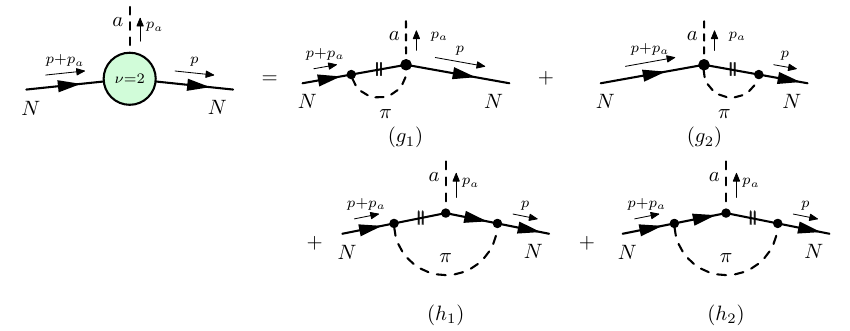}
    \caption{Corrections to the axion nucleon vertex at finite density at chiral order $\nu=2$. Filled dots denote $\Delta=0$ vertices according to \Eq{eq:pc}. Two lines on the nucleon propagator mean that the vacuum nucleon propagator has been replaced by the density insertion.}
    \label{fig:Nu2Dens}
\end{figure}

\subsubsection{Axion-nucleon vertex corrections} \label{sec:axion_vertex_at_finite_density}

Density effects start to contribute at chiral order $\nu=2$. As in the previous section we show diagrams that are non-zero in \Fig{fig:Nu2Dens}, while the explicit calculation is performed in \App{app:density_loops}.
Using the expressions defined in \App{ap:int_form}, the full result at this order reads
\begin{equation}
\label{eq:nu2Res}
\begin{aligned}
& \includegraphics[scale=1,valign=c]{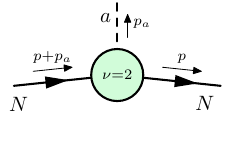} \quad \\ 
  = & - \left( \frac{g_A^2 c_{p}}{f_a f_\pi^2} \right) (3 ~\mathbb{1}-\tau^3) G_1^p(\vec{p},\vec{p}_a) + \left( \frac{\cm g_A}{f_a f_\pi^2} \right) (\mathbb{1}-\tau^3)  G_2^p(\vec{p},\vec{p}_a) \\
   &- \left( \frac{g_A^2 c_{n}}{f_a f_\pi^2} \right) (3 ~\mathbb{1}+\tau^3) G_1^n(\vec{p},\vec{p}_a) - \left( \frac{\cm g_A}{f_a f_\pi^2} \right) (\mathbb{1}+\tau^3)  G_2^n(\vec{p},\vec{p}_a),
\end{aligned}
\end{equation}
where $c_{n/p}$ are defined in \Eq{eq:tree_axion_couplings}, and $\tilde{m}_\pi^2=m_\pi^2-(p_a^0)^2$.
\\

\begin{figure}
    \centering
    \includegraphics[scale=1]{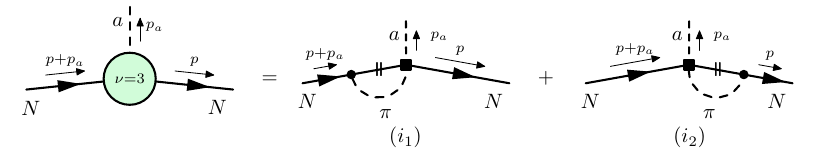}
    \caption{Dominant (\ie $\Delta$-enhanced) corrections to the axion nucleon vertex at finite density at chiral order $\nu=3$. Filled squares denote $\Delta=1$, filled circles denote $\Delta=0$ vertices according to \Eq{eq:pc}.}
    \label{fig:Nu3Dens}
\end{figure}

At the next higher order, \ie $\nu=3$, we again restrict ourselves to the diagrams proportional to the $\Delta(1232)$-enhanced coupling constants $\hat{c}_{3/4}$.
The corresponding diagrams are shown in figure \Fig{fig:Nu3Dens}.
Following the computation in \ref{app:density_loops}, we find 
\begin{equation}
\label{eq:nu3Res}
\begin{aligned}
  & \includegraphics[scale=1,valign=c]{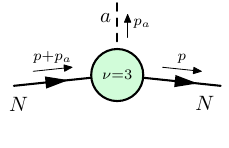} \\
=&  \left( \frac{\ch4 \cm g_A}{f_a f_\pi^2} \right) (\tau^3-\mathbb{1}) G_3^p(\vec{p},\vec{p}_a) + \left( \frac{\ch3 \cm g_A}{f_a f_\pi^2} \right) (\tau^3+\mathbb{1}) G_4^p(\vec{p},\vec{p}_a) \\
+ & \left( \frac{\ch4 \cm g_A}{f_a f_\pi^2} \right) (\tau^3+\mathbb{1}) G_3^n(\vec{p},\vec{p}_a) + \left( \frac{\ch3 \cm g_A}{f_a f_\pi^2} \right) (\tau^3-\mathbb{1}) G_4^n(\vec{p},\vec{p}_a). 
\end{aligned}
\end{equation}

The full result for the axion-nucleon vertex at finite density is then given by summing vacuum and density contributions denoted by gray and green blobs, respectively,
\begin{equation} \label{eq:full_vertex_result}
\begin{aligned}
    \includegraphics[scale=1,valign=c]{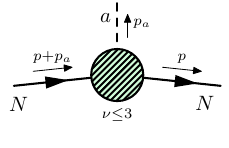} \; =&  \; \includegraphics[scale=1,valign=c]{figures/FullVacSymb.pdf} \; + \; \includegraphics[scale=1,valign=c]{figures/Nu2DensSymb.pdf} \\
    & + \; \includegraphics[scale=1,valign=c]{figures/Nu3DensSymb.pdf}.
\end{aligned}
\end{equation}

In order to gain some intuition of the parametric dependence of these expressions, we take a simplifying limit of \Eq{eq:nu2Res} and \Eq{eq:nu3Res}, namely $|\vec{p}\,|\to {k}_f$, $k_f^n=k_f^p=k_f$, $m_{\pi}/k_f\to 0$ and keep only the leading order in $p_a^0$. 
The result then simplifies to
\small
\begin{align} \label{eq:simplified}
    & \includegraphics[scale=1,valign=c]{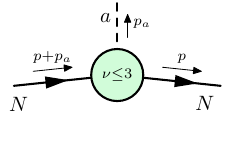} \,\,  = - \left( -\frac{
    g_A^2 
    \hat{c}_N
    }{9}\frac{k_f^2}{\left(4\pi f_\pi\right)^2} + \frac{16 g_A \cm \tau^3 }{9}\frac{k_f^3 (\hat{c}_3 -5 \hat{c}_4) }{ \left(4 \pi  f_\pi\right)^2} \right) \frac{S \cdot p_a}{f_a} \nonumber \\
    & \quad - \left( -
    \frac{g_A\left(17 g_A \hat{c}_N+108\cm\tau^3\right)}{27}
    \frac{k_f p_a^0}{ \left(4\pi f_\pi\right)^2}+ \frac{16 g_A \cm \tau^3 }{9}\frac{k_f^2 p_a^0( \hat{c}_3 + \hat{c}_4
   )}{\left(4\pi f_\pi\right)^2} \right)  \frac{S \cdot p}{f_a}+\ldots\,.
\end{align}
\normalsize
Except for very low densities, $n \ll n_0$, this simplified expression describes the result shown in \Fig{fig:axion_nucleon_coupling_KSVZ_sym_matter} quite well.

In the limit where tensor-like terms are neglected, $(\vec{k}\cdot \vec{\sigma}) \vec{k} \sim 1/3 k^2 \vec{\sigma} $, the result includes the terms of the chiral two-body current calculations in heavy nuclei, see \cite{Menendez:2011qq,Gysbers:2019uyb}. However, in Eq.~\eqref{eq:simplified} there are additional terms that in the formalism of \cite{Menendez:2011qq,Gysbers:2019uyb} arise from solving the Lippmann-Schwinger equation. Note also that we only keep terms consistent with our power counting, \ie we neglect terms $\propto \hat{c}_D$, relativistic corrections and terms that are kinematically suppressed.

\subsubsection{Results} \label{sec:results}

Finally, in analogy to \Eq{eq:vacuumcouplingfull}, we collect our results in terms of form factors, now including density effects, as
\begin{equation} \label{eq:vertex_result}
   \includegraphics[scale=1,valign=c]{figures/FullVertexSymb.pdf} \,\,= - \mathcal{A}(p, k_f^{p/n}, p_a) \, \frac{S\cdot p_a}{f_a} - \mathcal{B}(p, k_f^{p/n}, p_a) \, \frac{S\cdot p}{f_a},
\end{equation}
where the density dependent form factors $\mathcal{A}$ and $\mathcal{B}$ are again diagonal matrices in isospin space in analogy to \Eq{eq:def_A_B}.
In order to compare the two contributions it is useful to rescale the second form factor to $\,\mathcal{B}|\vec{p}\,|/|\Vec{p}_a|$, since to leading order $\mathcal{B}\sim p_a$.
In this case, they multiply a spin structure of similar size, \ie
\begin{equation}
|\Vec{p}_a|\left(\mathcal{A} \,S\cdot \hat{p}_a+ \frac{|\Vec{p}\,|}{|\Vec{p}_a|} \mathcal{B}\, S \cdot \hat{p}\right).
\end{equation}
We proceed by making motivated kinematic
approximations in order to plot the above results.
At low temperatures, nucleons are expected to have momenta close to the Fermi momentum $|\vec{p}\,| \sim k_f$.
Furthermore, the momentum of the axion is small compared to the momenta of the nucleons, \ie $|\vec{p}_a| \ll k_f \sim p$, and we only keep the leading order which is independent of $|\vec{p}_a|$. 
Additionally, we average over the direction of the outgoing axion $\vec{p}_a\cdot \vec{p}\sim k_f |\vec{p}_a|/3 $.

\begin{figure}[h] 
  \centering
  \begin{subfigure}{.49\textwidth}
  \centering
  % include first image
\includegraphics[width=1.\textwidth]{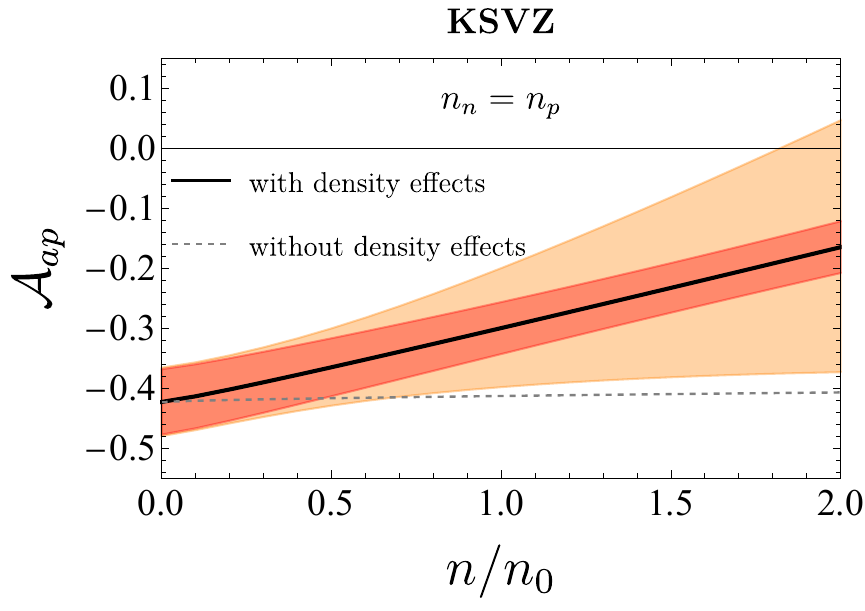}
\end{subfigure}
  \begin{subfigure}{.49\textwidth}
  \centering
  % include second image
\includegraphics[width=1.\textwidth]{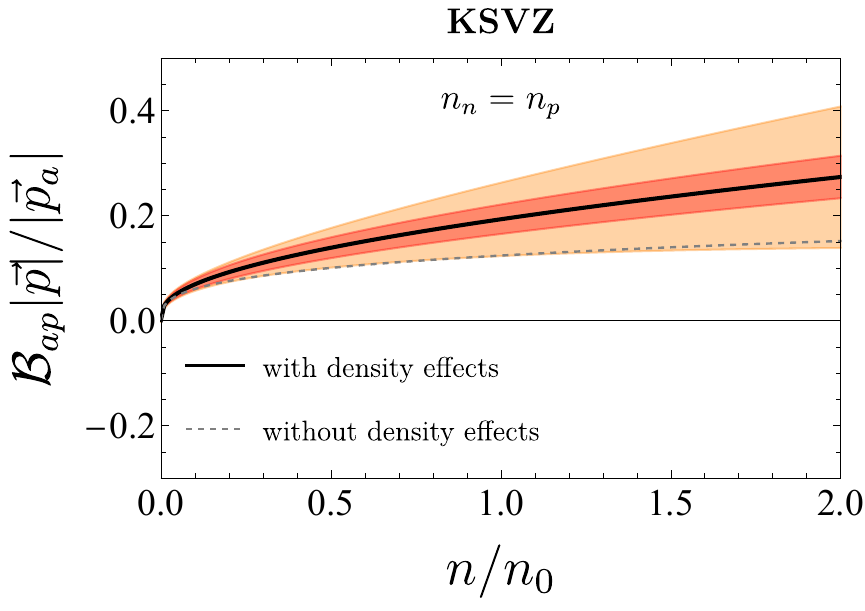}
\end{subfigure}
\begin{subfigure}{.49\textwidth}
  \centering
  % include first image
\includegraphics[width=1.\textwidth]{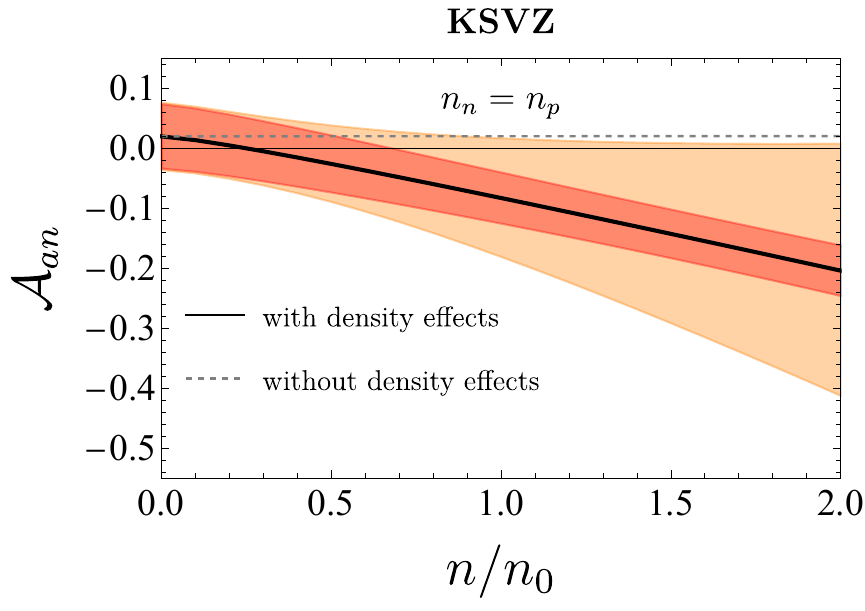}
\end{subfigure}
  \begin{subfigure}{.49\textwidth}
  \centering
  % include second image
\includegraphics[width=1.\textwidth]{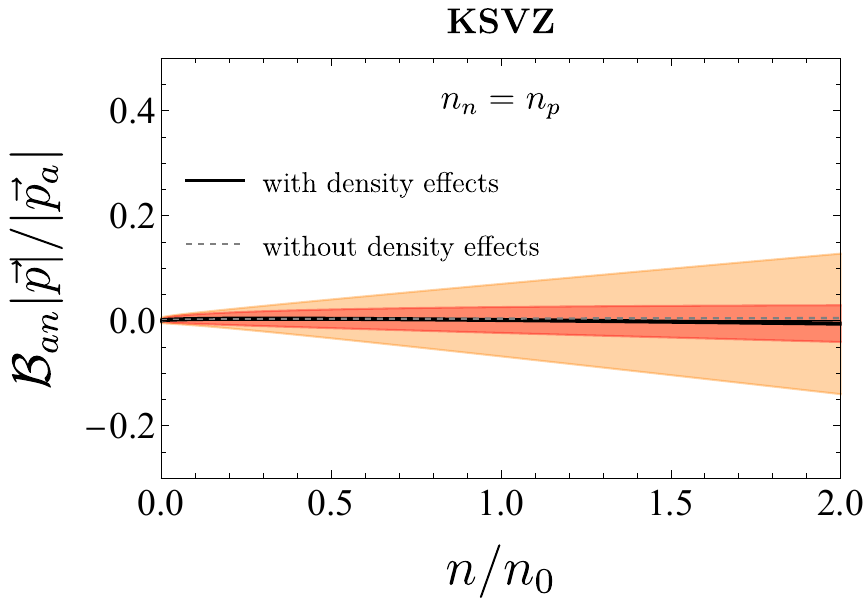}
\end{subfigure}
 \caption[]{Density dependence of the KSVZ axion-nucleon couplings in isospin symmetric nuclear matter as a function of number density in units of nuclear saturation density $n_0$. Error bars are shown as in \Fig{fig:axion_nucleon_coupling_KSVZ_vac}. The gray dashed line shows the vacuum results, see \Eq{eq:vertex_result}, \ie it neglects all density effects and only shows the effects of the momentum dependence evaluated at the corresponding Fermi momentum.}  \label{fig:axion_nucleon_coupling_KSVZ_sym_matter}
\end{figure}

\begin{figure}[h] 
  \centering
  \begin{subfigure}{.49\textwidth}
  \centering
  % include first image
\includegraphics[width=1.\textwidth]{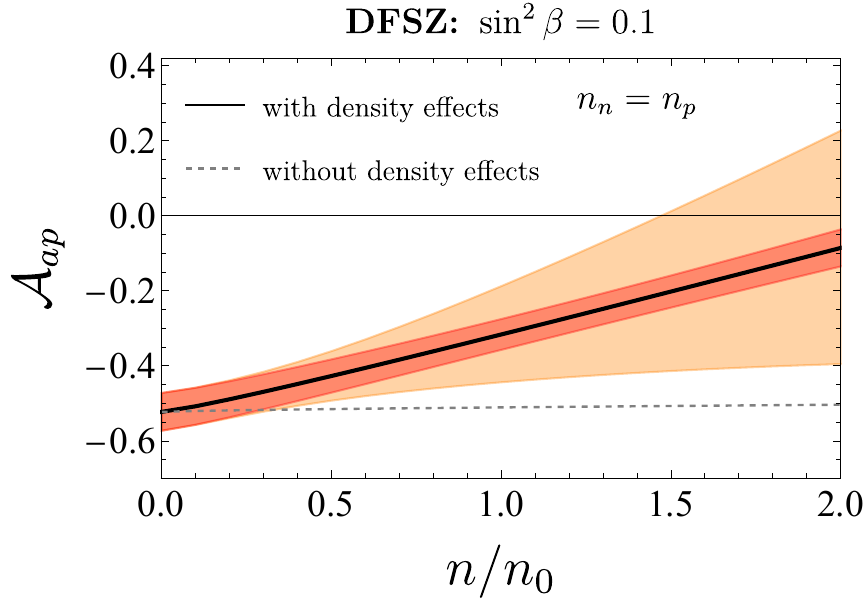}
\end{subfigure}
  \begin{subfigure}{.49\textwidth}
  \centering
  % include second image
\includegraphics[width=1.\textwidth]{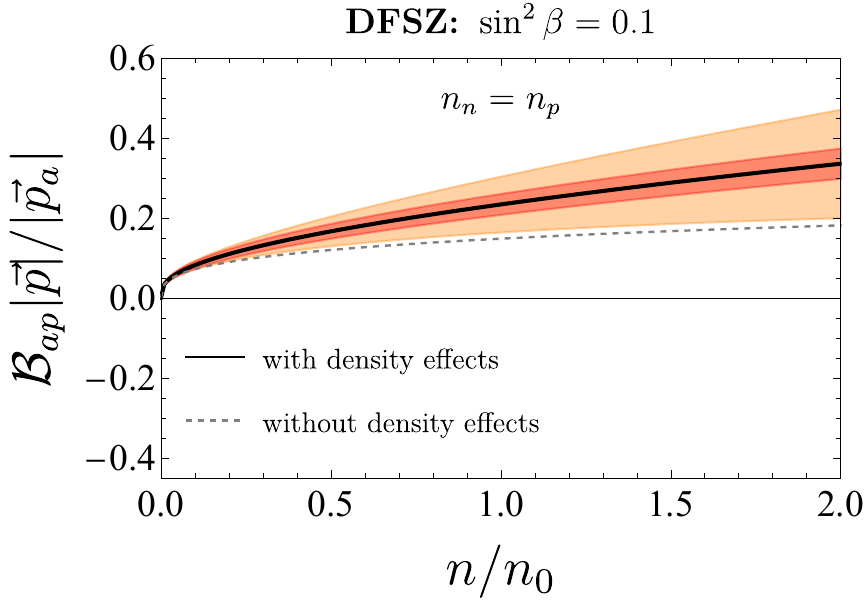}
\end{subfigure}
\begin{subfigure}{.49\textwidth}
  \centering
  % include first image
\includegraphics[width=1.\textwidth]{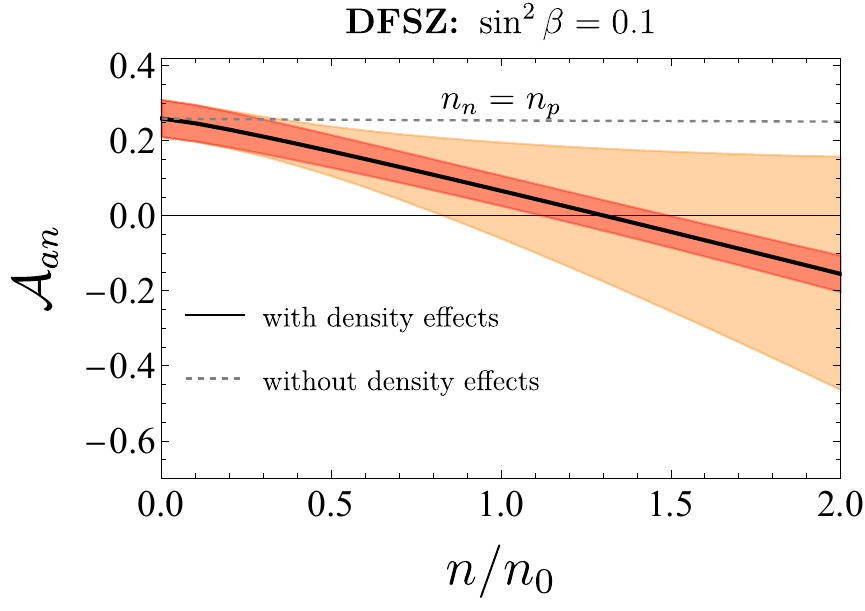}
\end{subfigure}
  \begin{subfigure}{.49\textwidth}
  \centering
  % include second image
\includegraphics[width=1.\textwidth]{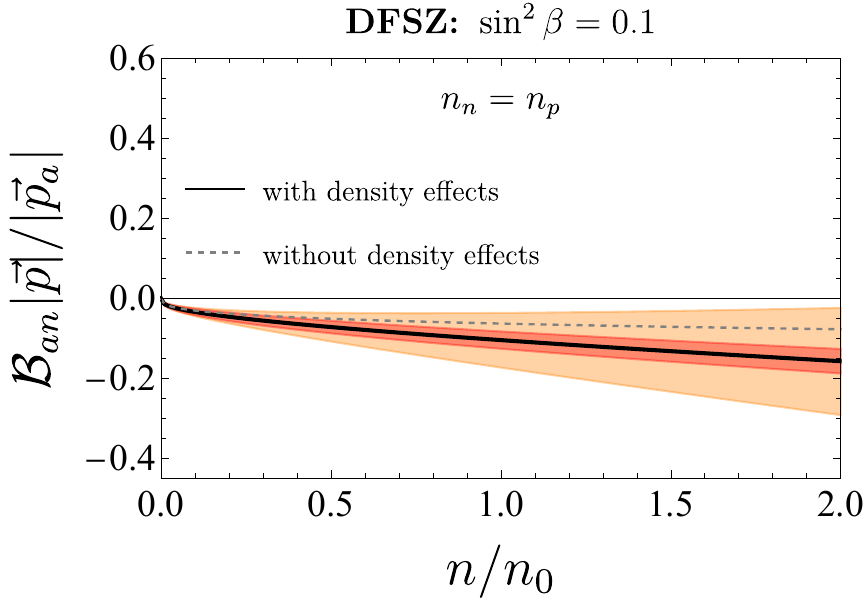}
\end{subfigure}
 \caption[]{Density dependence of the DFSZ axion-nucleon couplings with $\sin^2 \beta = 0.1$ in symmetric nuclear matter as a function of density in units of nuclear saturation density $n_0$. Error bars are shown as in \Fig{fig:axion_nucleon_coupling_KSVZ_vac}. 
 The gray dashed line shows the vacuum results (see \Eq{eq:vertex_result}).} 
 \label{fig:axion_nucleon_coupling_DFSZ_0_sym_matter}
\end{figure}

Note that typically $\mathcal{B}$ is dominated by the vacuum contribution, while $\mathcal{A}$ is dominated by $\nu=3$ density loops. The vacuum contribution to $\mathcal{B}$ is $\propto p_a$, which after rescaling results in $\mathcal{B}\propto n^{1/3}$, as shown in the right panels of Figs.~\ref{fig:axion_nucleon_coupling_KSVZ_sym_matter} and \ref{fig:axion_nucleon_coupling_DFSZ_0_sym_matter}.

Let us estimate the error coming from truncating the ChPT expansion. 
The first $\Delta(1232)$-enhanced diagrams appear at chiral order $\nu=3$, and we expect that higher order terms can be enhanced as well. 
We can estimate them, assuming that the next order has a similar numerical prefactor as the calculated diagrams but is suppressed by an additional factor of $k_f/m_N$ (or similarly $k_f/\Lambda_\chi$). As in \Sec{sec:vac_results}, we add an uncertainty to the result, which we take to be the $\nu =3$ result multiplied by the above suppression factor.

In the following, we show the density dependence of the axion-nucleon coupling, based on the systematic calculations within the validity of the ChPT expansion from the previous sections, for two benchmark models, the KSVZ and the DFSZ axion.

\subsubsection*{KSVZ axion}
\begin{figure}[h] 
  \centering
  \begin{subfigure}{.49\textwidth}
  \centering
\includegraphics[width=1.\textwidth]{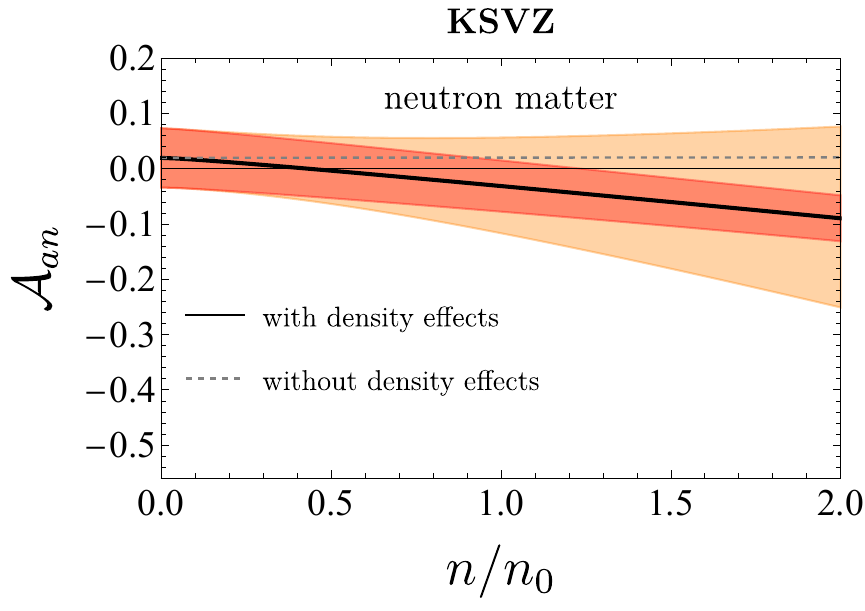}
\end{subfigure}
  \begin{subfigure}{.49\textwidth}
  \centering
\includegraphics[width=1.\textwidth]{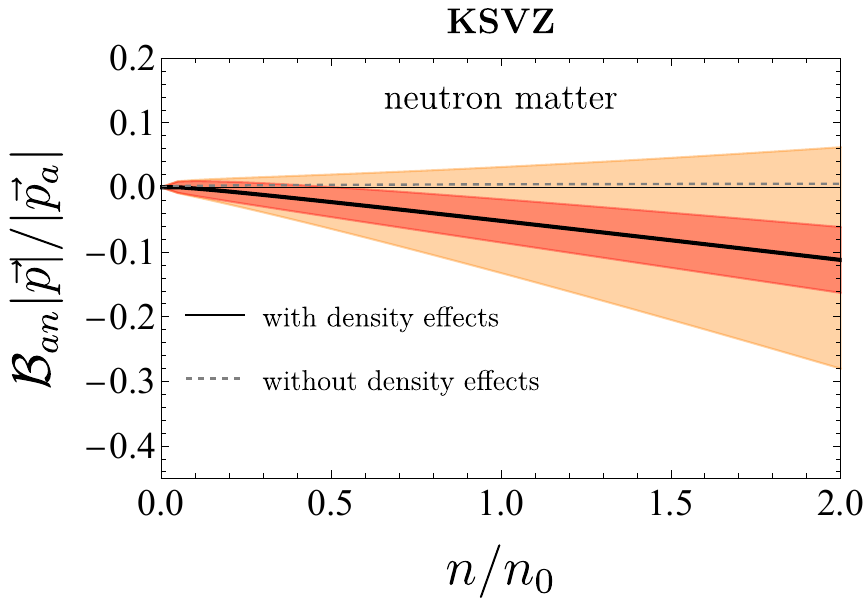}
\end{subfigure}
\begin{subfigure}{.49\textwidth}
  \centering
\includegraphics[width=1.\textwidth]{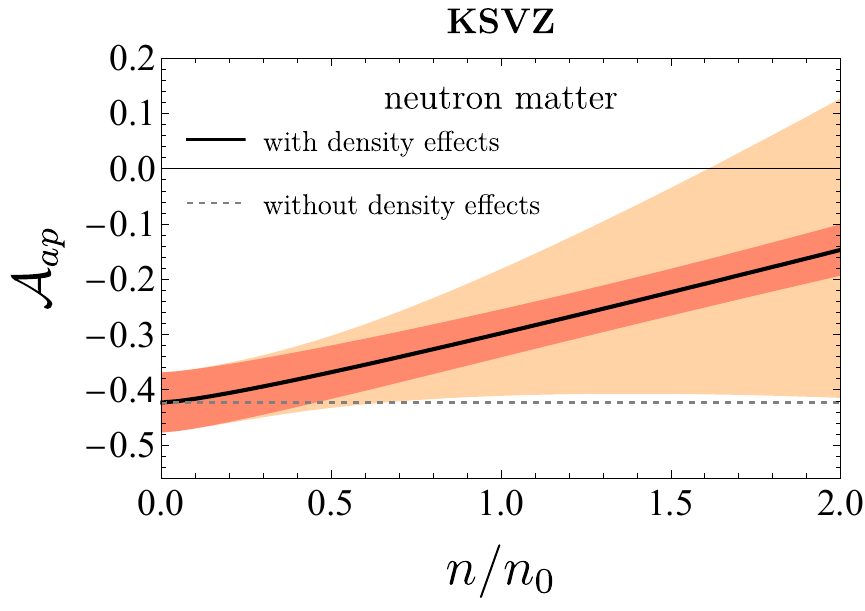}
\end{subfigure}
 \caption[]{Density dependence of the KSVZ axion-nucleon couplings in pure neutron matter as a function of density in units of nuclear saturation density $n_0$. Error bars are shown as in \Fig{fig:axion_nucleon_coupling_KSVZ_vac}. 
 Note that for pure neutron matter, the vacuum contribution to $\mathcal{B}_{an}$ is small (see \Eq{eq:vacuumcouplingfull}), and the dominant contribution comes from the $\nu=3$ density diagram $\propto \ch3$. 
 Therefore, we do not find the $\propto n^{1/3}$, but instead a $\propto n$ scaling.
 The gray dashed line shows the vacuum results (see \Eq{eq:vertex_result}). 
 Note that $\mathcal{B}_{ap}$ is zero in the limit of $|\vec{p}| \sim k_f^p \sim 0$ and therefore suppressed.} \label{fig:axion_nucleon_coupling_KSVZ_neu_matter}
\end{figure}

For the KSVZ axion, ($c_q^0= 0$), we find that the density corrections are crucial: the accidental cancellation in the axion neutron coupling, \Eq{eq:axioncouplingswithoutpions}, is eliminated by finite density effects. 
We show our results of the axion-nucleon coupling in isospin symmetric matter, $(n_p=n_n=n/2)$, in \Fig{fig:axion_nucleon_coupling_KSVZ_sym_matter}.
Around nuclear saturation density \hbox{$n = n_0 \simeq 0.16 \, \mathrm{fm}^{-3}$} to leading order in $p_a$, we find
\begin{subequations} \label{eq:axionnuclKSVZsaturation}
    \begin{align}
        \mathcal{A}_{ap}^{\text{\tiny KSVZ}}(n_0)=-0.299(43)(98),\quad & \frac{\left\lvert\Vec{p}\right\rvert}{\left\lvert\Vec{p}_a\right\rvert}\mathcal{B}_{ap}^{\text{\tiny KSVZ}}(n_0)=+0.193(28)(70),\\
         \mathcal{A}_{an}^{\text{\tiny KSVZ}}(n_0)=-0.083(43)(98),\quad & \frac{\left\lvert\Vec{p}\right\rvert}{\left\lvert\Vec{p}_a\right\rvert}\mathcal{B}_{an}^{\text{\tiny KSVZ}}(n_0)=0.001(25)(69),
    \end{align}
\end{subequations}
where the first error is due to the uncertainty of the low energy constants, while the second error is the total error estimate, including both the uncertainty of the ChPT expansion as well as the uncertainty of low energy constants.
In \Fig{fig:axion_nucleon_coupling_KSVZ_neu_matter}, we show the results for pure neutron matter ($n_p=0$ and $n_n=n$). 
At nuclear saturation density, we find the axion-nucleon couplings for pure neutron matter to be 
\begin{align}
 \mathcal{A}_{an}^{\text{\tiny KSVZ}}(n_0) & =-0.031(47)(86),\quad  \frac{\left\lvert\Vec{p}\right\rvert}{\left\lvert\Vec{p}_a\right\rvert}\mathcal{B}_{an}^{\text{\tiny KSVZ}}(n_0)=-0.051(33)(82), \\
 \mathcal{A}_{ap}^{\text{\tiny KSVZ}}(n_0) & =-0.298(43)(115).
\end{align}
At large densities, the dominant error comes from truncation of the ChPT series, while at low densities, it comes from uncertainties in the low energy constants. 

Note that the coupling to a proton in pure neutron matter is similarly modified by $\mathcal{O}(1)$ as shown in \Fig{fig:axion_nucleon_coupling_KSVZ_neu_matter}. 
This is because the neutron density alone can modify the proton vertex by contributing as an internal line to the density loop. 
Hence in a small proton but large neutron density, corrections to the proton vertex are important as well.

\subsubsection*{DFSZ axion}

Next, we consider the DFSZ axion model with $c_u^0 = \sin^2{\beta} / 3 $, $c_d^0 = 1- c_u^0$, where $\beta$ is a parameter in the IR theory that is related to the vacuum expectation values of the two Higgs fields in the UV theory. 
Here, we find $\mathcal{O}(1)$ changes in the couplings, which we show a benchmark point $\sin^2 \beta = 0.1$ for symmetric nuclear matter in \Fig{fig:axion_nucleon_coupling_DFSZ_0_sym_matter}.
In isospin symmetric matter, at nuclear saturation density, the axion-nucleon couplings $\mathcal{A}$ with error estimates are given by
\begin{subequations}
    \begin{align}
        &\mathcal{A}_{ap}^{\text{\tiny DFSZ}}(n_0)=-0.339(43)(98) + 0.208(22)(62)\sin^2\beta,\\
        &\mathcal{A}_{an}^{\text{\tiny DFSZ}}(n_0)=-0.087(43)(98) - 0.208(22)(62)\sin^2\beta,
    \end{align}
\end{subequations}
while we find the coupling $\mathcal{B}$ to be
\begin{subequations}
    \begin{align}
        &\frac{\left\lvert\Vec{p}\right\rvert}{\left\lvert\Vec{p}_a\right\rvert}\mathcal{B}_{ap}^{\text{\tiny DFSZ}}(n_0)=+0.253(27)(70) - 0.184(13)(60)\sin^2\beta,\\
        &\frac{\left\lvert\Vec{p}\right\rvert}{\left\lvert\Vec{p}_a\right\rvert}\mathcal{B}_{an}^{\text{\tiny DFSZ}}(n_0)=-0.123(22)(68) + 0.184(13)(60)\sin^2\beta.
    \end{align}
\end{subequations}
We show the results for another benchmark point, $\sin^2\beta=0.5$, as well as both benchmark points in pure neutron matter in \App{ap:DFSZresults}.

\FloatBarrier

\section{Implications of density and momentum dependent couplings}\label{sec:implications}
We now investigate the influence of our results on two of the strongest bounds on QCD axions: supernova and neutron star cooling bounds. 
We then comment on how these effects also influence terrestrial experiments seeking to detect a QCD axion as well as an ALP coupling to nucleons. 

\subsection{Supernova cooling bound}\label{sec:supernova}

\begin{figure}[t] 
  \centering
  \begin{subfigure}{.301\textwidth}
  \centering
      \includegraphics[width=.95\textwidth]{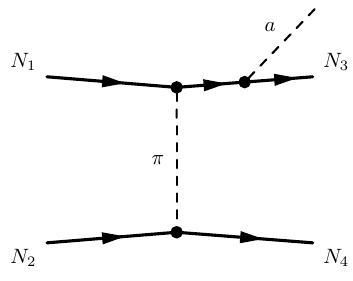}
      \caption{}
  \end{subfigure}
   \raisebox{14.5\height}{$\mathbf{\longrightarrow}$}
  \begin{subfigure}{.301\textwidth}
  \centering
      \includegraphics[width=0.95\textwidth]{figures/OPE_loop_wo_label.pdf}
      \caption{}
  \end{subfigure}
  \raisebox{9.2\height}{$+$}
  \begin{subfigure}{.301\textwidth}
  \centering
      \includegraphics[width=0.95\textwidth]{figures/CT_CS_ax_loop.pdf}
      \caption{}
  \end{subfigure}
 \caption[]{(a) Leading order OPE diagram. (b) OPE diagram with axion nucleon vertex corrections from finite and zero density. (c) Nucleon contact interaction with axion nucleon vertex corrections from finite and zero density.}
  \label{fig:OPE}
\end{figure}

The observed neutrinos during SN 1987A \cite{Hirata:1987hu, Bionta:1987qt} impose stringent upper limits on axion couplings, as axions escaping the SN would affect the neutrino burst \cite{Raffelt:1996wa}.
These limits are among the leading constraints on the axion-nucleon coupling and, thus, the axion decay constant.
In this work, we are only considering the case of free streaming axions and are not interested in the trapping regime.
The dominant contribution to the axion luminosity comes from the LO tree-level one-pion-exchange (OPE) diagram, shown in the left panel of \Fig{fig:OPE}, and was first calculated in \cite{Iwamoto:1984ir, Brinkmann:1988vi}.
In addition to the LO contribution, several partial corrections have recently been included in a phenomenological manner, see \eg~\cite{Carenza:2019pxu,Chang:2018rso}.
However, an evaluation using a systematic expansion, including an estimation of the errors of the axion emissivity, is still missing.

\begin{figure}[h] 
  \centering
%%%
  \begin{subfigure}{.24\textwidth}
  \centering
      \includegraphics[width=1\textwidth]{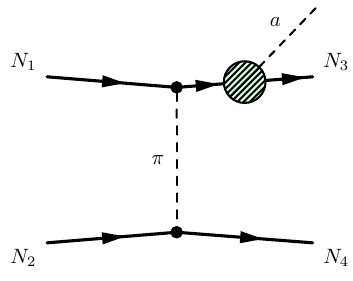}
      \caption{}
      \label{fig:a}
  \end{subfigure}
    \begin{subfigure}{.24\textwidth}
  \centering
      \includegraphics[width=1\textwidth]{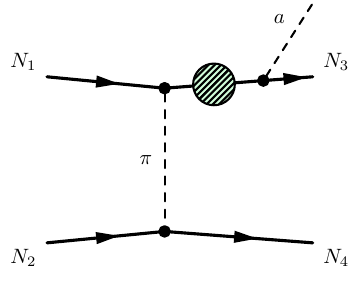}
      \caption{}
       \label{fig:b}
  \end{subfigure}
  \begin{subfigure}{.24\textwidth}
  \centering
      \includegraphics[width=1\textwidth]{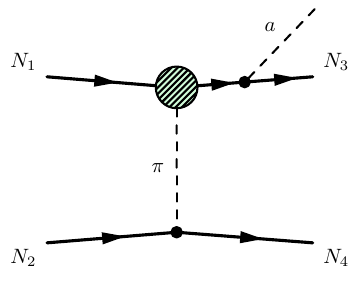}
      \caption{}
      \label{fig:c}
  \end{subfigure}
  \begin{subfigure}{.24\textwidth}
  \centering
      \includegraphics[width=1\textwidth]{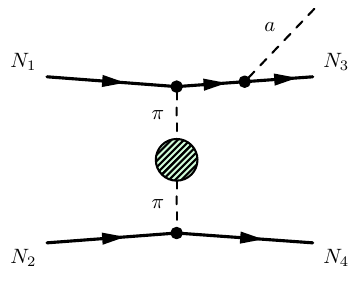}
      \caption{}
      \label{fig:d}
  \end{subfigure}
%%%
  \begin{subfigure}{.24\textwidth}
  \centering
      \includegraphics[width=1\textwidth]{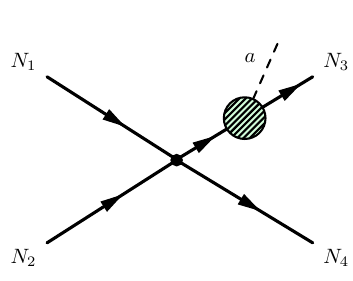}
      \caption{}
  \end{subfigure}
   \begin{subfigure}{.24\textwidth}
  \centering
      \includegraphics[width=1\textwidth]{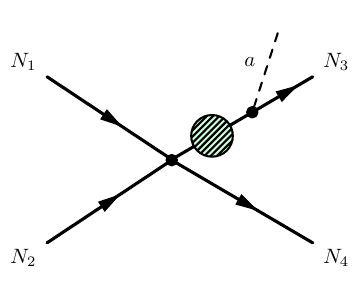}
      \caption{}
      \label{fig:e}
  \end{subfigure}
  \begin{subfigure}{.24\textwidth}
  \centering
      \includegraphics[width=1\textwidth]{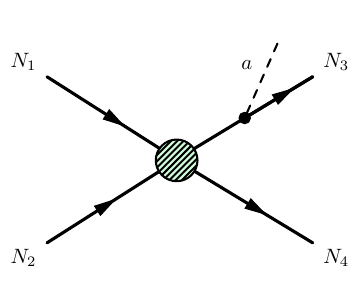}
      \caption{}
      \label{fig:f}
  \end{subfigure}
    \begin{subfigure}{.24\textwidth}
  \centering
      \includegraphics[width=1\textwidth]{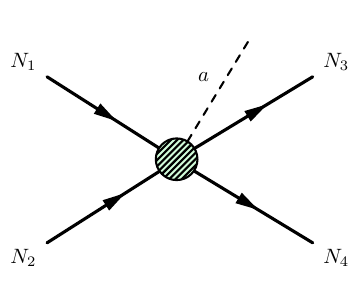}
      \caption{}
      \label{fig:g}
  \end{subfigure}
%%%
  \begin{subfigure}{.24\textwidth}
  \centering
      \includegraphics[width=1\textwidth]{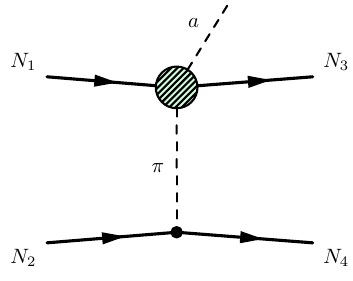}
      \caption{}
      \label{fig:h}
  \end{subfigure}
   \begin{subfigure}{.24\textwidth}
  \centering
      \includegraphics[width=1\textwidth]{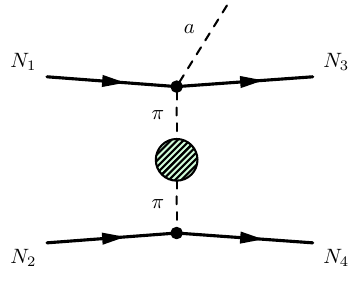}
      \caption{}
      \label{fig:j}
  \end{subfigure}
  \begin{subfigure}{.24\textwidth}
  \centering
      \includegraphics[width=1\textwidth]{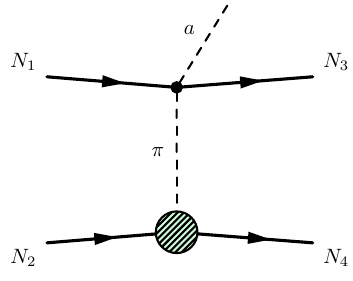}
      \caption{}
      \label{fig:k}
  \end{subfigure}
 \caption[]{Classes of diagrams contributing to the axion emission.
 Green blobs represent 1PI diagrams. See text for details. We do not show mixed contributions from multiple lower-order blobs in a single diagram.}
  \label{fig:alldiagramsSN}
\end{figure}

For a systematic treatment of the process $N N \to N N a$,
 it is crucial to consider all contributing diagrams up to a specified chiral order.
Although the chiral expansion is formally valid for small momenta, in typical SN environments the expansion parameter is non-negligible, $p/\Lambda \sim 1/3$. In particular, if there are accidental cancelations in the LO result, higher-order contributions can be of similar size.
For example, the NLO diagrams of type-\figref{fig:d}{d}, which are naively suppressed by $(p/\Lambda)^2$, are enhanced by an $\mathcal{O}(1)$ factor, such that they are only down by roughly $1/3$ compared to the leading order result.

We categorize all diagrams contributing to $N N \to N N a$ into distinct classes, see \Fig{fig:alldiagramsSN}.
Here, the green dashed blob represents all possible one-particle-irreducible (1PI) structures up to the chosen chiral order, including vacuum and finite density corrections.
For the nucleon-axion vertex, this blob was calculated in the previous sections.

Let us first investigate the matrix elements at leading order, \ie $\nu=0$.
In this case, the diagrams \figref{fig:a}{a}-\figref{fig:d}{d} reduce to the standard OPE diagram. 
Diagrams \figref{fig:e}{e}-\figref{fig:g}{g} at leading level come from nucleon contact terms, and have two contributions, one from $C_T$ and one from $C_S$, see \Eq{eq:nuclcontlag}. 
The contribution proportional to $C_S$ vanishes identically due to a pairwise cancellation of diagrams with the axion attached at different external legs. The low energy constant $C_T$ is numerically small and thus neglected. 
There are no diagrams contributing to \figref{fig:h}{h}-\figref{fig:k}{k} at leading order.

At chiral orders, $\nu\geq1$, the picture is more complicated, and a full evaluation of the axion emissivity is left for future work. 
Here, we take into account some of the diagrams (type \figref{fig:a}{a} and \figref{fig:e}{e}), and comment on some of the effects that will become important for a full evaluation at $\nu\leq 3$.

Diagrams of type \figref{fig:h}{h}-\figref{fig:k}{k} are suppressed by an additional power of $|\vec{p}_a|$, which in typical supernova environments is rather small compared to $|k_N|$ and of similar size as $\vec{k}_N^2/4\pi f_\pi$ corrections.
The $\nu=1$ diagrams of type (i) have been considered in \cite{Choi:2021ign}, with the result confirming our expectation of this additional suppression.

Diagrams of type \figref{fig:c}{c}, \figref{fig:d}{d}, \figref{fig:g}{g}, which modify the nucleon-nucleon interaction at high momentum and density, have been considered, \eg, in \cite{Bacca:2008yr, Lykasov:2008yz, Bartl:2014hoa, Bartl:2016iok}. For the axion supernova bound, they have been modeled by an effective $\rho$-exchange \cite{Carenza:2019pxu,Bottaro:2024ugp} or taken into account as a ChPT-inspired correction factor \cite{Chang:2018rso}.
These contributions seem to reduce the axion emissivity~\cite{Hanhart:2000ae,Schwenk:2003pj,Lykasov:2008yz,Bacca:2008yr} in the limit of $T\to 0$ and in the soft axion limit $\omega_a\to 0$. However, in supernova conditions high temperatures might spoil the suppression which we systematically explore in a forthcoming publication~\cite{springmann3}.

Diagrams of type \figref{fig:b}{b} and \figref{fig:f}{f} involve a resummed temperature and density-modified nucleon propagator. 
These corrections are interpreted as nucleon re-scatterings that can be sizable and have first been phenomenologically described in \cite{Raffelt:1991pw}, and subsequently used, \eg, in \cite{Raffelt:2006cw,Chang:2018rso,Carenza:2019pxu}. A systematic ChPT calculation of these contributions at finite density and temperature is still missing and is left for future work.

While we recognize the potential importance of all of these effects, our initial step towards a systematic calculation involves exclusively considering the effect that is separable into a change in the axion-nucleon coupling attached to a tree-level diagram, specifically types \figref{fig:a}{a} and \figref{fig:e}{e}. 
One must remember that this approach comes at the cost of losing a systematic expansion. However, one can combine our results with contributions arising from other topologies.

For diagram \figref{fig:e}{e}, higher-order corrections introduce a non-trivial dependence on the external nucleon momenta.
This lifts the cancellation that occurs at $\nu=0$ and hence the correction proportional to $C_S$ contributes $\mathcal{O}(1)$ to the $NN\to NNa$ process, while $C_T$ remains numerically subdominant.
Thus, when calculating the axion emissivity, we take into account the vertex-corrected OPE diagram as well as the vertex-corrected nucleon contact interaction, including all $\nu\leq2$ and $\Delta(1232)$-enhanced $\nu=3$ corrections as shown in \Fig{fig:OPE}.

\begin{figure}[b] 
  \centering
  \begin{subfigure}{.49\textwidth}
    % include second image
\includegraphics[width=1.\textwidth]{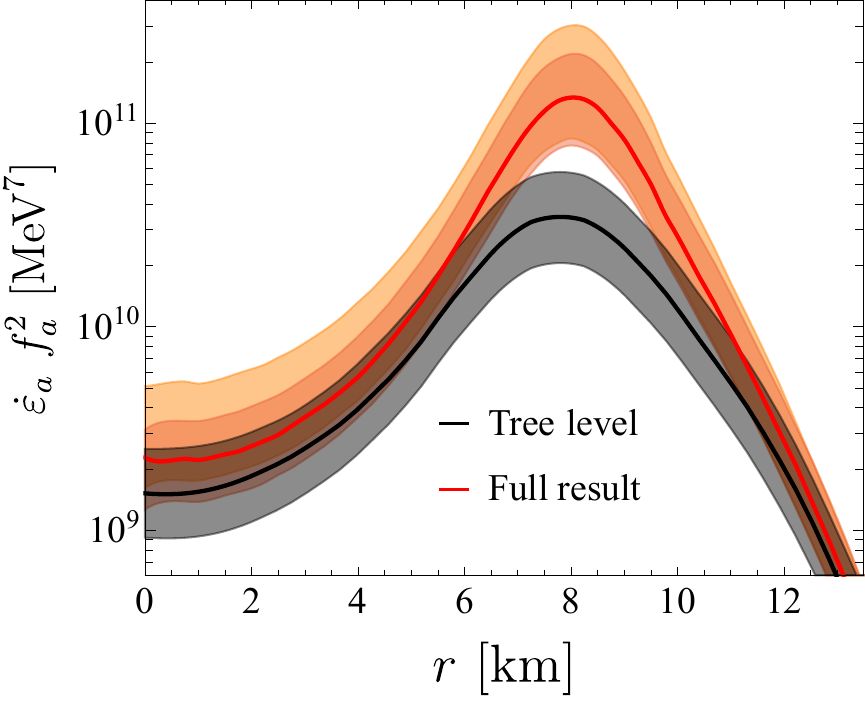}
\end{subfigure}
\begin{subfigure}{.49\textwidth}
    % include second image
\includegraphics[width=1.\textwidth]{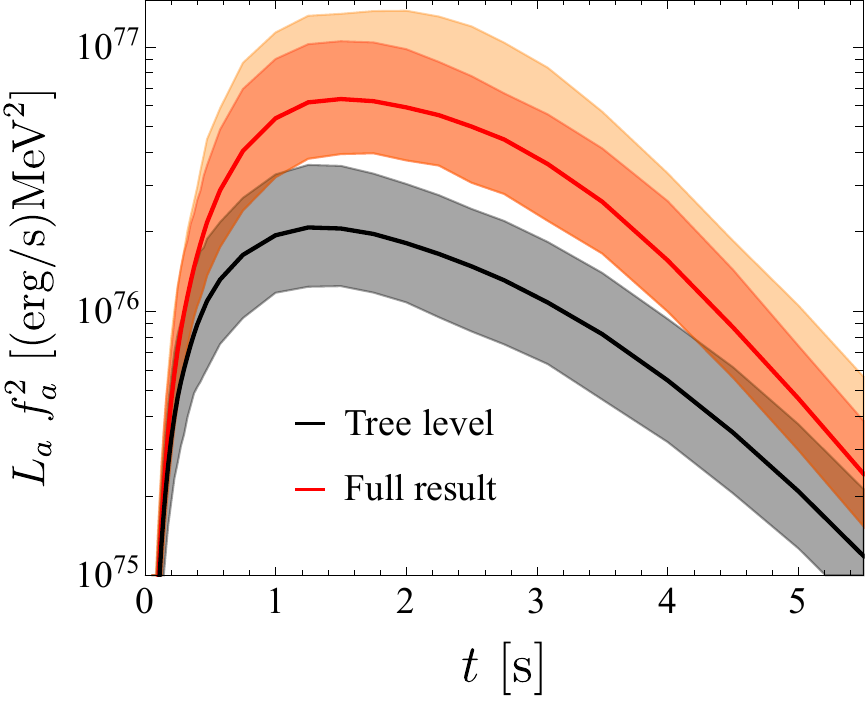}
\end{subfigure}
 \caption[]{Left: Axion emissivity over radius for the KSVZ axion evaluated at $t_{pb}=1\,\mathrm{s}$. Right: Axion luminosity over time integrated over the SN profile for the KSVZ axion. The shaded region represents the error bars resulting from the uncertainties of the constants given in, as well as the error emerging from higher-order terms in the chiral expansion.}
  \label{fig:axion_luminosity}
\end{figure}

The axion emissivity is given by
\begin{equation} \label{eq:emiss}
\dot{\varepsilon}_{a}=\int \prod_{i=1}^{4}d \Pi_{i} d \Pi_{a}(2 \pi)^{4} S |\mathcal{M}|^{2} \delta^{(4)}\left(\textstyle \sum_{i} p_i-p_{a}\right) E_{a} f_{1} f_{2}\left(1-f_{3}\right)\left(1-f_{4}\right),
\end{equation}
which describes the amount of energy emitted by axions per volume and time.
Here $f_i$ denotes the Fermi-Dirac distribution of nucleon $i$ and $d \Pi_i$ is the Lorentz invariant phase space measure. For details see \App{app:emissivity}.
The matrix element $\mathcal{M}$ is the sum of the modified OPE and nucleon contact interaction diagrams, \Fig{fig:OPE} (b) and (c). 

To perform the integration explicitly, specific details about the supernova (SN) environment are required. 
For the numerical evaluation of the emissivity, we utilize data from SN simulations from the Garching group, which include contributions from muons as outlined here~\cite{Garching}, based on \cite{Bollig:2020xdr}. 
Specifically, we use the $18.6\, M_\odot$ progenitor model with SFHo EOS.
Note that in the distribution functions $f_i$, we use the effective density-dependent nucleon mass provided by \cite{Garching}. A similar density-dependent nucleon mass also arises within ChPT, see \App{app:emissivity}.
In the left panel of \Fig{fig:axion_luminosity}, we show the emissivity \Eq{eq:emiss} for a KSVZ-type QCD axion evaluated at $t_{pb}=1\mathrm{s}$ post-bounce, (red) including axion-vertex corrections and (black) using only the leading order vacuum couplings.

\begin{figure}[t] 
\centering
  \begin{subfigure}{.55\textwidth}
  \vspace{-6mm}
  \centering
  % include first image
\includegraphics[width=1.\textwidth]{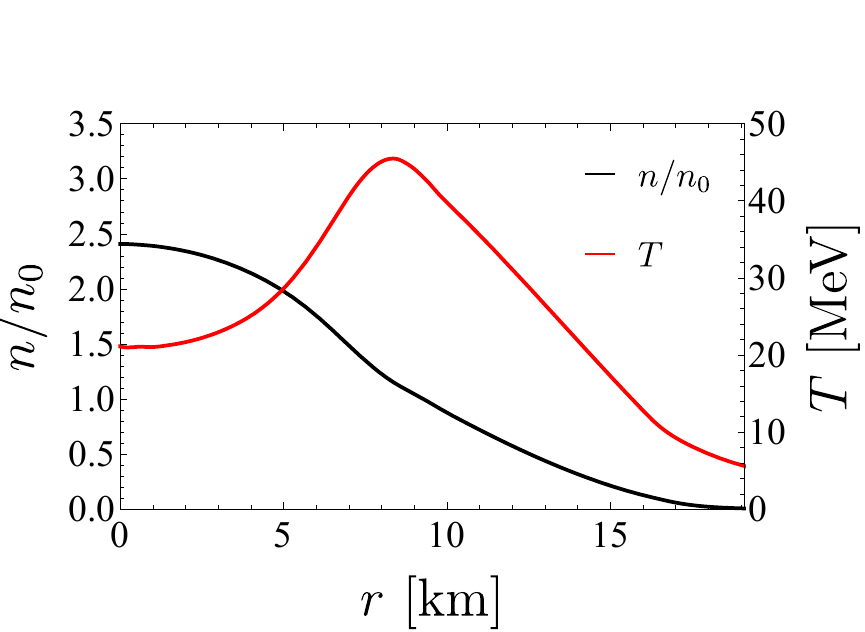}
\end{subfigure}
  \begin{subfigure}{.49\textwidth}
  \centering
  % include second image
\includegraphics[width=1.\textwidth]{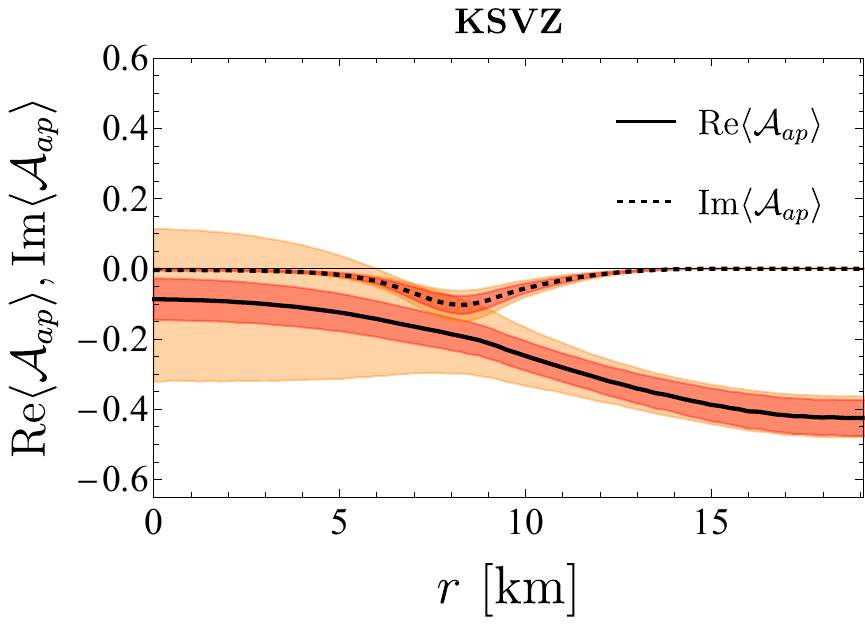}
\end{subfigure}
  \begin{subfigure}{.49\textwidth}
  \centering
  % include third image
\includegraphics[width=1.\textwidth]{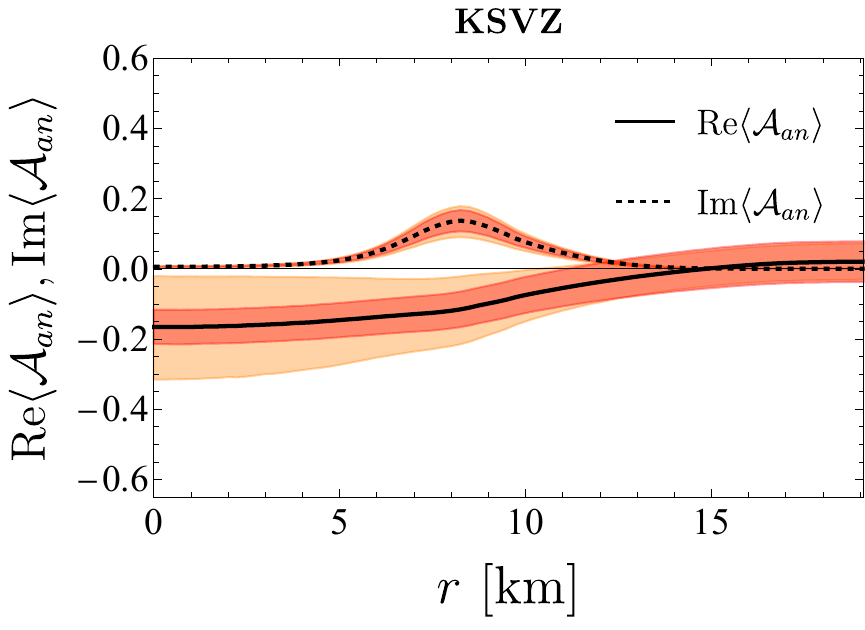}
\end{subfigure}
  \begin{subfigure}{.49\textwidth}
  \centering
  % include second image
\includegraphics[width=1.\textwidth]{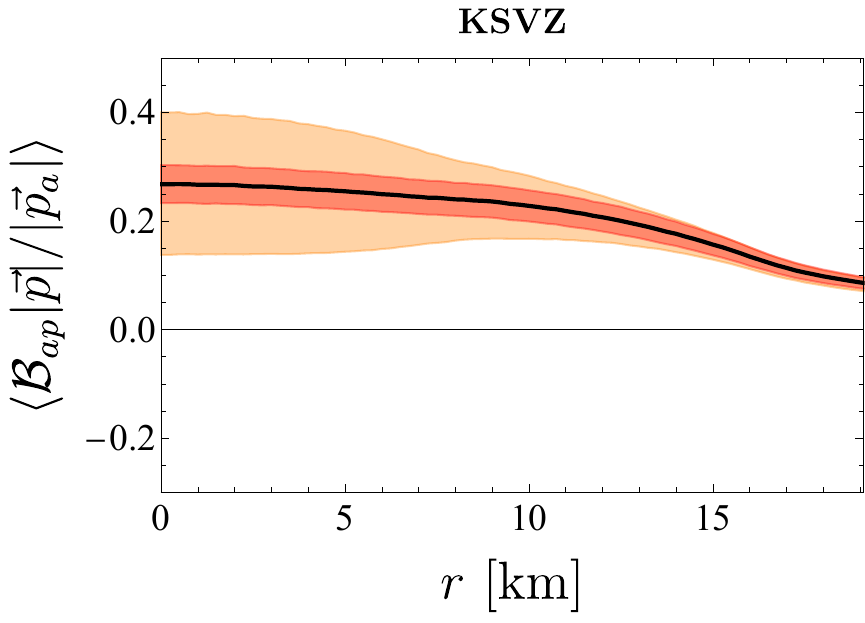}
\end{subfigure}
  \begin{subfigure}{.49\textwidth}
  \centering
  % include third image
\includegraphics[width=1.\textwidth]{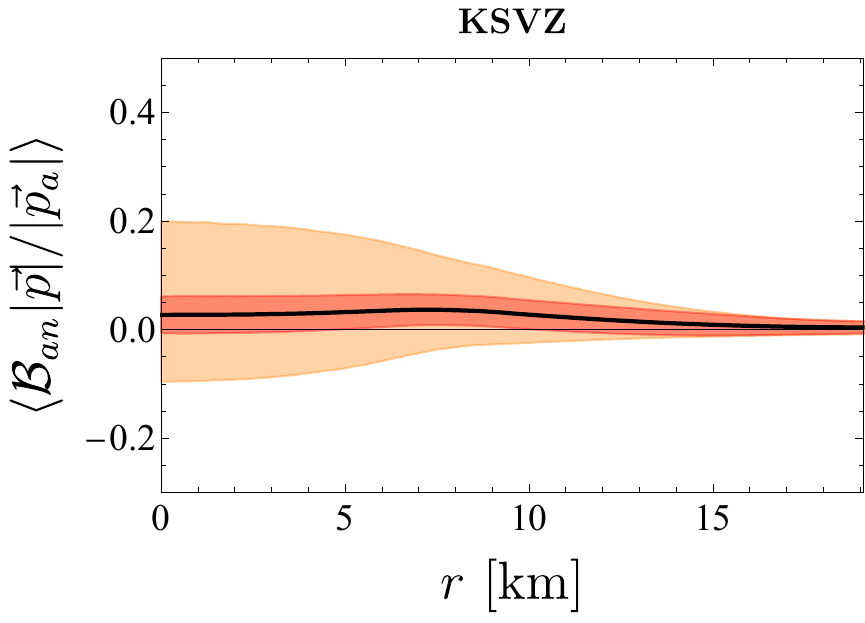}
\end{subfigure}
  \centering
 \caption[]{Top: Density and temperature profile taken from a SN simulation of \cite{Garching} at $t_{pb}=1\,\mathrm{s}$. Middle: Effective form factors $\mathcal{A}$ of the KSVZ axion evaluated at typical momenta dictated by a SN profile at $t_{pb}=1 \,\mathrm{s}$ taken from simulations of \cite{Garching}. The red-shaded region represents the error bars resulting from the uncertainties of the constants given in \Tab{tab:constants}, while the orange-shaded region represents the error emerging from higher-order terms in the chiral expansion. Note that the values of the real part of the couplings for large radii slightly deviate from the vacuum value in \Eq{eq:res_vac_KSVZ_constants} because $T \neq 0$. Bottom: Analogous results for the form factor $\mathcal{B}$. 
 }
  \label{fig:axion_SN_couplings}
\end{figure}

The axion luminosity, \ie the emitted energy by axions per time, is defined by the volume integral over the emissivity. When determining the luminosity observed by a distant observer from the supernova (SN), it is necessary to consider general-relativistic effects, as explained in \Refscite{Rampp:2002bq,Marek:2005if}. We adjust for the gravitational redshift experienced by axions as they are emitted from a radius 
$r$ within the SN and travel to spatial infinity by multiplying with the lapse function squared provided by \Refcite{Garching}. Additionally, the data of \Refcite{Garching} is given in the comoving reference frame of the emitting medium. Thus, the radial velocity $v_r$ of the SN causes another red- or blueshift due to the Doppler effect. In the limit of small radial velocity, this is accounted for by multiplying with a factor of $(1 + 2v_r)$ inside the volume integral. Hence, the axion luminosity is given by
\begin{equation}
L_{a}=\int d r 4 \pi r^{2} \dot{\varepsilon}_{a}(r) \text{(lapse)}^2 (1 + 2v_r).
\end{equation}
In the right panel of \Fig{fig:axion_luminosity}, we show the luminosity for the KSVZ axion as a function of time including axion-vertex corrections (red), and compare to the leading order vacuum couplings (black). 
As can be seen, both the emissivity and the luminosity are enhanced by up to a factor of three, depending on radius and time, respectively. 
It is important to highlight that our calculations account for the uncertainties 
arising from the truncation of the ChPT expansion as well as from the low-energy constants, see \Sec{sec:AxioncouplingsHBCHPT}.

In \Fig{fig:axion_SN_couplings}, we show the expectation value of the modified coupling of the axion to protons and neutrons as a function of the radius (lower four panels) for a given density and temperature profile (upper panel) taken from \Refcite{Garching}. 
The expectation value is defined as
\begin{equation} \label{eq:exp_value_mt}
\langle \mathcal{A} \,\rangle \equiv \mathcal{N} \int  \prod_{i=1}^4 d \Pi_{i}   d \Pi_{a} \mathcal{A}(p, k_f^{p/n}, p_a) \delta^{(4)}\left(\textstyle \sum_{i} p_i-p_{a}\right) f_{1} f_{2}\left(1-f_{3}\right)\left(1-f_{4}\right),
\end{equation}
where $\mathcal{N}$ is a normalization constant, and for simplicity we choose $p=p_{1}$.
As can be seen, also temperature effects strongly modify the effective coupling, see also \App{app:temp}.

In general, additional energy loss channels in a supernova can rival the neutrino channel, potentially conflicting with the observed neutrino signals from SN 1987A \cite{Hirata:1987hu, Bionta:1987qt}.
This can be translated into a constraint on weakly coupled particles.
In particular, if the luminosity in axions exceeds the neutrino luminosity, this would lead to a shortening of the neutrino signal \cite{Raffelt:2006cw}. As a criterion, we adopt comparing the luminosities $1\, \text{s}$ post-bounce. The supernova simulation employed in this work has a neutrino luminosity $1\, \text{s}$ post-bounce\footnote{Note that similar effects as considered for the axion luminosity can also affect neutrino luminosity. However, while the axion is free streaming, neutrinos are trapped at large densities. Because of this, the neutrino luminosity mainly depends on low-density regions, and we expect only small corrections, as seen when including the quenching of $g_A$ \cite{Buras:2005rp}.} of
\begin{equation}
    L_\nu\sim 5.5 \times 10^{52}\,\mathrm{erg}\,\mathrm{s}^{-1}.
\end{equation}
Demanding that the luminosity in axions is smaller than that leads to a bound of
\begin{equation}\label{eq:FullBoundLumi}
    f_a \gtrsim  1.0^{+0.5}_{-0.2} \times 10^9 \, \mathrm{GeV},\quad m_a \lesssim 5.9^{+1.8}_{-2.0} \, \mathrm{meV}. \qquad \text{(including vertex corrections)}
\end{equation}
This can be compared with the naive bound that is found with the same SN but using the leading order process only. In that case one finds
\begin{equation}\label{eq:OPEBoundLumi}
    f_a \gtrsim  6.1^{+1.7}_{-1.4} \times 10^8 \, \mathrm{GeV},\quad m_a \lesssim 9.8^{+3.0}_{-2.2} \, \mathrm{meV}. \qquad \text{(leading order)}
\end{equation}
On can see that including the vertex corrections strengthens the SN bound on the axion mass (or decay constant) by roughly a factor of two.

Let us briefly compare our findings to the recent literature.
We find a strengthening of the SN bound on $f_a$ as compared to \cite{Carenza:2019pxu,Chang:2018rso} by a factor of a few and up to an order of magnitude, respectively. 
This enhancement can be attributed to two main factors. First, in our case, the luminosity is enhanced compared to the OPE when including the axion form factors. 
And second, in both \cite{Carenza:2019pxu} and \cite{Chang:2018rso} the reduction of the axion luminosity due to a reduction of the nucleon-nucleon interaction, as well as due to multiple nucleon scatterings \cite{Raffelt:2006cw}, was modeled in a phenomenological way. 
A systematic analysis of these effects in SN environments is still missing~\cite{springmann3} (see discussion of diagrams \figref{fig:c}{c}, \figref{fig:d}{d}, \figref{fig:g}{g} above). Our focus has been on the previously unexplored and significant axion vertex corrections.
We expect the enhancement due to the modified couplings to remain. This underscores that our analysis should be viewed as a first step and emphasizes the need for a systematic evaluation of axion emission during SN core collapse within ChPT.

Finally, we emphasize that a systematic and consistent axion bound from supernovae cannot be achieved by combining this ChPT approach with additional phenomenological corrections.

\subsection{Neutron star cooling bound} \label{sec:neutron_star}

Axions can be produced in scattering processes within the neutron star core and can escape the star due to their weak interaction \cite{Iwamoto:1984ir,Iwamoto:1992jp}, the dominant contribution again being axion production via nucleon-nucleon Bremsstrahlung, and thereby modify the cooling of NSs~\cite{Buschmann:2021juv}.

The typical density in the bulk of neutron star matter is $n\sim \mathcal{O}(\text{few}) n_0$.
While a significant portion of a neutron star's luminosity arises from regions with densities around $(0.5-2) n_0$, the precise contribution depends on the neutron star's maximum mass and the chosen equation of state (EOS). 
However, the luminosity originating from high-density regions, where no controlled calculation exists and, in particular, the ChPT expansion is no longer applicable, is considerable and can exceed the low-density contribution.
This is especially pronounced for heavier NSs.

This limitation suggests two possible approaches for dealing with the high-density regions. 
The first approach sets the axion cooling bound by focusing solely on parts of the neutron star where controlled calculations are possible - specifically, regions with densities below nuclear saturation density. 
Since this approach neglects large parts of the star which contribute to the axion luminosity, we find that for vacuum couplings, it weakens the bound on the decay constant by a factor of three to six compared to \Refcite{Buschmann:2021juv}, depending on the exact mass of the star and to a small extend on the EOS used.

However, within this approach, the emissivity can be systematically calculated using finite-density couplings.
We compute the luminosity resulting from the region below nuclear saturation density $(n\lesssim 1n_0)$, including the full systematic density dependence of the axion-nucleon coupling.\footnote{
We adopt the same phenomenological suppression factors as in
\cite{Buschmann:2021juv} describing the short-range nucleon-nucleon scattering \cite{Yakovlev:2000jp,1995A&A...297..717Y}, the density modification of the pion nucleon coupling \cite{Mayle:1989yx} as well as multiple pion exchange.
While this approach sacrifices all systematic rigor, a more consistent evaluation of these terms would be desirable but exceeds the scope of this work.
We then compare these results with the luminosity from the entire star, however, using the phenomenological density dependence of the couplings from \cite{Mayle:1989yx} employed in \cite{Buschmann:2021juv}.}

We find that using systematically calculated axion-nucleon couplings further relaxes the bound—beyond the reduction already caused by neglecting a significant portion of the star—primarily due to the diminished axion-proton coupling at high densities (see \Eq{eq:axionnuclKSVZsaturation}). 
This effect reduces the $1\sigma$ lower bound on the axion mass by a factor of a bit more than two.
This result contrasts with the strengthening of the supernova bound discussed in \Sec{sec:supernova}, which arises because neutron star temperatures are significantly lower. Consequently, effects such as the energy dependence of the axion-nucleon coupling, which we include there for the first time, become negligible for NSs.

For concreteness, we consider a neutron star with $1.5 M_\odot$ using the BSk24 EOS \cite{Goriely:2013xba}. 
Comparing with \Refcite{Buschmann:2021juv}, we find the bound on the KSVZ axion model from neutron star cooling in the conservative approach to be
\begin{equation}\label{eq:FullBoundLumi1}
    f_a \gtrsim  1.0^{+0.3}_{-0.2} \times 10^8 \, \mathrm{GeV},\quad m_a \lesssim 60^{+28}_{-13} \, \mathrm{meV},\quad 68\%\,\text{C.L.,} \qquad \text{(KSVZ)}
\end{equation}
where we sample over the axion-nucleon couplings within their ChPT error bars, see \Sec{sec:AxioncouplingsHBCHPT}. 
This is a significant weakening of the bound on the axion mass and decay constant.

Another less rigorous approach estimates axion production in the high-density regime $(n\gtrsim 1n_0)$ by making use of naive dimensional analysis to determine the axion couplings,
\begin{equation}\label{eq:couplingNS}
    |\mathcal{A}_{aN}|\sim 0.3(3),
\end{equation}
where the range is inspired by \Eq{eq:axioncouplingswithoutpions} with order one coefficients.

Due to the density gradient inside the star, it is improbable that the couplings remain small at all densities. 
If the vacuum couplings are not tuned, their values have minimal impact on axion-induced neutron star cooling, and no distinction between different axion models is possible.
Assuming \Eq{eq:couplingNS} for the region of the NS where $n \gtrsim 1n_0$, we calculate the luminosity and set a more aggressive bound on the axion mass and decay constant,
\begin{equation} \label{eq:bound_NDA}
    f_a \gtrsim  5.3^{+1.8}_{-2.7} \times 10^8 \, \mathrm{GeV},\quad m_a \lesssim 11^{+15}_{-2} \, \mathrm{meV},\quad 68\%\,\text{C.L.} \quad \text{(model-independent)},
\end{equation}
which is independent of the axion model under consideration, assuming no tuning, and completely neglects any potential axion production in the low-density region of the NS.

If additionally the contribution to the luminosity from the low-density region of the star is taken into account, we find for \eg the KSVZ axion
\begin{equation}
    f_a \gtrsim  5.4^{+1.7}_{-2.7} \times 10^8 \, \mathrm{GeV},\quad m_a \lesssim 11^{+12}_{-3} \, \mathrm{meV},\quad  68\%\,\text{C.L.} \qquad \text{(KSVZ)}.
\end{equation}
Similar bounds can be derived for any axion model, with the main contribution to the luminosity coming from the region with large densities, see \Eq{eq:bound_NDA}, where a distinction between models is not possible.

However, there are QCD axion models, such as the astrophobic axion \cite{DiLuzio:2017ogq}, where the shift symmetric couplings of the axion are tuned. 
In such cases, the dominant contribution to the coupling $|\mathcal{A}_{aN}|$ comes from shift symmetry- and isospin-breaking terms, which are hence suppressed. 
This suppression should also hold at large densities, see \cite{DiLuzio:2024vzg} for a related study.
Therefore the NS cooling bound is weaker than expected from our NDA estimate. 

Nevertheless, we point out that there exists an additional process that arises from a shift-symmetry breaking operator and contributes to axion cooling both in supernovae and neutron stars that cannot be tuned small for any QCD axion model.
We explore this process in light of an additional energy loss channel during supernovae cooling in a companion paper \cite{Springmann:2024ret}.

For neutron star cooling, taking all derivative couplings to be zero, we calculate the axion luminosity, which gives rise to a bound from NS cooling on the axion mass of \hbox{$m_a\lesssim 149^{+87}_{-31} \, \mathrm{meV}$} and decay constant of \hbox{$f_a \gtrsim  4.1^{+1.0}_{-1.5} \times 10^7 \, \mathrm{GeV}$}.

Note that axions can also be produced in merger events, see \eg \cite{Harris:2020qim}, for which our findings of the previous sections can be significant.

\subsection{Terrestrial experiments} \label{sec:terrestrial}

Our findings are not only relevant for high-density astrophysical objects like supernovae and neutron stars, but also have important implications for axion dark matter detection experiments, which target the axion-nucleon derivative coupling~\cite{Brandenstein:2022eif,Jiang:2021dby,JEDI:2022hxa,Abel:2017rtm,Gao:2022nuq,Lee:2022vvb,Bloch:2019lcy,2009PhRvL.103z1801V,JacksonKimball:2017elr,Wu:2019exd,Garcon:2019inh,Wei:2023rzs,Xu:2023vfn,Chigusa:2023hmz,Bloch:2021vnn,Bloch:2022kjm,Graham:2020kai,Mostepanenko:2020lqe,Adelberger:2006dh,Bhusal:2020bvx}. Similarly higher order density corrections to non-derivative couplings, can be important for time-dependent new physics searches such as current and future nuclear clock experiments \cite{EPeik_2003,Flambaum2012,Caputo:2024doz,Kim:2022ype,Fuchs:2024edo}.
Nucleons inside a large nucleus do not have small momenta, and in the axion-nucleon interaction, the rest of the nucleus can be seen as a background density. 

This is in direct analogy to the well-known quenching of $G_A$ in the beta decay of large nuclei, see, \eg, \cite{Menendez:2011qq,Gysbers:2019uyb}. There, the rate of Gamov-Teller transitions is related to the reduced matrix element
\begin{equation}
    |\mathcal{M}_{GT}| = |\langle f | \vec{O}^{\pm}_{GT} | i \rangle|.
\end{equation}
with the usual Gamov-Teller operator,
\begin{equation}
    \vec{O}^{\pm}_{GT} = 2 G_A \vec{S} \tau^{\pm}.
\end{equation}
Inside nuclei however, there are higher order contributions to the Gamov-Teller operator induced by the background nucleons, leading to a quenching of $G_A$ as described \eg in \cite{Menendez:2011qq,Gysbers:2019uyb}. 

The interaction of a dark matter axion background wind with nuclei on Earth is an analogous problem. 
One has to evaluate a similar matrix element, however with $\vec{O}^{\pm}_{GT}$ replaced by $2  \vec{S} \left(G_A\tau^{3}+ G_0 \mathbb{1}\right) $. 
The higher-order contributions from a homogeneous background of the other nucleons are exactly the ones discussed in \Sec{sec:finite_density_couplings}.

However, the study of axion coupling with nuclei is complex and depends on the specific nucleus, posing a challenging nuclear physics problem beyond the scope of our work. To provide a preliminary estimate, we model the nucleus as a Fermi gas, similar to the approach used in~\cite{Menendez:2011qq} for the quenching of $G_A$ in beta decays. 

Then the axion-nucleon coupling \Eq{eq:full_vertex_result} has to be evaluated around nuclear saturation density and for mixed matter, leading for the KSVZ axion to \Eq{eq:axionnuclKSVZsaturation}.
While this changes the predictions of the QCD axion and ALP couplings by $\mathcal{O}(1)$, in case of detection, this can be crucial to discriminate UV completions. Even if a UV model predicts a small (or vanishing) derivative axion coupling in vacuum, experiments using large nuclei are still expected to be sensitive. A notable example of this is the KSVZ axion neutron coupling, where the mean value is estimated to be enhanced by a factor of $\sim 3$.

\section{Conclusions} \label{sec:conclusion}

In this work, we have conducted the first systematic study on the interactions of QCD axions with nucleons in the formalism of chiral perturbation theory, including non-relativistic baryons, at finite nuclear density and temperature.

We reviewed the effective theory of QCD axions across different energy scales.
We started with the effective axion theory at energies below electroweak symmetry breaking and matched to the chiral, two-flavor Lagrangian, including mesons and non-relativistic baryons, at energies below the QCD confinement scale.
We constructed the effective Lagrangian consistent with power counting up to next-to-leading order.
This theory describes the axion-nucleon and axion-meson interactions relevant in dense and hot environments, like those found in supernovae and neutron stars.

Our results reveal that the axion-nucleon couplings in these environments are highly sensitive to both momentum and energy, with contributions arising from ordinary loop corrections and finite density effects. 

At energies $\omega_a \simeq m_\pi$, where virtual pions can go on-shell, we found that the form factor acquires a significant imaginary part, which has to be included in phenomenological applications.
We have found that the axion-nucleon form factors generically change $\mathcal{O}(1)$ for axion energies up to $\omega_a\simeq 2 m_\pi$. 
Notably, for the KSVZ axion, an accidental cancellation at zero energy that results in a suppressed coupling to neutrons is lifted as the axion energy increases, with the form factor enhanced by up to an order of magnitude at $\omega_a \simeq 2 m_\pi$.

We calculated the density dependence of the axion-nucleon form factors systematically for the first time. 
We used the real-time formalism of thermal quantum field theory to do this.
In this approach, the nucleon propagator gets an additional component that depends on temperature and chemical potential, which represents the interaction with the nucleon background.
In the $T \to 0$ limit, this additional piece is usually referred to as a density insertion.
In practice, density effects are then calculated by evaluating loop diagrams where an internal nucleon propagator is replaced by the density insertion.

We found that the density corrections predominantly scale with the nucleon Fermi momentum, contrary to the pure loop effects, which scale with the axion energy.
Therefore, if the axion energies are negligible, the momentum dependence of the axion-nucleon couplings is dominated by density effects.
This is particularly important for environments with low temperatures and high densities.

We have calculated both loop and density corrections up to chiral order $\nu=3$, including large Wilson coefficients that occur due to the low-lying $\Delta(1232)$-resonance.
We found that higher-order and density-corrected couplings have significant implications for phenomenological applications both in astrophysical environments as well as for terrestrial experiments.

On the astrophysical side, supernova explosions present themselves as ideal laboratories to apply our result: the extreme densities and temperatures in supernova environments have a large influence on the axion couplings and, consequently, significantly change the axion production rate.
We evaluated the supernova bound on the QCD axion, focussing on the predictive KSVZ model, including our density-dependent couplings, and found that the axion luminosity during SN 1987A can be enhanced by up to one order of magnitude.
This translates to a stricter bound on the axion decay constant $f_a$ by a factor of three compared to previous works, \eg \cite{Carenza:2019pxu}.
We highlighted the systematic treatment of the axion couplings, allowing for the first time to estimate a concise uncertainty for the axion luminosity, which directly propagates to the bound, which we found as
\begin{equation}
    f_a \gtrsim  1.0^{+0.5}_{-0.2} \times 10^9 \, \mathrm{GeV},\quad m_a \lesssim 5.9^{+1.8}_{-2.0} \, \mathrm{meV}.\quad\text{(KSVZ)}
\end{equation}
For our analysis, we used the supernova profile \cite{Bollig:2020xdr} kindly provided by the Garching group of Thomas Janka \cite{Garching}.
While we found that the density correction due to the chemical potential alone could potentially reduce the axion luminosity, the large temperatures present during a supernova explosion, of order $\sim 30\,\text{MeV}$, leads to a large axion high energy component and consequently to an enhanced axion production rate.

We emphasize that the density-dependent axion-nucleon couplings evaluated in this work mark an important first step towards a consistent evaluation of the complete axion production during supernovae.
In order to get a consistent result up to a given order in the chiral expansion, one needs to evaluate every diagram that contributes to the axion production to that order.
We classified all such contributions up to chiral order $\nu=3$, which we will present in future publications.
While the bound provided here incorporates all existing perturbative corrections, a comprehensive analysis that includes a complete set of diagrams could yield a different axion constraint.
We stress that combining phenomenological effects, such as \eg rho meson exchange or multiple scatterings, with our systematic corrections is inconsistent and should be avoided.

Next, we studied the effect of density-dependent axion-nucleon couplings on axion production in neutron star environments, which are characterized by large densities but low temperatures. 
We re-evaluated the neutron star cooling bound, including density-dependent axion-nucleon couplings.

Within neutron stars, densities exceed nuclear saturation by a factor of a few, and the axion-nucleon coupling is dominantly influenced by high nucleon densities, contrary to supernovae.
At these large densities, the chiral series breaks down, and all perturbative control is lost.
However, we consistently calculated the axion luminosity, including our modified density-dependent couplings at low densities, where the chiral expansion is valid.

In order to place a conservative, fully controlled bound, we considered the axion luminosity arising from these low densities only.
By re-evaluating the most recent bound from \Refcite{Buschmann:2021juv}, we found that the axion bound from neutron star cooling is relaxed by up to one order of magnitude. 
In particular, for the KSVZ axion, we find
\begin{equation}
    f_a \gtrsim  1.0^{+0.3}_{-0.2} \times 10^8 \, \mathrm{GeV},\quad m_a \lesssim 60^{+28}_{-13} \, \mathrm{meV},\quad 68\%\,\text{C.L.}. \qquad \text{(KSVZ)}
\end{equation}
This relaxation of the bound comes from two effects: we restrict our analysis to regions of the star where the density is sufficiently low to permit a perturbative treatment, and the axion-proton coupling decreases at finite density, while temperature effects remain negligible.
As for the supernovae calculation, one should calculate all axion production channels and include corrections systematically. 
We leave this and further improvements for future work.

In order to place a more aggressive but less rigorous bound, we included the axion luminosity from high-density regions using order one axion-nucleon couplings estimated via naive dimensional analysis, which are expected for any axion model that is not tuned.
In this case, we find a similar bound as previous works \Refcite{Buschmann:2021juv}, which holds independent of the underlying (not-tuned) QCD axion model.
As a consequence, this more aggressive bound on the axion from neutron star cooling cannot distinguish between different axion models.

In the supplementary file, we provide numerical tables for density and temperature-dependent axion-nucleon couplings, including error bars as estimated within the ChPT expansion, which can be used in dedicated supernova simulations~\href{https://github.com/michael-stadlbauer/Axion-Couplings.git}{\faGithub}. 

Finally, we investigated the phenomenological consequences of density-dependent axion-nucleon couplings in axion experiments on Earth. 
Inside large nuclei, our findings correct the axion couplings to nucleons by an order one factor, which is relevant to all searches sensitive to axion-nucleon couplings, such as nucleon spin-precession experiments or the nuclear clock.
Especially in case of a discovery it will be important to precisely determine the fundamental couplings of the axion field to quarks and gluons, in which case our findings will be crucial.

Our work opens up multiple new avenues for future work. 
We have established a framework for a systematic and fully controlled axion precision phenomenology at finite density and temperature.
In future works, we will address the crucial task of calculating the missing diagrammatic contributions to axion production relevant to SNe and NSs, which we have classified in this work.
In a subsequent paper, we will investigate how modifications to the nucleon-nucleon interaction influence the axion production rate in supernova environments \cite{springmann3}.
Additionally, this framework should also be extended for key standard model processes, such as neutrino dynamics in the same environments, which we will explore in upcoming work \cite{caputo1}.

Furthermore, a fully self-consistent SNe simulation, taking into account the additional energy loss from axion production, is needed. 
Such a simulation can benefit from our findings of axion couplings derived in this work but eventually has to include neutrino and axion production rates calculated consistently at finite density and temperature.

Lastly, we note that our findings allowed us to calculate a model-independent contribution to the axion luminosity during supernovae which we present in \cite{Springmann:2024ret}.

Our findings provide a framework for revisiting axion phenomenology in extreme astrophysical environments, enabling precision studies and guiding future discoveries in both theoretical and experimental axion physics.

\section*{Acknowledgements}
We would like to thank Reuven Balkin, Alexander Bartl, Kai Bartnick, Andrea Caputo, Pierluca Carenza, Tim Cohen, Raffaele Del Grande, Majid Ekhterachian, Laura Fabbietti, Albert Feijoo, Tobias Fischer, Doron Gazit, Malte Heinlein, Anson Hook, Thomas Janka, Norbert Kaiser, Daniel Kresse, Valentina Mantovani Sarti, Georg Raffelt, Riccardo Rattazzi, Sanjay Reddy, Javi Serra, and Giovanni Villadoro for useful conversations and discussions.
We would especially like to thank Thomas Janka and his group for providing us with supernova simulation data and Tobias Fischer for helping us cross-check our numerical simulations.
This work has been supported by the Collaborative Research Center SFB1258, the Munich Institute for \hbox{Astro-, Particle} and BioPhysics (MIAPbP), and by the Excellence Cluster ORIGINS, which is funded by the Deutsche Forschungsgemeinschaft (DFG, German Research Foundation) under Germany's Excellence Strategy – EXC 2094 – 39078331. The research of MS is partially supported by the International Max Planck Research School (IMPRS) on “Elementary Particle Physics”.
KS is supported by a research grant from Mr. and Mrs. George Zbeda.
SS is partially supported by the Swiss National Science Foundation under contract 200020-213104. SS thanks the CERN theory group for its hospitality.

\clearpage

\appendix
\section{Construction of the effective axion HBChPT Lagrangian}\label{app:construction}

In this appendix, we briefly summarize the construction of the effective axion chiral Lagrangian in HBChPT up to $\text{N}^2\text{LO}$ and some selected pieces $\text{N}^3\text{LO}$, necessary for renormalization. 
We start by writing down the relativistic pion axion nucleon Lagrangians and projecting to HBChPT.
For reviews on the subject without the axion, see \eg \Refscite{Bernard:1995dp,Scherer:2002tk,Epelbaum:2008ga}, as well as \Refscite{Fettes:2000gb} for the HBChPT projection.  The LO axion HBChPT Lagrangian as well as some selected terms at $\text{N}^2\text{LO}$ are constructed in \Refcite{Vonk:2020zfh}.
\subsection{Chiral QCD Lagrangian}
We analyze the transformation properties of external sources, performing a spurion analysis.
Starting point is the QCD Lagrangian with external isovector axial vector $(a_{\mu})$, isoscalar axial vector $(a^s_{\mu})$, scalar $(s)$ and pseudo-scalar $(p)$ sources. 
In absence of isovector and isoscalar vector sources ($v_\mu^{(s)}=0$), which is the case for axion ChPT, we can write parity eigenstates in terms of left- and right-handed fields, \ie $r^{(s)}_{\mu}=a^{(s)}_{\mu}$ and $\ell^{(s)}_{\mu}=-a^{(s)}_{\mu}$. 
The QCD Lagrangian containing external sources then reads
\begin{equation}
    \Lag_{\text{ext}}=\bar{q}_L\gamma^{\mu}\left(\ell_{\mu}+\ell^{s}_{\mu}\right)q_L+\bar{q}_R\gamma^{\mu}\left(r_{\mu}+r^{s}_{\mu}\right)q_R-\bar{q}_L\left(s-ip\right)q_R-\bar{q}_R\left(s+ip\right)q_L,
\end{equation}
where $q_{L/R}$ are left- and right-handed quark fields. 
Isovector and isoscalar parts are associated with the $SU(2)$ and $U(1)$ parts of $U(2)$, respectively.
The Lagrangian $\Lag_{\text{QCD}}=\Lag_{\text{QCD},0}+\Lag_{\text{ext}}$, with $\Lag_{\text{QCD},0}$ given by \Eq{eq:LQCD0}, is invariant under local $U(2)_L\times U(2)_R$ transformations
\begin{equation}
    q_{L/R}\rightarrow\exp\left(-i\frac{\theta_{L/R}(x)}{3}\right)V_{L/R}(x)q_{L/R},\quad \theta_{L/R}\in U(1)_{L/R},\quad V_{L/R}\in SU(2)_{L/R},
\end{equation}
as long as the external fields satisfy
\begin{equation}
    \begin{aligned}
    r_{\mu}&\,\,\rightarrow\,\,V_R r_{\mu}V^{\dagger}_R+iV_R\partial_{\mu} V^{\dagger}_{R},\\
    \ell_{\mu}&\,\,\rightarrow\,\, V_L \ell_{\mu}V^{\dagger}_L+iV_L\partial_{\mu} V^{\dagger}_{L},\\
    \ell^{s}_{\mu}&\,\,\rightarrow\,\,\ell^{s}_{\mu}-\frac{1}{3}\partial_{\mu}\theta_{L},\\
    r^{s}_{\mu}&\,\,\rightarrow\,\, r^{s}_{\mu}-\frac{1}{3}\partial_{\mu}\theta_{R},\\
    \left(s-ip\right)&\,\,\rightarrow\,\, e^{-\frac{i}{3}\left(\theta_L-\theta_R\right)}V_L\left(s-ip\right)V^{\dagger}_R,\\
    \left(s+ip\right)&\,\,\rightarrow\,\, e^{-\frac{i}{3}\left(\theta_R-\theta_L\right)}V_R\left(s+ip\right)V^{\dagger}_L.
    \end{aligned}
    \label{eq:TransExtFields}
\end{equation}
The derivative terms cancel analogous terms originating from the kinetic terms in the Lagrangian and are later used to construct covariant derivatives.

\subsection{Effective Pion Lagrangian}

In the broken phase, pions are described by the unitary flavor matrix $U$, see \Eq{eq:Upion}. It transforms linearly under chiral and axial $U(1)_A$ transformations \begin{equation}
    U\rightarrow e^{-\frac{i}{3}\left(\theta_R-\theta_L\right)}V_R U V_L^{\dagger}.
\end{equation}
Since it collects the NGBs, associated with the broken generators of chiral symmetry breaking, it is charged under axial transformations only. 
Furthermore, under $\operatorname{C}$ and $\operatorname{P}$ transformations the NGBs and $U$ transform as
\begin{equation}
    \begin{aligned}
    \boldsymbol{\pi}(x)\overset{\operatorname{C}}{\longrightarrow}\,\boldsymbol{\pi}^{T}(x),& \quad\boldsymbol{\pi}\left(x\right)\overset{\operatorname{P}}{\longrightarrow}\,-\boldsymbol{\pi}\left(\mathcal{P}x\right)\\
    U(x)\overset{\operatorname{C}}{\longrightarrow}\,U^{T}(x),&\quad U(x)\overset{\operatorname{P}}{\longrightarrow}\,U^{\dagger}\left(\mathcal{P}x\right).
    \end{aligned}
\end{equation}
The covariant derivative is given by the linear representation
\begin{equation}
  	  \nabla_{\mu} U=\partial_{\mu}U-i\left(r_{\mu}+r^{s}_{\mu}\right)U+iU\left(\ell_{\mu}+\ell^{s}_{\mu}\right),
\end{equation}
resulting in \Eq{eq:covlin}. 
Another spurion in the pion sector can be defined as
\begin{equation}
    \chi=2B M_a^{\dagger}=2B(s+ip),
\end{equation}
see below \Eq{equ:lag_LpiN_2}, which transforms as
\begin{equation}
    \chi
    \rightarrow V_R\chi V_L^{\dagger}.
\end{equation}
In the meson sector, the list of fundamental building blocks that we are interested in \ie  up to $O(p^2)$ is
\begin{equation}
U,\nabla_{\mu}U,\nabla_{\mu}\nabla_{\nu}U,\chi.
\end{equation}
Out of these, the only non-trivial, hermitian, $\operatorname{C}$ and $\operatorname{P}$ invariant scalars under Lorentz are collected in the Lagrangian \Eq{equ:lag_LpiN_2}.

At NLO one finds the following Lagrangian \cite{Scherer:2002tk} with the covariant derivative and $\chi$ now including the axion, 
\begin{equation}
\begin{aligned}
\mathcal{L}^{(4)}_{\pi\pi}= & \frac{l_1}{4}\left\{\operatorname{Tr}\left[D_\mu U\left(D^\mu U\right)^{\dagger}\right]\right\}^2+\frac{l_2}{4} \operatorname{Tr}\left[D_\mu U\left(D_\nu U\right)^{\dagger}\right] \operatorname{Tr}\left[D^\mu U\left(D^\nu U\right)^{\dagger}\right] \\
& +\frac{l_3}{16}\left[\operatorname{Tr}\left(\chi U^{\dagger}+U \chi^{\dagger}\right)\right]^2+\frac{l_4}{4} \operatorname{Tr}\left[D_\mu U\left(D^\mu \chi\right)^{\dagger}+D_\mu \chi\left(D^\mu U\right)^{\dagger}\right] \\
& -\frac{l_7}{16}\left[\operatorname{Tr}\left(\chi U^{\dagger}-U \chi^{\dagger}\right)\right]^2 +\frac{h_1+h_3}{4} \operatorname{Tr}\left(\chi \chi^{\dagger}\right)\\
& +\frac{h_1-h_3}{16}\left\{\left[\operatorname{Tr}\left(\chi U^{\dagger}+U \chi^{\dagger}\right)\right]^2+\left[\operatorname{Tr}\left(\chi U^{\dagger}-U \chi^{\dagger}\right)\right]^2\right. \\
& \left.-2 \operatorname{Tr}\left(\chi U^{\dagger} \chi U^{\dagger}+U \chi^{\dagger} U \chi^{\dagger}\right)\right\}.
\end{aligned}
\end{equation}
We use this Lagrangian to calculate the axion-pion mixing at NLO as well as for the matching of the pion mass and decay constant.\subsection{Effective Baryon Lagrangian}\label{App:EffBaryonLag}
The baryon field transforms non-linearly under chiral transformations
\begin{equation}
    \Psi'=e^{-i\theta_V(x)}K(V_L(x),V_R(x),U(x))\Psi,
\end{equation}
where $K(V_L,V_R,U)$ is the so-called compensator field and defines a non-linear $SU(N_f)$ valued function of $U$ and $V_L,V_R$, defined by
\begin{equation}
    U(x)=u^2(x),\quad u(x)\rightarrow u'(x)=\sqrt{V_R U V_L^{\dagger}}\equiv V_R u K^{-1}(V_L,V_R,U),
\end{equation}
such that the explicit form of $K$ is easily evaluated to
\begin{equation}
    K(V_L,V_R,U)=(u')^{-1}V_Ru=\left(\sqrt{V_R U V^{\dagger}_L}\right)^{-1}V_R \sqrt{U}.
\end{equation}
The covariant derivative is given by
\begin{equation}
    \begin{aligned}
    &D_{\mu}\Psi=\left(\partial_{\mu}+\Gamma_{\mu}\right)\Psi,\\
    &\Gamma_{\mu}=\frac{1}{2}\left[u^{\dagger}(\partial_{\mu}-i r_{\mu})u+u(\partial_{\mu}-i \ell_{\mu})u^{\dagger}\right],
    \end{aligned}
\end{equation}
where $\Gamma_{\mu}$ is the chiral connection. Under parity and charge conjugation it transforms as
\beq
D_{\mu}\overset{\operatorname{P}}{\longrightarrow}+{\mathcal{P}_{\mu}^{}}^{\nu}D_{\nu},\quad D_{\mu}\overset{\operatorname{C}}{\longrightarrow}-D_{\mu}^{T}.
\eeq
with ${\mathcal{P}_{\mu}^{}}^{\nu}= \text{Diag}(1,-1,-1,-1)$.
One can furthermore construct a hermitian, isovector, axial vector object containing one derivative, called the vielbein
\begin{equation}
    u_{\mu}=i\left[u^{\dagger}(\partial_{\mu}-i r_{\mu})u-u(\partial_{\mu}-i \ell_{\mu})u^{\dagger}\right]=iu^{\dagger}\left(\partial_{\mu}U-ir_{\mu}U+iU\ell_{\mu}\right)u^{\dagger}.
\end{equation}
The transformation behavior under $SU(2)_L\times SU(2)_R$, parity ($\operatorname{P}$) and charge conjugation ($\operatorname{C}$) is
\begin{equation}
    u_{\mu}\overset{SU(2)_L\times SU(2)_R}{\longrightarrow}Ku_{\mu}K^{\dagger},\quad u_{\mu}\overset{\operatorname{P}}{\longrightarrow}-{\mathcal{P}_{\mu}^{}}^{\nu}u_{\nu},\quad u_{\mu}\overset{\operatorname{C}}{\longrightarrow}u_{\mu}^{T}.
\end{equation}
$u_{\mu}$ sandwiched between a baryon bilinear stays invariant under local chiral gauge transformations. There exist two identities for the covariant derivative,
\begin{subequations}
\begin{align}
\left[D_{\mu},D_{\nu}\right]&=\frac{1}{4}u_{\mu}u_{\nu},\label{eq:curvrel}\\
\left[D_{\mu},u_{\nu}\right]&=\left[D_{\nu},u_{\mu}\right].
\end{align}
\end{subequations}
The first is known as curvature relation and shows that only completely symmetrized products of $D_{\mu}$ with itself as well as products of $D_{\mu}$ with $u_{\mu}$ have to be considered. The second relation shows in a similar way that covariant derivatives on the vielbein have to be taken into account in a completely symmetrized combination only.
We can construct an analogous quantity to the vielbein from the isoscalar part of the linearly realized covariant derivative
\begin{equation}
    \hat{u}_{\mu}=iu^{\dagger}\left(-ir^{s}_{\mu}U+iU\ell^{s}_{\mu}\right)u^{\dagger}=iu^{\dagger}(-2i\as_{\mu}U)u^{\dagger}=2\as_{\mu},
\end{equation}
which under parity and charge conjugation has exactly the same transformation properties as $u_{\mu}$ but only transforms under $U(1)_A$
\begin{equation}
    \hat{u}_{\mu}\overset{U(1)_A}{\longrightarrow}\hat{u}_{\mu}-\partial_{\mu}\alpha_A,\quad \hat{u}_{\mu}\overset{\operatorname{P}}{\longrightarrow}-{\mathcal{P}_{\mu}^{}}^{\nu}\hat{u}_{\nu},\quad \hat{u}_{\mu}\overset{\operatorname{C}}{\longrightarrow}\hat{u}_{\mu}^{T}.
\end{equation}
Furthermore, we need a non-linearly transforming analog to the field $\chi$, which we construct as hermitian and anti-hermitian combinations
\beq
\chi_{\pm}=u^{\dagger}\chi u^{\dagger}\pm u\chi^{\dagger}u,
\eeq
with transformation properties
\beq
\chi_{\pm}\overset{SU(2)_L\times SU(2)_R}{\longrightarrow}K\chi_{\pm}K^{\dagger},\quad \chi_{\pm}\overset{\operatorname{P}}{\longrightarrow}\pm\chi_{\pm},\quad \chi_{\pm}\overset{\operatorname{C}}{\longrightarrow}\chi_{\pm}^{T}.
\eeq
For the construction of the chiral Lagrangian, it will be useful to separate every field into isovector and isoscalar components. Therefore we define
\beq
\tilde{X}=X-\frac{1}{2}\braket{X},
\eeq
where $\braket{\cdot}$ denotes the flavor trace such that $\tilde{X}$ projects out the isovector part of $X$. The basic building blocks in this form are therefore $u_{\mu},\hat{u}_{\mu},\tilde{\chi}_{\pm},\braket{\chi_{\pm}}$. Note that in the special case of Pauli matrices, one finds
\beq\label{eq:TraceRel}
\braket{\tau_i \tau_j} \mathbb{1}=\left\lbrace \tau_i,\tau_j\right\rbrace.
\eeq
We are now in the position to construct the chiral nucleon Lagrangian containing the axion. In the following we, write down the minimal set of operators for which we closely follow \Refcite{Fettes:2000gb}. Invariant monomials take the generic form
\beq
\bar{\Psi}A^{\mu\nu\dots}\Theta_{\mu\nu\dots}\Psi+\hc\,\,.
\label{eq:genericmonomials}
\eeq
The operator $A^{\mu\nu\dots}$ is a product of pion and/or external fields and their covariant derivatives, all of course in the non-linear representation. $\Theta_{\mu\nu\dots}$ is a product of Clifford algebra elements $\Gamma_{\mu\nu\dots}$ and the symmetrized product of $n$ covariant derivatives $D^{n}_{\alpha\beta\dots}=\left\lbrace D_{\alpha},\left\lbrace D_{\beta},\left\lbrace\dots,D_{\omega}\right\rbrace\right\rbrace\right\rbrace$ acting on the nucleon field. 

$\Gamma_{\mu\nu\dots}$  contains Clifford algebra elements which are understood to be expanded in the basis $(\mathbb{1},\gamma_5,\gamma_{\mu},\gamma_{5}\gamma_{\mu},\sigma_{\mu\nu})$ and the metric $\eta_{\mu\nu}$ as well as Levi-Civita symbols $\varepsilon_{\mu\nu\alpha\beta}$. \Eq{eq:genericmonomials} is not the most general form. The fact that $\Theta_{\mu\nu\dots}$ contains only the totally symmetrized product of covariant derivatives acting on $\Psi$ is due to the curvature relation, see \Eq{eq:curvrel}. Another property, namely that no two indices of $\Theta_{\mu\nu\dots}$ are contracted (except for Levi-Civita symbols) is due the fact that at a given chiral order $\slashed{D}\Psi$ can always be replaced by $-i\mn\Psi$. This explains why a couple of Clifford algebra structures do not need to be considered, \eg~the LHS of $\bar{\Psi}A^{\alpha\beta\dots}\gamma^{\lambda}D^{n}_{\lambda\alpha\beta\dots}\Psi+\hc=-i\mn\bar{\Psi}A^{\alpha\beta\dots}D^{n-1}_{\alpha\beta\dots}\Psi$ is replaced by the RHS. More such eliminations due to the use of the EOM can be made and the minimal set is shown, ordered by ascending chiral power, given in \Refcite{Fettes:2000gb} to be
\beq
\begin{aligned}
&\mathbb{1},\\
&\gamma_5\gamma_{\mu},D_{\mu},\\
&\eta_{\mu\nu},\sigma_{\mu\nu},\gamma_5\gamma_{\mu}D_{\nu},D_{\mu\nu},
\end{aligned}
\label{eq:minimalTheta}
\eeq
where $D_{\mu\nu}=\left\lbrace D_{\mu},D_{\nu}\right\rbrace$.

For the operator $A^{\mu\nu\dots}$ one writes down combinations of fields in every ordering. The number of Lorentz indices gives the chiral power. Furthermore, it is instructive to write them in terms of commutator and anticommutator. This has the advantage to obtain simple transformation properties for charge conjugation and parity. In the special case of $SU(2)$ generators this is equivalent to write as isovector and isoscalar combinations respectively, see \Eq{eq:TraceRel}. The transformation behaviour of the Clifford algebra elements under parity-, charge- and hermitian conjugation in the form
\beq
\begin{aligned}
\Gamma^p_{\mu\nu\dots}&=(\pm1){\mathcal{P}_{\mu}^{}}^{\alpha}{\mathcal{P}_{\nu}^{}}^{\beta}\dots\Gamma_{\alpha\beta\dots}\,,\quad \Gamma^c_{\mu\nu\dots}=(\pm1) \Gamma_{\mu\nu\dots},\\ 
\Gamma^{\dagger}_{\mu\nu\dots}&=(\pm1)^{\gamma_{\dagger}} \gamma_0\Gamma_{\mu\nu\dots}
\end{aligned}
\eeq
and the chiral order (including the metric and the Levi-Civita) are shown in \Tab{tab:Cliff}. Note that the full transformation behavior of $\bar{\Psi}\Theta_{\mu\nu\dots\alpha\beta\dots}\Psi=\bar{\Psi}\Gamma_{\mu\nu\dots}D^n_{\alpha\beta\dots}\Psi$ may include an additional sign due to integration by parts, but is included in \Tab{tab:Cliff}. All relevant operators from the minimal list of $\Theta_{\mu\nu\dots}$ are shown, see \Eq{eq:minimalTheta}.
\begin{table}[H]
    \centering
    \begin{tabular}{|c|c|c|c|c|c|c|c|c|c|c|}
    \hline
         $\Theta_{\mu\nu\dots}$& $\mathbb{1}$&$\gamma_5$ & $\gamma_{\mu}$ & $\gamma_5\gamma_{\mu}$ & $\sigma_{\mu\nu}$& $\eta_{\mu\nu}$ & $\epsilon_{\mu\nu\lambda\rho}$ & $D_{\mu}$&$\gamma_5\gamma_{\mu}D_{\nu}$ &$D_{\mu\nu}$  \\ \hline
        $O(p^n)$ & $O(1)$& $O(p)$ & $O(1)$&$O(1)$&$O(1)$&$O(1)$&$O(1)$&$O(1)$ &$O(1)$ &$O(1)$ \\
         P &$+$& $-$&$+$&$-$&$+$ &$+$ & $-$& $+$&$-$&$+$\\
         C &$+$& $+$&$-$&$+$&$-$&$+$&$+$&$-$&$-$&$+$\\
         $\dagger$&$+$ & $-$&$+$&$+$&$+$&$+$&$+$&$-$&$-$&$+$\\
         \hline
    \end{tabular}
    \caption{Transformation properties of $\Theta_{\mu\nu\dots}$.}
    \label{tab:Cliff}
\end{table}
The transformation behaviour of the fields $A^{\mu\nu\dots}$ under parity, charge and hermitian conjugation in the form
\beq
\begin{aligned}
A^p_{\mu\nu\dots}&=(\pm1)^{a_p}{\mathcal{P}_{\mu}^{}}^{\alpha}{\mathcal{P}_{\nu}^{}}^{\beta}\dots A_{\alpha\beta\dots}\quad A^c_{\mu\nu\dots}=(\pm1)^{a_c} A_{\mu\nu\dots}^T\quad A^{\dagger}_{\mu\nu\dots}=(\pm1)^{a_{\dagger}} A_{\mu\nu\dots}
\end{aligned}
\eeq
as well as the chiral order (including the metric and Levi-Civita) and the combinations of covariant derivatives acting on nucleons are shown in \Tab{tab:fields}. 
\begin{table}[H]
    \centering
    \begin{tabular}{|c|c|c|c|}
    \hline
        $ A_{\mu\nu\dots}$ & $\operatorname{P}$ &$\operatorname{C}$& $\dagger$   \\ \hline \hline
	\multicolumn{4}{|c|}{$O(p)$} \\
	\hline
         $u_{\mu}$&$-$&$+$&$+$\\ 
          $\hat{u}_{\mu}$&$-$&$+$&$+$\\ \hline
	\multicolumn{4}{|c|}{$O(p^2)$} \\
	\hline
	$\tilde{\chi}_{+}$&$+$&$+$&$+$\\
	$\braket{\chi_{+}}$&$+$&$+$&$+$\\
	$\tilde{\chi}_{-}$&$-$&$+$&$-$\\
	$\braket{\chi_{-}}$&$-$&$+$&$-$\\
	$\left[u_{\mu},u_{\nu}\right]$&$+$&$-$&$-$\\
	$\braket{u_{\mu}u_{\nu}}$&$+$&$+$&$+$\\
	$\left\lbrace\hat{u}_{\mu},u_{\nu}\right\rbrace$&$+$&$+$&$+$\\
	$\left\lbrace\hat{u}_{\mu},\hat{u}_{\nu}\right\rbrace$&$+$&$+$&$+$\\
	$\left[D_{\mu},u_{\nu}\right]$&$-$&$+$&$+$\\
	$\left[D_{\mu},\hat{u}_{\nu}\right]$&$-$&$+$&$+$\\
	\hline
    \end{tabular}
    \caption{Transformation properties of $A_{\mu\nu\dots}$.}
    \label{tab:fields}
\end{table}
With the transformation properties at hand we can write down all operators invariant under parity, charge conjugation and hermitian conjugation at a given chiral order. At first chiral order $O(p)$ we find the relativistic pion nucleon Lagrangian, including the axion
\beq
\LOneRel=\bar{\Psi}\left(i\slashed{D}-\mn+\frac{\ga}{2}\slashed{u}\gamma_5+\frac{\go}{2}\slashed{\hat{u}}\gamma_5\right)\Psi,
\eeq
given in \Eq{eq:baryon_lag_ChPT}.
At the next higher chiral order $O(p^2)$ we find the most general Lagrangian to be
\beq
\label{eq:LTwoRel}
\begin{aligned}
\LTwoRel=&\,\bar{\Psi}\left[c_1\braket{\chi_{+}}-\left(\frac{c_2}{8\mn^2}\braket{u_{\mu}u_{\nu}}D^{\mu\nu}+\hc\right)+\frac{c_3}{2}u\cdot u+c_4\frac{i}{4}\left[u_{\mu},u_{\nu}\right]\sigma^{\mu\nu}\right.\\ 
&\,\left.+c_5\tilde{\chi}_{+} -\left(\frac{c_8}{8\mn^2}\left\lbrace\hat{u}_{\mu},u_{\nu}\right\rbrace D^{\mu\nu}+\hc\right) +\frac{c_{9}}{2}\hat{u}\cdot u
\right]\Psi.
\end{aligned}
\eeq
Note that we dropped all $1/f_a^2$ terms, all terms $\propto\left[\hat{u}_{\mu},u_{\nu}\right]\sim\partial_{\mu}a\partial_{\nu}\pi^a\left[\mathbb{1},\tau^a\right]=0$, as well as terms $\propto\left[\hat{u},\hat{u}\right]=0$. Note that $c_{6,7}$ are terms proportional to the field strength tensor, which we do not include here.

Next, we treat the nucleons in the HBChPT limit, as described in \Sec{sec:HBChPT}.
We define projection operators according
\begin{equation}
    \Ppm=\frac{1\pm\slashed{v}}{2},
\end{equation}
and introduce velocity-dependent fields, \ie velocity eigenstates, $\Nv$ and $\Hv$ 
\begin{equation}
    \Nv\equiv e^{i\mn v\cdot x}\Pp \Psi,\quad \Hv\equiv e^{i\mn v\cdot x}\Pm \Psi,
\end{equation}
satisfying
\beq
\slashed{v}\Nv=+\Nv,\quad\slashed{v}\Hv=-\Hv.
\eeq
Note that we drop the subscript in the main text however it should also there be understood as velocity-dependent.
Therefore the nucleon field is decomposed as
\begin{equation}
    \Psi(x)=e^{-i\mn v\cdot x}\left[\Nv(x)+\Hv(x)\right].
    \label{eq:FieldDecomposition}
\end{equation}
Let us consider a positive plane wave solution to the free Dirac equation with momentum~$\vec{p}$
\begin{align}
    \psi_{\vec{p}}^{(+)(\alpha)}(\vec{x},t)&=u^{(\alpha)}(\vec{p})e^{-i p\cdot x},\\
    u^{(\alpha)}(\vec{p})&=\sqrt{E(\vec{p})+\mn}\begin{pmatrix}\chi^{(\alpha)}\\ \frac{\vec{\sigma}\cdot\vec{p}}{E(\vec{p})+\mn}\chi^{(\alpha)}\end{pmatrix}
\end{align}
where $E(\vec{p})=p^{0}=\sqrt{\vec{p}^2+\mn^2}$ and with our choice of $v^{\mu}$, they are given by
\begin{equation}
    \Pp=\begin{pmatrix}\mathbb{1}_{2\times2}& 0_{2\times2}\\ 0_{2\times2}&0_{2\times2}\end{pmatrix},\quad \Pm=\begin{pmatrix}0_{2\times2}& 0_{2\times2}\\ 0_{2\times2}&\mathbb{1}_{2\times2}\end{pmatrix}.
\end{equation}
Therefore, the velocity eigenstates take the form
\begin{align}
    \Nv^{(\alpha)}(x)&=\sqrt{E(\vec{p})+\mn}
    \begin{pmatrix}
        \chi^{(\alpha)}\\ 0_{2\times 1}
    \end{pmatrix}
    e^{-i\left[E(\vec{p})-\mn\right]t+i \vec{p}\cdot\vec{x}},\\
    \Hv^{(\alpha)}(x)&=\sqrt{E(\vec{p})+\mn}
    \begin{pmatrix}
        0_{2\times 1}\\ 
        \frac{\vec{\sigma}\cdot\vec{p}}{E(\vec{p})+\mn}\chi^{(\alpha)}
    \end{pmatrix}
    e^{-i\left[E(\vec{p})-\mn\right]t+i\vec{p}\cdot\vec{x}}.
\end{align}
Next, we integrate out the heavy component. We plug the decomposed nucleon field, \Eq{eq:FieldDecomposition} into the leading order EOM from \Eq{eq:baryon_lag_ChPT} and get rid of the phase. Then we project out $\Pp$ and $\Pm$ components, solve one formally for $\Hv$ using the definitions
\begin{equation}
    A^{\mu}_{\perp}=A^{\mu}-v\cdot A v^{\mu},\quad v\cdot A_{\perp}=0, \quad \slashed{A}_{\perp}=A^{\mu}_{\perp}\gamma_{\mu}.
\end{equation}
and plug it into the remaining equation, which leads to the following Lagrangian
\begin{equation}
\begin{aligned}
     \hat{\Lag}=&\,\Nvbar\left(iv \cdot D+\frac{\ga}{2}\slashed{u}_{\perp}\gamma_5+\frac{\go}{2}\slashed{\hat{u}}_{\perp}\gamma_5\right)\Nv+\Nvbar\left(i\slashed{D}_{\perp}+\frac{\ga}{2}v\cdot u\gamma_5+\frac{\go}{2}v\cdot \hat{u}\gamma_5\right)\\
     &\times\left(2\mn+i v\cdot D - \frac{\ga}{2}\slashed{u}_{\perp}\gamma_5- \frac{\go}{2}\slashed{\hat{u}}_{\perp}\gamma_5\right)^{-1}\left(i\slashed{D}_{\perp}-\frac{\ga}{2}v\cdot u\gamma_5-\frac{\go}{2}v\cdot \hat{u}\gamma_5\right)\Nv.
    \label{eq:LagEff}
\end{aligned}
\end{equation}
Now we define the velocity-dependent spin operator 
\begin{equation}
    S^{\mu}_{v}\equiv\frac{i}{2}\gamma_5\sigma^{\mu\nu}v_{\nu}=-\frac{1}{2}\left(\gamma^{\mu}\slashed{v}-v^{\mu}\right),\quad \left(S^{\mu}_{v}\right)^{\dagger}=\gamma_0 S^{\mu}_{v}\gamma_0,
\end{equation}
where again we drop the $v$ sub- or superscript in the main text. It has the properties
\begin{equation}
    v\cdot S_{v}=0,\quad\left\lbrace S_{v}^{\mu},S_{v}^{\nu}\right\rbrace=\frac{1}{2}\left(v^{\mu}v^{\nu}-\eta^{\mu\nu}\right),\quad \left[S_{v}^{\mu},S_{v}^{\nu}\right]=i\epsilon^{\mu\nu\rho\sigma}v_{\rho}S_{\sigma}^{v}.
\end{equation}
This definition is very useful since we can trade the 6 different Dirac structures that appear in \Eq{eq:LagEff} with two $4\times 4$ matrices, $\mathbb{1}_{4\times 4}$ and $S_{v}$. To that end we use the relations
\begin{subequations}
\label{eq:NvRelations}
    \begin{align}
    \Nvbar \mathbb{1}_{4\times 4}\Nv&=\Nvbar \mathbb{1}_{4\times 4}\Nv,\\
    \Nvbar \gamma_5\Nv&=0,\\
    \Nvbar \gamma^{\mu}\Nv&=v^{\mu}\Nvbar\Nv,\\
    \Nvbar \gamma^{\mu}\gamma_5\Nv&=2\Nvbar S_{v}^{\mu}\Nv,\\
    \Nvbar \sigma^{\mu\nu}\Nv&=2\epsilon^{\mu\nu\sigma\rho}v_{\rho}\Nvbar S_{\sigma}^{v}\Nv,\\
    \Nvbar \sigma^{\mu\nu}\gamma_5\Nv&=2i\left(v^{\mu}\Nvbar S^{\nu}_{v}\Nv-v^{\nu}\Nvbar S^{\mu}_{v}\Nv\right).
\end{align}
\end{subequations}
We expand the second bracket in the last term of \Eq{eq:LagEff} in orders of $1/\mn^n$ and write
\begin{equation}
\label{eq:LhatFormal}
    \hat{\Lag}=\LOneHB+\sum_{n=1}^{\infty}\frac{1}{\left(2\mn\right)^n}\hat{\Lag}^{(n)},
\end{equation}
where the first term is exactly given by
\begin{equation}
    \LOneHB=\Nvbar\left(iv\cdot D+\ga S_{v}\cdot u +\go S \cdot \hat{u}\right)\Nv,
\end{equation}
which we recognize as the leading pion nucleon HBChPT Lagrangian including the axion, see \Eq{eq:leading_order_HBChPT_lag}. The first term in the sum leads to relativistic corrections at chiral order $O(p^2)$.

Despite relativistic corrections from integrating out the heavy component from the leading order Lagrangian, the other contribution comes from the Lagrangian \Eq{eq:LTwoRel}, and we write at chiral order $O(p^2)$ 
\begin{equation}
     \LagHBn{2}= \LagHBnc{2}+ \LagHBnm{2}.
\end{equation}
Note that relativistic corrections from integrating out $\Hv$ from the EOM of $\LTwoRel$ will enter at $O(p^3)$, which we do not need for our analysis in full generality. Therefore, it is sufficient to consider the $\Pp$ projection of the $O(p^2)$ EOM and only keep the term proportional to the light component $\Nv$, since the term $\propto\Hv$ will only generate relativistic $1/\mn^2$ corrections of chiral order $O(p^3)$. Then, using the relations \Eq{eq:NvRelations}, one finds
\begin{equation}
\label{eq:LTwoHBcApp}
    \begin{aligned}
         \LagHBnc{2}=& \Nvbar \bigg( \ci1 \left\langle\chi_{+}\right\rangle + \frac{\ci2}{2} (v \cdot u)^{2} + \ci3 \left(u\cdot u\right) + \ci4  \left[S^{\mu}, S^{\nu}\right] u_{\mu} u_{\nu} + \ci5 \widetilde{\chi}_{+} + \bigg. \\
 & \bigg. + \frac{\ci8}{4}(v \cdot u)(v \cdot \hat{u}) + \ci9 \left(u\cdot\hat{u}\right) \bigg)\Nv.
    \end{aligned}
\end{equation}
The relativistic corrections come from the first term in the sum of \Eq{eq:LhatFormal}, namely
\begin{equation}
\label{eq:LTwoHBm}
\begin{aligned}
     \LagHBnm{2}=&\,\,\frac{1}{2\mn}\hat{\Lag}^{(1)}\\
    \simeq&\,\,\frac{1}{2\mn}\Nvbar\left(i\slashed{D}_{\perp}+\frac{\ga}{2}v\cdot u\gamma_5+\frac{g_0}{2}v\cdot \hat{u}\gamma_5\right)\left(i\slashed{D}_{\perp}-\frac{\ga}{2}v\cdot u\gamma_5-\frac{g_0}{2}v\cdot \hat{u}\gamma_5\right)\Nv\\
    =&\,\,\frac{1}{2\mn}\Nvbar\bigg[-D^2+(v\cdot D)^2-i\ga\left\lbrace D\cdot S_v,v\cdot u\right\rbrace-ig_0\left\lbrace D\cdot S_v,v\cdot \hat{u}\right\rbrace \bigg.\\
    &\bigg.-\frac{\ga^2}{4}\left(v\cdot u\right)^2-\frac{g_0^2}{4}\left(v\cdot \hat{u}\right)^2-\frac{\ga g_0}{2}\left(v\cdot u\right)\left(v\cdot\hat{u}\right)+\frac{i}{4}\epsilon^{\mu\nu\rho\sigma}\left[u_{\mu},u_{\nu}\right]v_{\rho}S^{v}_{\sigma}\bigg]\Nv,
\end{aligned}
\end{equation}
where we used
\begin{subequations}
\begin{align}
    \slashed{D}_{\perp}\slashed{D}_{\perp}&=D^2-\left(v\cdot D\right)^2-\frac{i}{4}\epsilon^{\mu\nu\rho\sigma}\left[u_{\mu},u_{\nu}\right]v_{\rho}S^{v}_{\sigma},\\
    \slashed{D}_{\perp}\gamma_5&=2D\cdot S_{v}.
\end{align}
\end{subequations}
The result is therefore
\begin{equation}\label{eq:NLOLagApp}
    \begin{aligned}
         \LagHBn{2}=&\LagHBnc{2}+ \LagHBnm{2}=\\
         =&\Nvbar\left[-\frac{1}{2\mn}\left(D^2-\left(v\cdot D\right)^2+i\ga\left\lbrace S_v\cdot D,v\cdot u\right\rbrace+i\go\left\lbrace S_v\cdot D,v\cdot\hat{u}\right\rbrace\right)\right.\\
        &+\ch1\braket{\chi_{+}}+\frac{\ch2}{2}\left(v\cdot u\right)^2+\ch3\left(u\cdot u\right)+\frac{\ch4}{2} i\epsilon^{\mu\nu\rho\sigma}\left[u_{\mu},u_{\nu}\right]v_{\rho}S^v_{\sigma}\\
        &+\left.\ch5\tilde{\chi}_{+}+\frac{\ch8}{4}\left(v\cdot u\right)\left(v\cdot\hat{u}\right)+\ch9\left(u\cdot\hat{u}\right)\right]\Nv,
    \end{aligned}
\end{equation}
where $\ch{i}$ include relativistic $1/\mn$ corrections,
\begin{equation}
    \arraycolsep=1.4pt\def\arraystretch{2}
    \begin{array}{cccc}
    \displaystyle{
    \ch1=\ci1},\quad&\displaystyle{\ch2=\ci2-\frac{\ga^2}{4\mn},\quad }&\displaystyle{\ch3=\ci3,\quad}&\displaystyle{\ch4=\ci4+\frac{1}{4\mn},}\\ \vspace{5pt}
    \displaystyle{\ch5=\ci5,}\quad&\displaystyle{\ch8=\ci8-\frac{\ga\go}{\mn},\quad}&\displaystyle{\ch9=\ci9.}\quad&
\end{array}
\end{equation}
Now there are still terms in $\hat{\Lag}^{(2)}_{\pi N,1/\mn}$ that can be eliminated by appropriate field re-definitions. One can, for instance, eliminate terms $\propto \left(v\cdot D\right)^2$. However, since this term does not contribute to our process in practice we do not eliminate it. For completeness, we define
\begin{equation}
     \Nv\rightarrow\left[1+\frac{iv \cdot D}{4\mn}-\frac{\ga S_{v}\cdot u}{4\mn}-\frac{g_0 S_{v}\cdot \hat{u}}{4\mn}\right]\Nv,
\end{equation}
and insert it into \Eq{eq:leading_order_HBChPT_lag} gives, up to total derivatives,
\begin{equation}
\begin{aligned}
    \LOneHB\rightarrow&\,\,\LOneHB+\Delta \LagHBn{2}\\
    =&\,\,\LOneHB+\frac{1}{2\mn}\Nvbar\left[-(v\cdot D)^2-\frac{\ga^2}{4}i\epsilon^{\mu\nu\rho\sigma}\left[u_{\mu},u_{\nu}\right]v_{\rho}S^v_{\sigma}-\frac{\ga^2}{2}\left(v\cdot u\right)^2\right.\\
    &\left.+\frac{\ga^2}{4}\left(u\cdot u\right)-\frac{\ga\go}{4}\left\lbrace v\cdot u,v\cdot \hat{u}\right\rbrace+\ga\go\left(u\cdot\hat{u}\right)\right]\Nv,
\end{aligned}
\end{equation}
and we neglected terms $O(1/\mn^2,1/f_a^2)$. The first term cancels the EOM term in \Eq{eq:LTwoHBm}, while the remaining terms shift the constants in \Eq{eq:LTwoHBcApp}. In total, we find the HBChPT Lagrangian at chiral order $O(p^2)$ as the sum $\LagHBnc{2}+ \LagHBnm{2}+\Delta \LagHBn{2}$ to be
\begin{equation}
    \begin{aligned}
        & \LagHBnc{2}+ \LagHBnm{2}+\Delta \LagHBn{2}\\
         &=\Nvbar\left[-\frac{1}{2\mn}\left(D^2+i\ga\left\lbrace S_v\cdot D,v\cdot u\right\rbrace+i\go\left\lbrace S_v\cdot D,v\cdot\hat{u}\right\rbrace\right)\right.\\
        &\quad+\frac{\ai1}{\mn}\braket{\chi_{+}}+\frac{\ai2}{2\mn}\left(v\cdot u\right)^2+\frac{\ai3}{\mn}\left(u\cdot u\right)+\frac{\ai4}{2\mn} i\epsilon^{\mu\nu\rho\sigma}\left[u_{\mu},u_{\nu}\right]v_{\rho}S^v_{\sigma}\\
        &\quad+\left.\frac{\ai5}{\mn}\tilde{\chi}_{+}+\frac{\ai8}{8\mn}\left\lbrace v\cdot u,v\cdot\hat{u}\right\rbrace+\frac{\ai9}{\mn}\left(u\cdot\hat{u}\right)\right]\Nv
    \end{aligned}
\end{equation}
where
\begin{equation}
    \arraycolsep=1.4pt\def\arraystretch{2}
    \begin{array}{cccc}
    \displaystyle{\frac{\ai1}{\mn}=\ci1,\quad}&\displaystyle{\frac{\ai2}{\mn}=\ci2-\frac{\ga^2}{8\mn},\quad }&\displaystyle{\frac{\ai3}{\mn}=\ci3+\frac{\ga^2}{8\mn},\quad}&\displaystyle{\frac{\ai4}{\mn}=\ci4+\frac{1-\ga^2}{4\mn},}\\
    \displaystyle{\frac{\ai5}{\mn}=\ci5,}\quad&\displaystyle{\frac{\ai8}{\mn}=\ci8-\frac{\ga\go}{2\mn},\quad}&\displaystyle{\frac{\ai9}{\mn}=\ci9+\frac{\ga\go}{2\mn}.\quad}&
\end{array}
\end{equation}
Note, however, that for convenience in the main text, we will work with the redundant basis of \Eq{eq:NLOLagApp}.
\subsection{Construction of nucleon contact terms}
We now construct terms that have four nucleon fields instead of two. We split the construction into pure nucleon contact terms and terms with four nucleons as well as pions or axions.
\subsubsection{Pure nucleon contact terms}

We start by constructing all possible contact terms containing not more than four nucleons and no other fields. According to \Tab{tab:Cliff}, the only terms with correct transformation properties are
\begin{equation} \label{eq:basis_contact}
\bar{N} {N} \bar{N} {N}, \quad \bar{N} S^\mu {N} \bar{N} S_\mu {N}.
\end{equation}
Further we can add any number of isospin matrices $\tau^a$, \eg $\bar{N} \tau^a {N} \bar{N} \tau^a {N}$ or one including spin matrices such as $\bar{N} S^\mu \tau^a {N} \bar{N} S_\mu \tau^a {N}$, as long as all their indices are contracted in the end. It can be shown by doing simple algebraic manipulations that all of them can be written as a linear combination of the operators given in \Eq{eq:basis_contact}, which are, therefore, the only two linear independent operators one can construct. Hence, the Lagrangian for the 2-nucleon contact terms is given by
\begin{equation}
\mathcal{L}_{N N}=-\frac{1}{2} C_{S}(\bar{N} N)(\bar{N} N)+2 C_{T}(\bar{N} S N) \cdot(\bar{N} S N),
\end{equation}
with the two well known constants $C_{S}$ and $C_{T}$. Note that the constant $F_\pi^2 C_T \approx 0.02$ is very small \cite{Epelbaum:2008ga} and its contribution therefore typically negligible. This is, however, not necessarily the case for the $C_S$ term.

\subsubsection{Higher order nucleon contact terms}

Next, we aim to construct all possible 2-nucleon contact terms, which include either the field $u_\mu$ or $\hat{u}_\mu$. Those are the only fields with chiral dimension one from table \Tab{tab:fields}. We start with the ones including $u_\mu$. From  \Tab{tab:fields} and \Tab{tab:Cliff} the only terms one can construct to fulfill the right transformation properties are
\begin{equation}
\bar{N} S^\mu {N} \bar{N} u_\mu {N}, \quad \bar{N} {N} \bar{N} S^\mu u_\mu {N}, \quad \epsilon^{a b c} \epsilon^{\mu \nu \rho \sigma} v_\sigma \bar{N} S_\nu \tau^b {N} \bar{N} S_\rho \tau^c {N} u^a_\mu.
\end{equation}
It can easily be shown that taking the non-relativistic limit and applying Fierz and other basic identities, these are in fact not independent of each other. The explicit relations are $ \bar{N} S^\mu {N} \bar{N} u_\mu {N} = - \bar{N} {N} \bar{N} S^\mu u_\mu {N}$ and $ 8 \bar{N} S^\mu {N} \bar{N} u_\mu {N} =  \epsilon^{a b c} \epsilon^{\mu \nu \rho \sigma} v_\sigma \bar{N} S_\nu \tau^b {N} \bar{N} S_\rho \tau^c {N} u^a_\mu$. Therefore we only need one of them in the Lagrangian which following the literature (\eg~\cite{Epelbaum:2002vt}) we write as
\begin{equation}\label{eq:NLOLagApp2}
\mathcal{L}_{\pi N N} \supset \frac{c_{D}}{2 f_{\pi}^{2} \Lambda_{\chi}}(\bar{N} N)\left(\bar{N} S^{\mu} u_{\mu} N\right).
\end{equation}

For the terms involving $\hat{u}_\mu$ things are even simpler because $\hat{u}_\mu$ is proportional to the identity matrix in isospin space. There are only two possible combinations satisfying the transformation properties given by
\begin{equation}
\bar{N} {N} \bar{N} S^\mu \hat{u}_\mu {N}, \quad \bar{N} \tau^a {N} \bar{N} S^\mu \hat{u}_\mu \tau^a {N}.
\end{equation}
Again, it can easily be seen that  $ \bar{N} \tau^a {N} \bar{N} S^\mu \hat{u}_\mu \tau^a {N} = - 3 \bar{N} {N} \bar{N} S^\mu \hat{u}_\mu {N} $ and so we find for the Lagrangian
\begin{equation}\label{eq:NLOLagApp3}
\mathcal{L}_{\pi N N} \supset \frac{\ca}{2 f_{\pi}^{2} \Lambda_{\chi}}(\bar{N} N)\left(\bar{N} S^{\mu} \hat{u}_{\mu} N\right).
\end{equation}
This concludes the construction of all terms of the Lagrangian necessary to calculate the corrections to the axion-nucleon coupling. Note, however, that the $c_D$ and $\tilde{c}_D$ terms are not $\Delta$-enhanced, and thus, we do not include them in our calculation of the vertex corrections.

\section{Cooling rates at finite density}\label{app:ratedensity}

In this appendix, we review how to calculate emissivities in thermal field theory. 
In \App{app:nuclen_prop_FD} we set the stage with basics about thermal field theory in the real-time formalism and eventually derive the density modified nucleon propagator in.
We note that this propagator can be derived without using thermal field theory, see App. A of \Refcite{Ghosh:2022nzo}.
We continue to use the previously derived propagator to review thermal particle production rates in \App{app:thermal_rates}, have a look at cutting rules at finite density in \App{sec:cutting_rules_FD}, and eventually discuss how to calculate an emissivity for \eg a SN in \App{app:emissivity}.

We would like to note that the results in the subsequent sections can be equally derived in thermal field theory using the imaginary time formalism, see \eg \cite{Weldon:1983jn,Keil:1988sp}.

\subsection{Nucleon propagator in thermal field theory} \label{app:nuclen_prop_FD}

The goal of this subsection is to review the derivation of the propagator of a fermion field interacting with a background. In this section we closely follow the review \Refscite{Landsman:1986uw,Furnstahl:1991vk}.
In thermal field theory one is mainly concerned with generating functionals of the form 
\begin{equation}
    \text{Tr}\left[e^{-\beta \left(\hat{H}-\mu\hat{N}\right)}T\exp\left(i\int_{-t_0}^{t_0}dt\int d^3\Vec{x}\,\,\bar{j}(x)\psi(x)+j(x)\bar{\psi}(x)\right)\right],
\end{equation}
where $t_0\in \mathbb{R}$, $j(x)$ is a classical source and $T$ denotes time ordering. 
Propagators are calculated by taking functional derivatives with respect to the sources.
The apparent problem is a systematic way to evaluate the above trace, which includes real-time exponentials as well as imaginary exponentials.
The solution is to extend the time variable on a contour in the complex plane and to introduce contour ordering $T_c$.

We write the generating functional in its path integral representation
\begin{equation}
    Z\left[j,\bar{j}\right]=\mathcal{N}\int\mathcal{D}\left[\psi,\bar{\psi}\right]\exp\left(i\int_{\mathcal{C}}d^4 x\,(\Lag+\mu\bar{\psi}\gamma_0\psi+\bar{j}\psi+j\bar{\psi})\right),
\end{equation}
subject to fermionic boundary conditions.
We split the Lagrangian into fermion bilinears $\Lag_0+\mu\bar{\psi}\gamma_0\psi$ and the interacting part $\Lag_\text{I}$ and formally solve the generating functional
\begin{equation}
\begin{aligned}
    Z\left[j,\bar{j}\right]=&\,Z_0\exp\left(i\int d^4 y\,\Lag_{\text{I}}\left[\frac{\delta}{i\delta \bar{j}},\frac{\delta}{-i\delta j}\right]
    \right)\\
    &\times\exp\left(-i\int d^4 x\int d^4 z\bar{j}(x)G_{0}^{(c)}(x-z)j(z)\right),
\end{aligned}
\end{equation}
such that the free, contour ordered propagator satisfies
\begin{equation}
    \left[i\slashed{\partial}+\mu\gamma_0-m\right]G_0^{(c)}(x)=\delta^{(4)}(x).
\end{equation}
In terms of retarded and advanced functions we can write the propagator as
\begin{equation}
\label{eq:RetAdvFn}
    iG_0^{(c)}(x)=\theta_c(t)\braket{\psi(x)\bar{\psi}(0)}_0-\theta_c(-t)\braket{\bar{\psi}(0)\psi(x)}_0,
\end{equation}
where $\braket{\cdot}$ denotes the expectation value with respect to the generating functional.
The propagator can be explicitly calculated using
\begin{equation}
\label{eq:FermiStates}
    \begin{aligned}
        \psi(x)&=\sum_s\int\frac{d^3k}{(2\pi)^3}\frac{1}{\sqrt{2\omega_k}}\left(a^s_ku^s_ke^{-ikx}+b^{s\dagger}_kv^s_ke^{+ikx}\right)e^{i\mu t},\\
        \bar{\psi}(x)&=\sum_s\int\frac{d^3k}{(2\pi)^3}\frac{1}{\sqrt{2\omega_k}}\left(a^{s\dagger}_k\bar{u}^s_ke^{+ikx}+b^{s}_k\bar{v}^s_ke^{-ikx}\right)e^{-i\mu t},
    \end{aligned}
\end{equation}
where we use
\begin{equation}
\label{eq:KMSCond}
    \begin{aligned}
        \braket{a_k^sa_p^{s'\dagger}}&=(2\pi)^3\delta^{ss'}\delta^{(3)}(\vec{k}-\vec{p})\left[1-f_{FD}(\omega_k,\mu)\right],\\
        \braket{b_k^{s\dagger}b_p^{s'}}&=(2\pi)^3\delta^{ss'}\delta^{(3)}(\vec{k}-\vec{p})f_{FD}(\omega_k,-\mu),
    \end{aligned}
\end{equation}
with the Fermi-Dirac distribution
\begin{equation} \label{eq:FD_dist}
    f_{FD}(\omega,\mu)=\frac{1}{e^{\beta(\omega-\mu)}+1}.
\end{equation}
One can now explicitly calculate the retarded and advanced functions from \Eq{eq:RetAdvFn}, using \Eqs{eq:FermiStates}{eq:KMSCond}, to give the contour dependent particle propagator
\begin{equation}
\label{eq:Prop1}
    iG^{(c)}_0(x)=\left(i\slashed{\partial}+m\right)\int\frac{d^4k}{(2\pi)^4}e^{-ikx}\rho_0(k)\left[f_{FD}(-\omega_k,-\mu)\theta_c(t)-f_{FD}(\omega_k,\mu)\theta_c(-t)\right],
\end{equation}
with the spectral density
\begin{equation}
\label{eq:SpecDens}
\begin{aligned}
     \rho_0(k)&=2\pi\left[\theta(k_0)-\theta(-k_0)\right]\delta(k^2-m^2)\\
     &=i\left[\theta(k_0)-\theta(-k_0)\right]\left(\Delta_{0F}(k)-\Delta^{\dagger}_{0F}(k)\right),
\end{aligned}
\end{equation}
where 
\begin{equation}
    \Delta_{0F}(k)=\frac{1}{k^2-m^2+i\epsilon}.
\end{equation}
Taking the Fourier transform, the propagator can be rewritten as
\begin{equation}
    iG^{(c)}_0(k)=iG_{0F}(k)-2\pi(\slashed{k}+m)\delta(k^2-m^2)F_{FD}(\omega_k,\mu),
\end{equation}
where $iG_{0F}(k)=i(\slashed{k}+m)\Delta_{0F}(k)$ and
\begin{equation}
    F_{FD}(\omega_k, \mu) \equiv \theta\left(k^0\right) f_{FD}(\omega_k, \mu)+\theta\left(-k^0\right) f_{FD}(-\omega_k,-\mu).
\end{equation}

\begin{figure}[h!] 
  \centering
\includegraphics[width=0.6\textwidth]{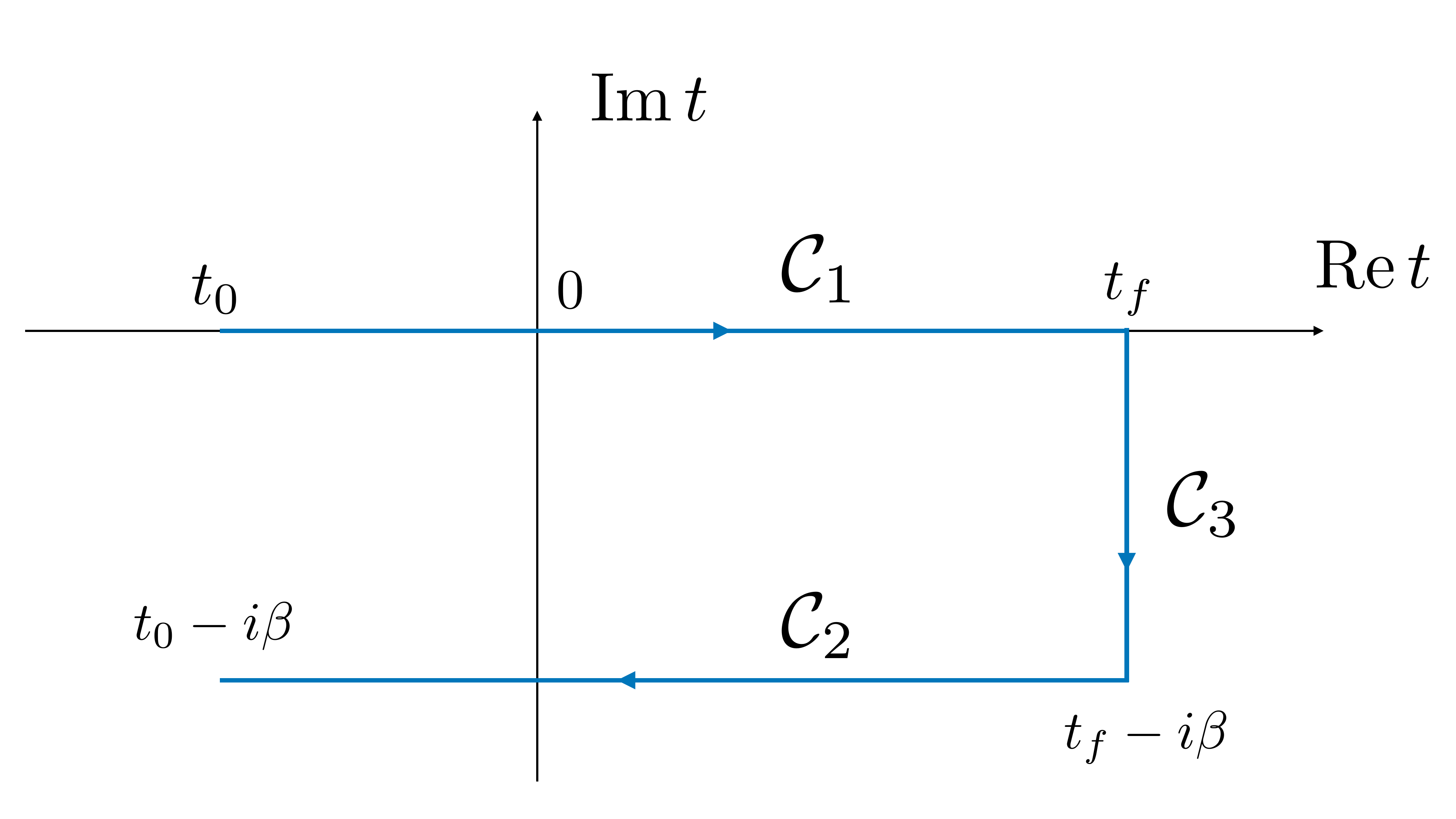}
 \caption[]{Real-time contour $\mathcal{C}=\mathcal{C}_1\cup\mathcal{C}_2\cup\mathcal{C}_3$.}
  \label{fig:contour}
\end{figure}

At this point we need to specify a contour $\mathcal{C}$ for the complex time $t$. 
Instead of taking the obvious purely imaginary contour from $t_0\to t_0-i\beta$ which gives rise to the imaginary time formalism, we will be interested in the real-time contour shown in \Fig{fig:contour}.
As discussed in \Refscite{Landsman:1986uw,Furnstahl:1991vk}, it can be shown that for the contour $\mathcal{C}=\mathcal{C}_1\cup\mathcal{C}_2\cup\mathcal{C}_3$ the generating functional factorizes in the limit of infinite real times $t_0,t_f$
\begin{equation}
    \lim_{t_f\to\infty}\lim_{t_0\to-\infty}Z[\bar{j},j]=Z_{12}[\bar{j},j]Z_{3}[\bar{j},j],
\end{equation}
with the important implication that all convolutions of operators from $\mathcal{C}_{1,2}$ with $\mathcal{C}_{3}$ vanish.
Therefore, we can focus on $Z_{12}[\bar{j},j]$ and write
\begin{equation}
\begin{aligned}
    Z_{12}\left[j,\bar{j}\right]=&\,Z_0\exp\left(i\int d^4 y\,\left\lbrace\Lag_{\text{I}}\left[\frac{\delta}{i\delta \bar{j}_1},\frac{\delta}{-i\delta j_1}\right]-\Lag_{\text{I}}\left[\frac{\delta}{i\delta \bar{j}_2},\frac{\delta}{-i\delta j_2}\right]\right\rbrace\right)\\
    &\times\exp\left(-i\int d^4 x\int d^4 z\bar{j}_{k}(x)G_{0}^{(kl)}(x-z)j_{l}(z)\right),
\end{aligned}
\end{equation}
where the indices indicate one of the contours, $k,l=1,2$, and the second term in the exponential comes with a minus since time runs opposite on $\mathcal{C}_2$ with respect to $\mathcal{C}_1$.

Now we can construct the propagator for arbitrary times $t\in\mathcal{C}_1\cup\mathcal{C}_2$, and importantly convolutions of operators thereof.
Using \Eq{eq:Prop1} together with \Eq{eq:SpecDens} we find
\begin{equation}
\begin{aligned}
     i\boldsymbol{G}(k)&=i\boldsymbol{G}_{0F}(k)+i\boldsymbol{G}_{0T}(k)\\
     &=\begin{pmatrix}
    iG_{0F}(k)& \\
    &-iG_{0F}^{\dagger}(k)
    \end{pmatrix}
    -2\pi(\slashed{k}+m)\delta(k^2-m^2)\sin\Theta(k)
    \begin{pmatrix}
        \sin\Theta(k) & \cos\Theta(k)\\
        \cos\Theta(k) & \sin\Theta(k)
    \end{pmatrix},
\end{aligned}
\end{equation}
where $\sin^2\Theta(k)=F_{FD}(\omega_k,\mu)$.
Therefore one finds that the propagator becomes a $2\times 2$ dimensional matrix at finite temperature.
For instance the 11-component is given by
\begin{equation}\label{eq:finiteTPropRel}
    i\boldsymbol{G}^{(11)}(k)=(\slashed{k}+m)\left[\frac{i}{k^2-m^2+i\epsilon}-2\pi\delta(k^2-m^2)F_{FD}(\omega_k,\mu)\right].
\end{equation}
In the limit $\beta\to\infty$ the situation simplifies since the propagator diagonalizes \ie  since $\cos\Theta(k)\to0$. 
Furthermore, since the external states of any diagram are always fixed to be on the (real) contour $\mathcal{C}_1$ all one ever needs at zero temperature is
\begin{equation}
\label{eq:nuc}
\begin{aligned}
    i\boldsymbol{G}^{(11)}(k)&=(\slashed{k}+m)\left[\frac{i}{k^2-m^2+i\epsilon}-2\pi\delta(k^2-m^2)\theta(k^0)\theta(k_f-|\Vec{k}|)\right]\\
    &\simeq\frac{i}{l^0+i\epsilon}-2\pi\delta(l^0)\theta(k^0)\theta(k_f-|\Vec{l}|)
\end{aligned}
\end{equation}
where we expanded the propagators for non-relativistic nucleons by decomposing their momenta as before $k^{\mu}=mv^{\mu}+l^{\mu}$.

In the above analysis, we considered the background of a single Dirac fermion. In case of nucleons, an isospin doublet, the above derivation goes through analogously with the difference that the chemical potentials of the nuclei $\mu_N$ (and therefore also their Fermi momenta in the limit $T \rightarrow 0$) are different for protons and neutrons. One finds, at finite $T$ in the non-relativistic limit
\begin{equation}\label{eq:NRfiniteTnucprop}
    i G(k) = \frac{i}{k^0 + i \epsilon} - 2 \pi \delta(k^0) \left[ \frac{1+\tau^{3}}{2} F_{FD}(\omega,\mu_p) + \frac{1-\tau^{3}}{2} F_{FD}(\omega,\mu_n)\right],
\end{equation}
with the proton and neutron chemical potentials $\mu_p$ and $\mu_n$. The generalization to $T\to0$ is straightforward.

\subsection{Particle production rates}\label{app:thermal_rates}

In vacuum, QFT parameters are related to observable quantities in a straightforward manner. 
The amplitude squared $|M|^2$ of a certain process is linked to an experimentally observable cross section $\sigma$.
The matrix element $M$ is, per construction, defined as the scattering process of $n$ incoming asymptotic states to $m$ outgoing asymptotic states.
In thermal field theory, instead of scattering asymptotic states, we are interested in the evolution of particle distribution functions.

We are interested in the scenario where a particle, say an axion, is produced due to scattering of nucleons within the thermal bath, described by their distribution function, and then is able to free stream out of the bath.
This scenario exactly describes a particle production rate. In vacuum, decay rates are related to the imaginary part of the self-energy.
Similarly, the axion production rate in a nucleon density background is related to the imaginary part of the axion self-energy, see \eg \cite{Weldon:1983jn},
\begin{equation}\label{eq:emiss_1}
    \Gamma_i(p_a)=\frac{1}{E_a}\text{Im}\,\Pi(p_a),
\end{equation}
which is given by the pole of the full propagator
\begin{equation}
    i \boldsymbol{\mathcal{D}}^{(11)}({p_a})=i \mathcal{D}({p_a})-F_{BE}(\omega_a)\left[i \mathcal{D}(p_a)-i \mathcal{D}^{\ast}(p_a)\right],
\end{equation}
with
\begin{equation}
    i\mathcal{D}(p_a)=\frac{i}{p_a^2-m_a^2-\Pi(p_a)+i\epsilon},
\end{equation}
where $p_a$ denotes the axion momentum and $E_a = p_a^0$ its energy. The Bose-Einstein distribution $F_{\rm BE}$ is defined by
\begin{equation} \label{eq:BE_distribution}
    F_{\rm BE}(\omega_k) \equiv \theta(k^0) f_{\rm BE}(\omega_k)+\theta(-k^0) f_{\rm BE}(-\omega_k) \quad \text{and} \quad f_{\rm BE}(\omega_k) \equiv \frac{1}{e^{\beta \omega_k}-1}.
\end{equation}

Note that without a density background, the imaginary part of the self-energy would of course vanish, \ie the axion kinematically cannot decay to nucleons. Note also that as we are assuming the axion is free streaming, it has negligible density, thus the second part of the propagator vanishes, $f_B(\omega_a)=0$\footnote{Dropping this assumption, one would need to be careful, as $\Pi(p_a)$ and the 11-component of the self-energy matrix $\Pi(p_a)_{ab}$ are non-trivially related. It can be shown (see \eg \cite{Keil:1988sp, Kobes:1986za}) that $\operatorname{Im\Pi }(p_a)= \tanh (\beta p_a^0 / 2) \mathrm{Im}\Pi_{11}(p_a)$, which of course in the case of zero axion density reduces to be trivial.}. The imaginary part of the self-energy gets a contribution from the (nucleon) density background. 

$\Gamma_i(p_a)$ describes the production rate of axions with on-shell momentum $p_a$, \ie the probability per time to produce an axion with momentum $p_a$.
To obtain the production rate summed over all possible momenta, $\Gamma$, we have to integrate over the whole phase space and normalize accordingly. As $p_a$ is on-shell and $p_a^0>0$, this is given by
\begin{equation} \label{eq:emiss_2}
d \Gamma = d \Gamma_i(p_a) \frac{V}{(2 \pi)^3} d^3 p_a,
\end{equation}
where $\int V/(2 \pi)^3 d^3 p_a = 1$. We are not only interested in $\Gamma$, our main goal is to calculate the axion emissivity $\dot{\epsilon}_a$ which is the emitted energy per volume and time. In other words
\begin{equation}
d \dot{\epsilon}_a = \frac{E_a}{V T} dP,
\end{equation}
whereas $\Gamma$ is given is the probability per time \ie 
\begin{equation}
d \Gamma = \frac{1}{T} dP,
\end{equation}
and thus
\begin{equation} \label{eq:emiss_3}
d \dot{\epsilon}_a = \frac{E_a}{V} d \Gamma.
\end{equation}
Combining equations \eqref{eq:emiss_1}, \eqref{eq:emiss_2} and \eqref{eq:emiss_3} the emissivity reads
\begin{equation} \label{eq:emissivity_def}
\dot{\epsilon}_a = \int \Gamma_i(p_a) E_a \frac{d^3 p_a}{(2 \pi)^3} = \int 2 \mathrm{Im} \Pi_{11}(p_a) E_a d \Pi_{a},
\end{equation}
where we defined $d \Pi_{a}=d^{3} p_{a} /\left[(2 \pi)^{3} 2 E_{a}\right]$. \\

\noindent In conclusion, calculating the emissivity \Eq{eq:emissivity_def} boils down to calculating $\mathrm{Im} \Pi_{11}(p_a)$ in the real-time formalism of thermal field theory, which is the problem we are going to review next.

\subsection{Cutting rules at finite density} \label{sec:cutting_rules_FD}

In quantum field theory in vacuum, one can use the so-called Cutkosky rules \cite{Cutkosky:1960sp, Veltman:1994wz} to calculate the imaginary part of amplitudes. One can show that twice the imaginary part of an amplitude is equal to the sum of its cuts. Cuts are nothing else than phase space integrals over internal on-shell particles.  This was formulated in a systematic way in \cite{tHooft:1973wag}. A very similar approach can be used within thermal field theory as we will review in the following. \\
The calculation of decay/production rates within thermal field theory was pioneered by Weldon \cite{Weldon:1983jn} using the imaginary time formalism. Not long after, it was shown by Kobes and Semenoff \cite{Kobes:1985kc, Kobes:1986za} that one can derive a set of rules along the lines of \cite{tHooft:1973wag}, which simplify the calculation of the imaginary part of the self-energy function. However, it does not always provide a set of cutting rules as in the vacuum, as there are diagrams left which do not vanish and do not have a simple representation as a cut diagram.

Let us consider the cutting rules at finite density following \cite{Kobes:1985kc, Kobes:1986za, Keil:1988sp, Bedaque:1996af} in the case where we just have diagrams with external states of type 1 (physical states). In such a case it is sufficient to only consider internal vertices of type 1, as was shown in \cite{Kobes:1986za, Keil:1988sp, Bedaque:1996af}. 

As we know from section \Sec{app:nuclen_prop_FD}, the boson/fermion propagator is given by $2 \times 2$ matrix $i \mathbf{D}_{(a b)}$/$i \mathbf{G}_{(a b)}$. The components relevant in the following are for the bosons
\begin{equation}
i \boldsymbol{{D}}_{(11)}(p)  =-i \boldsymbol{{D}}_{(22)}^\star(p) = \frac{i}{p^2-m^2+i \epsilon}+2 \pi F_{\rm BE}(\omega_p) \delta\left(p^2-m^2\right),
\end{equation}
and we define the so-called "circled" propagators
\begin{equation}
i \boldsymbol{{D}}^{ \pm}(p) = 2 \pi\left[\theta ( \pm p^0)+F_{\rm BE}(\omega_p)\right] \delta\left(p^2-m^2\right).
\end{equation}
For fermions the relevant expressions are
\begin{equation}
\begin{aligned}
i \mathbf{G}_{(11)}(p) & =-i \mathbf{G}_{(22)}^*(p) = \frac{i}{\slashed{p}-m+i \epsilon}-2 \pi(\slashed{p}+m) F_{FD \rm}(\omega_p) \delta\left(p^2-m^2\right),
\end{aligned}
\end{equation}
and
\begin{equation} \label{eq:ident_prop}
i \mathbf{G}^{ \pm}(p)=2 \pi(\slashed{p}+m)\left[\theta( \pm p^0)-F_{FD \rm}(\omega_p)\right] \delta\left(p^2-m^2\right).
\end{equation}
For later convenience let us also define $\Sigma_F(p) \equiv \slashed{p} + m$. \\

We do not review the full set of rules, but only the ones needed to calculate the axion production rate at finite density. In the $T\to0$ limit, all propagators are diagonal, simplifying the problem. The rules to obtain two times the imaginary part of any two-point function in the real-time formalism at finite density read as follows:
\begin{enumerate}
\item Draw all Feynman diagrams with the two corresponding external legs with all possible combinations where at least one vertex has a circle around it but at most all except one are circled.
\item Reverse the sign of each circled vertex.
\item Propagators $P_{(11)}$ connecting circled vertices are changed to $P_{(22)}$.
\item Propagators $P_{(11)}$ connecting circled and uncircled vertices are changed to the circled progator $P^{+}$ if the momentum flows from uncircled to circled and to $P^{-}$ if the momentum flows from circled to uncircled. 
\end{enumerate}
Further, it has been shown in \cite{Bedaque:1996af} that diagrams with internal clusters of circled or uncircled vertices, which are not connected to an external line, vanish. Thus, we are left only with diagrams that have only type 1 vertices and the circled and uncircled vertices clustered around one of the external legs, which also has to be circled or uncircled respectively.
For self-energy functions this means that we can draw a line through the diagram with one external leg on each side, where all vertices on one side are circled and on the other side are uncircled or the other way round (see \eg \Figs{fig:self_E_emiss_example}{fig:self_E_emiss_example_2}). In vacuum this mentioned "line" represents the usual on-shell cut obeying the Cutkosky rules.

Examples for such diagrams can be seen on the RHS of \Figs{fig:self_E_emiss_example}{fig:self_E_emiss_example_2}. This sets the stage for systematically calculating the axion emissivity from a SN within the real-time formalism of thermal field theory.

\subsection{Axion emissivity from supernova explosions}
\label{app:emissivity}

\begin{figure}[htbp] 
  \centering
  \begin{subfigure}{1.\textwidth}
  \centering
  % include first image
\includegraphics[width=.45\textwidth]{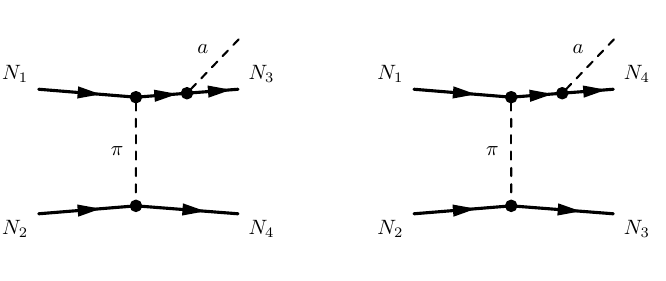}
\end{subfigure}
\vspace{-5mm}
  \begin{subfigure}{1.\textwidth}
  \centering
  % include second image
\includegraphics[width=.45\textwidth]{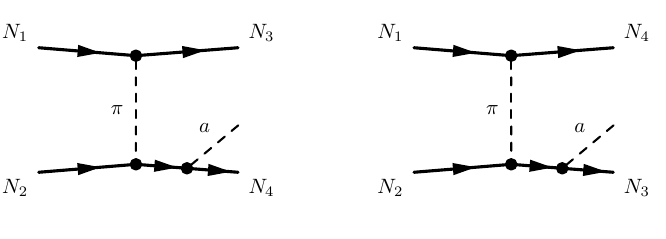}
\end{subfigure} 
\vspace{-5mm}
  \begin{subfigure}{1.\textwidth}
  \centering
  % include first image
\includegraphics[width=.45\textwidth]{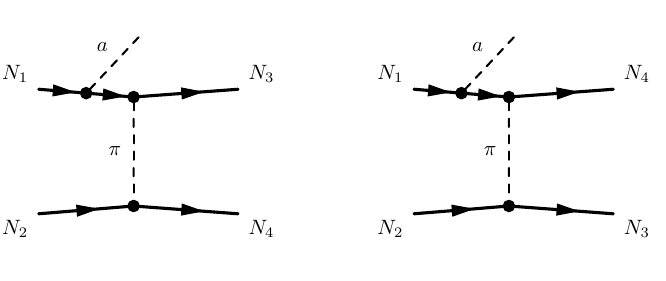}
\end{subfigure}
  \begin{subfigure}{1.\textwidth}
  \centering
  % include second image
\includegraphics[width=.45\textwidth]{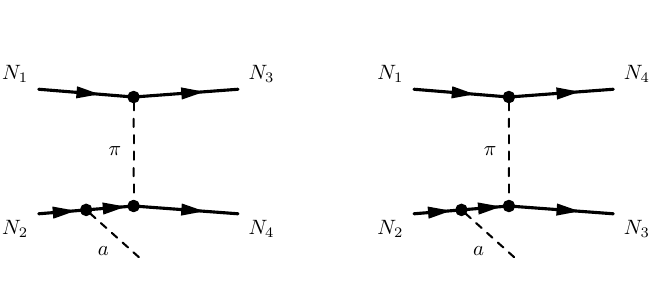}
\end{subfigure}
 \caption[Eight OPE diagrams]{OPE diagrams contributing at leading order to the axion emissivity from SN.}
  \label{fig:8_OPE_dias}
\end{figure}

\begin{figure}[t] 
  \centering
  \begin{subfigure}{.49\textwidth}
  \centering
  % include first image
\includegraphics[width=0.7\textwidth]{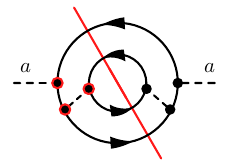}
\end{subfigure}
  \begin{subfigure}{.49\textwidth}
  \centering
  % include second image
\includegraphics[width=0.65\textwidth]{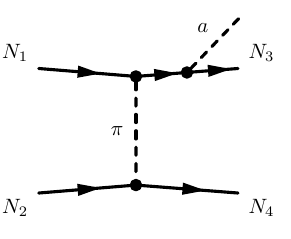}
\end{subfigure}
 \caption[]{The left figure shows the self energy diagram whose imaginary part corresponds to the emissivity of the example OPE diagram $\mathcal{A}$ on the right hand side.}
  \label{fig:self_E_emiss_example}
\end{figure}

According to the previous sections, in order to calculate the axion emissivity from a SN we have to calculate $\mathrm{Im} \Pi_{11}(p_a)$ using the cutting rules given above.
For SN environments the dominant contribution comes from diagrams where only nucleon lines are cut\footnote{If the pion 
density during the supernova is large, a diagram where one pion line is cut might also be important or even dominant, leading to the process discussed in \cite{Carenza:2020cis}}. As diagrams with two nucleon lines are kinematically forbidden, and no diagrams with three nucleon lines cut can exist due to baryon number conservation, the leading order diagrams must have at least four nucleon lines cut. An example for such a diagram with the above rules applied is shown in \Fig{fig:self_E_emiss_example}. 
Applying now the rules summarized in the previous section, we find that the imaginary part of the self-energy induced by the specific diagram shown in \Fig{fig:self_E_emiss_example} is equal to
\begin{equation} \label{eq:2Im_11}
\int d \Pi_{1} d \Pi_{2} d \Pi_{3} d \Pi_{4} (2 \pi)^4 \delta^{(4)}\left(\textstyle \sum_{i} p_i-p_{a}\right) |\mathcal{A}|^2 f_1 f_2 \left(1-f_3\right) \left(1-f_4\right),
\end{equation}
where $\mathcal{A}$ is the matrix element of the diagram, $d \Pi_{i}=d^{3} p_{i} /\left[(2 \pi)^{3} 2 E_{i}\right]$, and $f_{i}$ is the Fermi-Dirac distribution of the $i$th nucleon defined in \Eq{eq:FD_dist} and $E_i$ the energy of the $i$th nucleon.
Drawing consistently all possible diagrams at the same order, we arrive at
\begin{equation}
\dot{\epsilon}^{LO \rm}_{a}=\int d \Pi_{1} d \Pi_{2} d \Pi_{3} d \Pi_{4} d \Pi_{a}(2 \pi)^{4} S |\mathcal{M}^{LO \rm}|^{2} \delta^{(4)}\left(\textstyle \sum_{i} p_i-p_{a}\right) E_{a} f_{1} f_{2}\left(1-f_{3}\right)\left(1-f_{4}\right),
\end{equation}
where $\mathcal{M}^{LO \rm}$ is the sum of all possible tree-level OPE diagrams shown in figure \Fig{fig:8_OPE_dias}. 
This shows that the emissivity obtained in \cite{Brinkmann:1988vi} follows directly from the imaginary part of the self-energy of the axion propagator within the real-time formalism of thermal quantum field theory. 

\begin{figure}[t] 
  \centering
  \begin{subfigure}{.49\textwidth}
  \centering
  % include first image
\includegraphics[width=.7\textwidth]{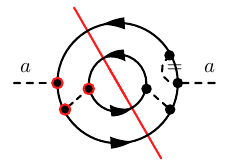}
\end{subfigure}
  \begin{subfigure}{.49\textwidth}
  \centering
  % include second image
\hspace{-15mm}
\raisebox{1.3\height}{$\left(\includegraphics[width=.42\textwidth,valign=c]{figures/OPE_emiss_tree.pdf}\right)^\star \times \includegraphics[width=.42\textwidth,valign=c]{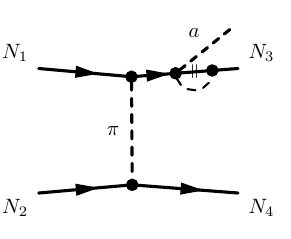}$}
\end{subfigure}
 \caption[]{The left figure shows the self-energy diagram whose imaginary part corresponds to the emissivity of the product of the two diagrams shown on the right-hand side.}
  \label{fig:self_E_emiss_example_2}
\end{figure}

At next higher order, which in the language of self-energy diagrams means to add a $\nu=1$ vertex instead of a $\nu=0$ one.
Going another order higher, we can add another loop as shown on the left-hand side of \Fig{fig:self_E_emiss_example_2}. 
Applying the above rules leads to the conclusion that this is equal to multiplying the complex conjugate of the tree-level diagram with the 1-loop diagram, as shown on the right-hand side of \Fig{fig:self_E_emiss_example_2}, and integrate over the same phase space as in the example before. Thus, as one would expect, the leading order result is equivalent to calculating the corresponding matrix element at leading order in (vacuum) QFT, averaging over incoming momenta and summing over outgoing momenta, both subject to the corresponding particle distribution functions. However, at higher order, density insertions become important and the result starts deviating from the corresponding vacuum calculation.

Thus, denoting by $|\mathcal{M}|^2 = |\mathcal{M}^{LO \rm} + \mathcal{M}^{NLO \rm} + \ldots|^2$ the squared sum of all diagrams, the full emissivity is given by
\begin{equation}
\dot{\epsilon}_{a}=\int d \Pi_{1} d \Pi_{2} d \Pi_{3} d \Pi_{4} d \Pi_{a}(2 \pi)^{4} S |\mathcal{M}|^{2} \delta^{(4)}\left(\textstyle \sum_{i} p_i-p_{a}\right) E_{a} f_{1} f_{2}\left(1-f_{3}\right)\left(1-f_{4}\right).
\end{equation}
As discussed in \Sec{sec:supernova}, in order to systematically calculate the axion emissivity from a SN at one loop order, one would, in principle, have to include all possible one loop diagrams of the process $N + N \rightarrow N + N + a$, which we leave for future work. 
As a first step, however, we include in this work all the diagrams correcting the nucleon-axion vertex at 1-loop order in finite density and relevant terms from the nucleon contact interaction.

\section{Explicit loop calculations}\label{app:explicit_calc}

\subsection{Useful integrals} \label{ap:useful_integrals}
In this appendix, we want to summarize the integrals needed to evaluate the diagrams contribution to the axion form factor. 

\subsubsection{Integrals for vacuum form factor}

For the vacuum loops, the following standard integrals defined in \cite{Scherer:2002tk} are needed
\begin{align}
\Delta_{\pi}=-\frac{1}{i} \int \frac{\mathrm{d}^{d} k}{(2 \pi)^{d}} \frac{1}{k^{2}-m_{\pi}^{2}+i \eta}  =2 m_{\pi}^{2}\left(L(\lambda)+\frac{1}{(4 \pi)^{2}} \ln \frac{m_{\pi}}{\lambda}\right)+\mathcal{O}(d-4),\end{align}
\begin{equation}\label{eq:Loflambda}
L(\lambda)=\frac{\lambda^{d-4}}{(4 \pi)^{2}}\left(\frac{1}{d-4}-\frac{1}{2}\left[\ln (4 \pi)+\Gamma^{\prime}(1)+1\right]\right),
\end{equation}
and
\begin{align}
\frac{1}{i} \int \frac{\mathrm{d}^{d} k}{(2 \pi)^{d}} \frac{\left\{1, k_{\mu}, k_{\mu} k_{\nu}\right\}}{\left(k^{2}-m_{\pi}^{2}+i \eta\right)(\omega-v \cdot k+i \eta)}  =\left\{J_{0}(\omega), v_{\mu} J_{1}(\omega), g_{\mu \nu} J_{2}(\omega)+v_{\mu} v_{\nu} J_{3}(\omega)\right\},
\end{align}
where
\begin{equation}
\begin{aligned}
&J_{0}(\omega)=-4 \omega L+\frac{2 \omega}{(4 \pi)^{2}}\left(1-2 \ln \frac{m_{\pi}}{\lambda}\right)-\frac{1}{8 \pi^{2}} \frac{m_\pi^3}{(m_\pi^2-\omega^2)} F \left( \frac{-\omega}{m_{\pi}} \right) + \mathcal{O}(d-4), \\
&J_{1}(\omega)=w J_{0}(\omega)+\Delta_{\pi}, \\
&J_{2}(\omega)=\frac{1}{d-1}\left[\left(m_{\pi}^{2}-\omega^{2}\right) J_{0}(\omega)-\omega \Delta_{\pi}\right], \\
&J_{3}(\omega)=w J_{1}(\omega)-J_{2}(\omega).
\end{aligned}
\end{equation}
Here we defined
\begin{equation}
    {F}(x)= - ({x}^2-1+i \varepsilon)^{3/2}\left[\log \left({x}-\sqrt{{x}^2-1+i \varepsilon}\right)-\log \left({x}+\sqrt{{x}^2-1+i \varepsilon}\right)\right],
\end{equation}
with $\varepsilon \rightarrow 0$ and $\varepsilon>0$, in order to properly account for the imaginary part of the loop integrals above the pion productions threshold $\omega>m_\pi$. \\
\subsubsection{Integrals for finite density form factor} \label{ap:int_form}

At finite density, the relevant standard integrals, given in \cite{Holt:2009ty, Holt:2019bah}, read
\begin{subequations} \label{eq:int_1}
\begin{align}
&{\Gamma}_{0}(m_\pi, |\vec{p}\,|)=\int_{0}^{k_{f}} d k \int_{-1}^{1} d x \frac{k^{2}}{m_\pi^2+|\vec{p}\,|^{2}+k^{2}+2 |\vec{p}\,| k x}, \\
&{\Gamma}_{1}(m_\pi, |\vec{p}\,|)=\int_{0}^{k_{f}} d k \int_{-1}^{1} d x \frac{k^{3} x / |\vec{p}\,|}{m_\pi^2+|\vec{p}\,|^{2}+k^{2}+2 |\vec{p}\,| k x}, \\
&{\Gamma}_{2}(m_\pi, |\vec{p}\,|)=\int_{0}^{k_{f}} d k \int_{-1}^{1} d x \frac{k^{4}\left(1-x^{2}\right) / 2}{m_\pi^2+|\vec{p}\,|^{2}+k^{2}+2 |\vec{p}\,| k x}, \\
&{\Gamma}_{3}(m_\pi, |\vec{p}\,|)=\int_{0}^{k_{f}} d k \int_{-1}^{1} d x \frac{k^{4}\left(3 x^{2}-1\right) /\left(2 |\vec{p}\,|^{2}\right)}{m_\pi^2+|\vec{p}\,|^{2}+k^{2}+2 |\vec{p}\,| k x}.
\end{align} 
\end{subequations}
Evaluating the integrals, they read as
\small
\begin{subequations}
\begin{align}
\Gamma_0(m_\pi,|\vec{p}\,|)  = &k_f-m_\pi\left[\tan^{-1} \frac{k_f+|\vec{p}\,|}{m_\pi}+\tan^{-1} \frac{k_f-|\vec{p}\,|}{m_\pi}\right]+\frac{m_\pi^2+k_f^2-|\vec{p}\,|^2}{4 |\vec{p}\,|} \log \frac{m_\pi^2+\left(k_f+|\vec{p}\,|\right)^2}{m_\pi^2+\left(k_f-|\vec{p}\,|\right)^2} \\
\Gamma_1(m_\pi,|\vec{p}\,|) = & \frac{k_f}{4 |\vec{p}\,|^2}\left(m_\pi^2+k_f^2+|\vec{p}\,|^2\right)-\Gamma_0(|\vec{p}\,|) \nonumber \\
&-\frac{1}{16 |\vec{p}\,|^3}\left[m_\pi^2+\left(k_f+|\vec{p}\,|\right)^2\right]\left[m_\pi^2+\left(k_f-|\vec{p}\,|\right)^2\right] \log \frac{m_\pi^2+\left(k_f+|\vec{p}\,|\right)^2}{m_\pi^2+\left(k_f-|\vec{p}\,|\right)^2}, \\
\Gamma_2(m_\pi,|\vec{p}\,|) = & \frac{k_f^3}{9}+\frac{1}{6}\left(k_f^2-m_\pi^2-|\vec{p}\,|^2\right) \Gamma_0(|\vec{p}\,|)+\frac{1}{6}\left(m_\pi^2+k_f^2-|\vec{p}\,|^2\right) \Gamma_1(|\vec{p}\,|), \\
\Gamma_3(m_\pi,|\vec{p}\,|) = & \frac{k_f^3}{3 |\vec{p}\,|^2}-\frac{m_\pi^2+k_f^2+|\vec{p}\,|^2}{2 |\vec{p}\,|^2} \Gamma_0(|\vec{p}\,|)-\frac{m_\pi^2+k_f^2+3 |\vec{p}\,|^2}{2 |\vec{p}\,|^2} \Gamma_1(|\vec{p}\,|) .
\end{align}
\end{subequations}
\normalsize
For the calculation provided in \Sec{sec:axion_vertex_at_finite_density} and \App{app:density_loops} it additionally turns out to be useful to define
\begin{subequations} \label{eq:int_2}
\begin{align}
& I_1 \left(m_\pi, \vec{p}, \vec{q} \right)  = \int_{0}^{k_{f}} \frac{d^3 k}{( 2 \pi )^3} \frac{ \vec{\sigma} \cdot (\vec{k} + \vec{p}) \vec{q} \cdot (\vec{k} + \vec{p}) }{m_{\pi}^{2}+(\vec{k} + \vec{p})^2} \\
 & = \frac{1}{4 \pi^2} \left[ \left( \vec{\sigma} \cdot \vec{q} \right) \Gamma_2(m_\pi, |\vec{p}\,|) + \left( \vec{\sigma} \cdot \vec{p} \right) \left( \vec{q} \cdot \vec{p} \right) \left( \Gamma_0(m_\pi, |\vec{p}\,|) + 2 \Gamma_1(m_\pi, |\vec{p}\,|) + \Gamma_3(m_\pi, |\vec{p}\,|) \right) \right],\nonumber \\
& I_2 \left(m_\pi, \vec{p} \right) = \int_{0}^{k_{f}} \frac{d^3 k}{( 2 \pi )^3} \frac{ (\vec{k} + \vec{p}) \cdot (\vec{k} + \vec{p}) }{m_{\pi}^{2}+(\vec{k} + \vec{p})^2} = \frac{1}{4 \pi^2} \left( \frac{2 k_f^3}{3} - m_\pi^2 \Gamma_0(m_\pi, |\vec{p}\,|) \right), \\
& I_3 \left(m_\pi, \vec{p} \right) = \int_{0}^{k_{f}} \frac{d^3 k}{( 2 \pi )^3} \frac{ \vec{\sigma} \cdot (\vec{k} + \vec{p}) }{m_{\pi}^{2}+(\vec{k} + \vec{p})^2} = (\vec{\sigma} \cdot \vec{p}) \frac{1}{4 \pi^2} \left( {\Gamma}_1(m_\pi, |\vec{p}\,|) + {\Gamma}_0(m_\pi, |\vec{p}\,|) \right).
\end{align} 
\end{subequations}
To further simplify the notation in the main text we define the following expressions
\small
\begin{subequations}
\begin{align}
   G_1^N(\vec{p},\vec{p}_a) & = \frac{1}{16 |\vec{p}_a|} \left[ (\vec{\sigma} \cdot \vec{p}_a) I^N_2(m_\pi,\vec{p}) - 2 ~ I^N_1(m_\pi,\vec{p},\vec{p}_a) - \left\{ \begin{array}{ll}  m_\pi \rightarrow \tilde{m}_\pi \\ \vec{p} \rightarrow \vec{p} + \vec{p}_a \end{array} \right\} \right], \\
    G_2^N(\vec{p},\vec{p}_a) & = \frac{1}{4 |\vec{p}_a|} \left[ I^N_3(m_\pi,\vec{p}) + I^N_3(\tilde{m}_\pi,\vec{p}+\vec{p}_a) \right], \\
    G_3^N(\vec{p},\vec{p}_a) & = \frac{1}{2} \left[I^N_{1}(\tilde{m}_\pi, \vec{p}+\vec{p}_a, \vec{p}_a) + I^N_{1}(m_\pi, \vec{p}, \vec{p}_a) - (\vec{\sigma} \cdot \vec{p}_a) (I^N_{2}(\tilde{m}_\pi, \vec{p} + \vec{p}_a ) + I^N_{2}(m_\pi, \vec{p} )) \right], \\
    G_4^N(\vec{p},\vec{p}_a) & = \frac{1}{2} \left[ I^N_1 \left( \tilde{m}_\pi, \vec{p}+\vec{p}_a, \vec{p}_a \right) + I^N_1 \left(m_\pi, \vec{p}, \vec{p}_a \right) -  |\vec{p}_a|^2 I^N_3(\tilde{m}_\pi, \vec{p}+\vec{p}_a) \right],
\end{align}
\end{subequations}
\normalsize
with $N=\{p,n\}$.

\subsection{Loop corrections in vacuum} \label{ap:vac_loops}

The loop diagrams at chiral order $\nu = 2$ given in \Fig{fig:Nu2Vac} have been calculated in \cite{Vonk:2020zfh}. The first two diagrams both vanish,
\begin{equation}
    (a_1) = (a_2) = 0.
\end{equation}
For the third diagram in \Fig{fig:Nu2Vac} we find
\begin{equation}
    (b) = \frac{2 g_A c_{u-d} m_\pi^2}{f_a f_\pi^2} \tau^3 \left(L(\lambda)+\frac{1}{(4 \pi)^2} \ln \frac{m_\pi}{\lambda}\right) S \cdot p_a,
\end{equation}
which agrees with \cite{Vonk:2020zfh} up to a typo. On the other hand, for the loop diagram $(c)$ we find
\begin{equation}
    (c) = \frac{\hat{c}_N g_A^2}{f_a f_\pi^2} \left( \frac{D-3}{4} \right) \frac{1}{v \cdot p_a} ( J_2(0) - J_2(v \cdot p_a) ) S \cdot p_a.
\end{equation}
where $D=4-2\epsilon$ in dimensional regularization.
In \cite{Vonk:2020zfh}, they find $1/4$ instead of the factor $(D-3)/4$, which we believe is necessary in dimensional regularization to obtain the correct finite contributions.
Renormalizing these diagrams appropriately by adding the counterterms given in diagram $(d)$, we arrive at the result given in the main text in \Eq{eq:Nu2VacResult} and following. \\

Following the nomenclature of \Fig{fig:Nu3Vac} we calculate the two non-vanishing one loop diagrams, $(e_1)$ and $(e_2)$, and give the results in terms of the standard integrals defined in \App{ap:useful_integrals}.

We find the results for the loop diagrams to be given by
\begin{equation}
    (e_1) = \left( \frac{g_A \cm}{f_\pi^2 f_a \Lambda} \right) \left( c_3  - 2  c_4  \right)  \tau^3 \vec{\sigma} \cdot \vec{p}_a J_2(v \cdot p_a),
\end{equation}
as well as
\begin{equation}
    (e_2) = \left( \frac{g_A \cm}{f_\pi^2 f_a \Lambda} \right) \left( c_3  - 2  c_4  \right)  \tau^3 \vec{\sigma} \cdot \vec{p}_a J_2(0).
\end{equation}
Summing them up and expanding the standard integrals leads to 
\begin{equation}
\begin{aligned}
& (e_1) +(e_2) 
=  -\frac{2 g_{A} \cm}{f_{a} f_{\pi}^{2}} \left(\hat{c}_{3}- 2\hat{c}_{4}\right) \tau^{3}\left[J_{2}\left(v \cdot p_a\right)+J_{2}\left(0\right)\right] S \cdot p_{a} \\
& =- \frac{4 g_{A} \cm}{3f_{a} f_{\pi}^{2}} \left(\hat{c}_{3}- 2\hat{c}_{4}\right) \tau^{3} \left\{\frac { 1 } { 16 \pi ^ { 2 } } \bigg[m_{\pi}^{2} (v \cdot p_a) -(v \cdot p_a)^3 - m_\pi^2 F\left(\frac{-v \cdot p_a}{m_{\pi}}\right) -m_\pi^2 F\left(0\right) \right] \\
& - \Lambda(\lambda)\left(3 m_{\pi}^{2}(v \cdot p_a)-2((v \cdot p_a))^3\right)\bigg\} S \cdot p_{a}.
\end{aligned}
\end{equation}
Here $\Lambda(\lambda)$ is a function collecting the scale-dependent and divergent pieces in dimensional regularization given by
\begin{equation}
\Lambda(\lambda)=L(\lambda)+\frac{1}{16 \pi^{2}} \log \frac{m_{\pi}}{\lambda}.
\end{equation}
As described in the main text, we neglect all contributions that are not $\Delta$-enhanced, \ie not coming from $\ch3$ or $\ch4$ at this loop order and thus drop the $\ch9$ as well as $\ch2$ which is in principle $\Delta$-enhanced but turns out to be kinematically suppressed.

\subsubsection{Renormalization}\label{app:reno}

In order to arrive at the result at \Eq{eq:Anu3}, we renormalized the divergent pieces with terms at $\Delta=3$ given by \cite{Meissner:1998rw}
\begin{align}
&e_{117}(\lambda) \bar{N}\left(\left\langle\chi_{+}\right\rangle S \cdot u \; i v \cdot D+\text { h.c.}\right) N, \\
&e_{193}(\lambda) \bar{N}\left(S \cdot u \; i(v \cdot D)^{3}+\text { h.c.}\right) N,
\end{align}
which give rise to the following tree level contributions
\begin{align}
-\frac{4 e_{117}}{f_{a}} c_{u-d} \tau^{3} m_{\pi}^{2}(v \cdot p_a) S \cdot p_{a}, \\
\frac{e_{193}}{f_{a}} c_{u-d} \tau^{3}(v \cdot p_a)^3 S \cdot p_{a} .
\end{align}
As one can easily check this renormalizes the loop contribution if the beta functions
\begin{equation}
e_{i}(\lambda)=\bar{e}_{i}+\frac{\beta_{i}}{f_{\pi}^{2}} \Lambda(\lambda),
\end{equation}
have the following property
\begin{align}
&\beta_{117} \supset g_{A}\left( \hat{c}_{3} -2 \hat{c}_{4}\right) , \\
&\beta_{193} \supset \frac{8 g_{A}}{3}\left(\hat{c}_{3}-2 \hat{c}_{4}\right),
\end{align}
which exactly agrees with \cite{Meissner:1998rw}.
Note that the function collecting the divergent and scale-dependent pieces is given by
\begin{equation}
    \Lambda(\lambda)=L(\lambda)+\frac{1}{16\pi^2}\log\frac{m_\pi}{\lambda},
\end{equation}
with $L(\lambda)$ given in \Eq{eq:Loflambda}.
 Note that as is customary in HBChPT, instead of the usual $\overline{\text{MS}}$ scheme, we use the modified $\widetilde{\text{MS}}$ scheme \cite{Gasser:1983yg}.
A priori there are finite pieces which however are zero for operators that can be eliminated by an appropriate field redefinition (proportional to $v \cdot D$), see \eg \cite{Fettes:1998ud}.

\subsection{Density corrections} \label{app:density_loops}

The relevant non-vanishing vertex diagrams at order $\nu=2$ and $\nu=3$ are the ones shown in \Fig{fig:Nu2Dens} and \Fig{fig:Nu3Dens}.  We calculate them explicitly in the following first in the case of isospin symmetric nuclear matter and then in the case of arbitrarily mixed matter and give the results in terms of the standard integrals defined in \ref{ap:useful_integrals}.

\subsubsection{Isospin symmetric matter} \label{app:symmetricmatter}

In this section, we explicitly calculate some of the diagrams shown in \Fig{fig:Nu2Dens} and \Fig{fig:Nu3Dens} in the isospin symmetric limit and give the results for the others.
In the isospin symmetric case, the propagator is given by \Eq{eq:HBChPT_propagator_isopin_Symetric_T0}. 
Let us consider the diagrams $(h_1) + (h_2)$ as an example. The ansatz for two loop integrals reads
\begin{equation}
\begin{aligned}
 &(h_1) + (h_2)  =  \int \frac{d^4k}{(2 \pi)^4}  \left[ - \frac{g_A}{2 f_\pi} \vec{\sigma} \cdot ( \vec{k} - \vec{p} ) \tau^a \right] \left[ \frac{i}{k^0} - 2 \pi \delta(k^0) \theta(k_f - | \vec{k} |) \right] \left[ \frac{c_N}{2 f_a} \vec{\sigma} \cdot \vec{p}_a \right] 
 \\
 & \times \left[ \frac{i}{k^0 +p_a^0} - 2 \pi \delta(k^0 + p_a^0) \theta(k_f - | \vec{k} + \vec{p}_a |) \right] \left[ \frac{g_A}{2 f_\pi} \vec{\sigma} \cdot ( \vec{k} - \vec{p} ) \tau^b \right] \left[ \frac{- i \delta^{a b}}{m_\pi^2 - ({k}-{p})^2} \right].
\end{aligned}
\end{equation}
Taking care of the spin as well as the isospin structures and only keeping the density-dependent part (we already calculated the vacuum loop in diagram in \App{ap:vac_loops}) leaves us, after simplifying and evaluating the $k^0$ integrals over the $\delta$ functions, with
\begin{align}
    (h_1) + (h_2)  = &\left( \frac{g_A^2 \hat{c}_N}{8 f_\pi^2 f_a}  \right) \frac{1}{p_a^0}\int_0^{k_f} \frac{d^{3} k}{(2 \pi)^{3}}  \left\{   \left[ \frac{ 2 \vec{\sigma} \cdot ( \vec{k} - \vec{p}) \vec{p}_a \cdot ( \vec{k} - \vec{p}) - \vec{\sigma} \cdot \vec{p}_a ( \vec{k} - \vec{p})^2 }{\tilde{m}_{\pi}^{2} + ( \vec{k} - \vec{p} )^2 } \right]  \right. \nonumber \\
    & \left. - \left[ \frac{ 2 \vec{\sigma} \cdot ( \vec{k} - \vec{p} - \vec{p}_a ) \vec{p}_a \cdot ( \vec{k} - \vec{p} - \vec{p}_a ) - \vec{\sigma} \cdot \vec{p}_a ( \vec{k} - \vec{p} - \vec{p}_a )^2 }{\tilde{m}_{\pi}^{2} + ( \vec{k} - \vec{p} - \vec{p}_a )^2} \right] \right\},
\end{align}
where we used that $m_\pi^2-(k-p_a)^2\simeq\tilde{m}_\pi^2+(\vec{k}-\vec{p}_a)^2$, since $k_0 \sim \vec{k}^2/2m_N \ll |\vec{k}|$ and defined $\tilde{m}_\pi^2 = m_\pi^2 - (p_a^0)^2$. With the help of the standard integrals defined in \App{ap:useful_integrals} we arrive at the final result
\begin{equation}
\begin{aligned}
    \hspace{-3mm} (h_1) + (h_2) =  -\left( \frac{g_A^2 \hat{c}_N}{8 f_\pi^2 f_a} \right) \frac{1}{|\vec{p}_a|} & \bigg[ 2 \left[ {I}_1(\tilde{m}_\pi,\vec{p}+\vec{p}_a,\vec{p}_a) - I_1(m_\pi,\vec{p},\vec{p}_a)  \right]  \bigg. \\
    & \hspace{2cm} \bigg. - (\vec{\sigma} \cdot \vec{p}_a )\left[ {I}_2(\tilde{m}_\pi,\vec{p} + \vec{p}_a) - I_2(m_\pi,\vec{p}) \right] \bigg].
\end{aligned}
\end{equation}

The calculation of the diagrams $(g_1)$ and $(g_2)$ is very similar and yields
\begin{subequations}
\begin{align}
    (g_1) 
    & = -\frac{\cm g_A p_a^0}{2 f_\pi^2 f_a} \tau^3 I_3(\tilde{m}_\pi,\vec{p}+\vec{p}_a), \\
    (g_2) & = -\frac{\cm g_A p_a^0}{2 f_\pi^2 f_a} \tau^3 {I}_3(m_\pi,\vec{p}).
\end{align}
\end{subequations}
For the diagrams at $\nu = 3$ we have different vertices contributing to the same diagram, therefore we split up the calculation for the diagram into parts. Further, the diagrams shown in \Fig{fig:Nu3Dens} are not the only non-vanishing ones, as mentioned in the main text, and are shown in \Fig{fig:vertex_diagrams_FD_nu3}. For diagram $(i_1)$ we find
\begin{equation}
\begin{aligned}
(i_1)  =& +\left(\frac{g_A \cm c_3 }{f_{\pi}^2 f_{a} \Lambda_{\chi}} \right) \tau^{3} \left[ {I}_1 ( \tilde{m}_\pi, \vec{p}+\vec{p}_a, \vec{p}_a )  - |\vec{p}_a|^2 I_3(\tilde{m}_\pi, \vec{p}+\vec{p}_a)\right] \\
& + \left(\frac{3 g_A c_9 \cp}{2 f_{\pi}^2 f_{a} \Lambda_{\chi}}\right) \left[  {I}_1 ( \tilde{m}_\pi, \vec{p}+\vec{p}_a, \vec{p}_a ) - |\vec{p}_a|^2 I_3(\tilde{m}_\pi, \vec{p}+\vec{p}_a)  \right] \\
& + \left( \frac{g_A \cm c_4}{f_{\pi}^2 f_{a} \Lambda_{\chi}} \right) \tau^3 \left[ {I}_{1}(\tilde{m}_\pi, \vec{p}+\vec{p}_a, \vec{p}_a) - (\vec{\sigma} \cdot \vec{p}_a) {I}_{2}(\tilde{m}_\pi, \vec{p} + \vec{p}_a ) \right] \\
& - \left( \frac{6 g_A c_5 g_A^2 f_a}{f_\pi^4 \Lambda_{\chi}} \right)I_3(\tilde{m}_\pi,\vec{p}+\vec{p}_a),
\end{aligned}
\end{equation}
and for diagram $(i_2)$ we obtain
\begin{equation}
\begin{aligned}
(i_2)  =&  +\left(\frac{g_A \cm c_3 }{f_{\pi}^2 f_{a} \Lambda_{\chi}}\right) \tau^{3} I_1 \left(m_\pi, \vec{p}, \vec{p}_a \right) + \left(\frac{3 g_A c_9 \cp}{2 f_{\pi}^2 f_{a} \Lambda_{\chi}}\right) I_1 \left(m_\pi, \vec{p}, \vec{p}_a \right) \\
 & + \left( \frac{g_A \cm c_4}{f_{\pi}^2 f_{a} \Lambda_{\chi}} \right) \tau^3 \left[ I_{1}(m_\pi, \vec{p}, \vec{p}_a) - (\vec{\sigma} \cdot \vec{p}_a) I_{2}(m_\pi, \vec{p} ) \right] + \left( \frac{6 g_A c_5 g_A^2 f_a}{f_\pi^4 \Lambda_{\chi}} \right){I}_3(m_\pi,\vec{p}).
\end{aligned}
\end{equation}

\subsubsection{Arbitrarily mixed matter}\label{app:mixedmatter}

For an arbitrary matter configuration, we also need vertex corrections where the nucleon propagator is altered as
\begin{equation} \label{eq:sub_propagator}
 -2 \pi \delta\left(k_{0}\right) \theta\left(k_{f}-|\vec{k}|\right) \rightarrow -2 \pi \delta\left(k_{0}\right) \theta\left(k_{f}-|\vec{k}|\right) \tau^3,
\end{equation}
Together with the above results, we can then construct the full vertex corrections for arbitrary mixed matter using \Eq{eq:HBChPT_propagator_isopin_Asymetric_T0}. Substituting the propagator as in \Eq{eq:sub_propagator} and defining $\tilde{c}_N = 3 g_A c_{u-d} \mathbb{1} - g_0 c_{u+d} \tau^{3}$ leads to:

\begin{equation}
(g_1)+(g_2) \rightarrow \left(\frac{\cm g_A}{2 f_\pi^2 f_a} \right) |\vec{p}_a| \left[ I_3(m_\pi,\vec{p}) + I_3(\tilde{m}_\pi,\vec{p}+\vec{p}_a) \right].
\end{equation}

\begin{equation}
\begin{aligned}
(h_1)+(h_2) \rightarrow - \left( \frac{g_A^2 \tilde{c}_N}{8 f_\pi^2 f_a} \right) \frac{1}{|\vec{p}_a|} & \left[ 2 \left( {I}_1(\tilde{m}_\pi,\vec{p}+\vec{p}_a,\vec{p}_a) - I_1(m_\pi,\vec{p},\vec{p}_a) \right) \right. \\
 & \, \, \left. - (\vec{\sigma} \cdot \vec{p}_a) \left( {I}_2(\tilde{m}_\pi,\vec{p} + \vec{p}_a) - I_2(m_\pi,\vec{p}) \right) \right].
\end{aligned}
\end{equation}

\begin{equation}
\begin{aligned}
(i_1) \rightarrow & \left(\frac{g_A \cm c_3 }{f_{\pi}^2 f_{a} \Lambda_{\chi}}\right) \left[ {I}_1 \left( \tilde{m}_\pi, \vec{p}+\vec{p}_a,\vec{p}_a \right) -  |\vec{p}_a|^2 I_3(\tilde{m}_\pi, \vec{p}+\vec{p}_a) \right] \\
& - \left(\frac{g_A c_9 \cp}{2 f_{\pi}^2 f_{a} \Lambda_{\chi}}\right) \tau^3 \left[{I}_1 \left( \tilde{m}_\pi, \vec{p}+\vec{p}_a,\vec{p}_a \right) -   |\vec{p}_a|^2 I_3(\tilde{m}_\pi, \vec{p}+\vec{p}_a)  \right] \\
& - \left( \frac{g_A \cm c_4}{f_{\pi}^2 f_{a} \Lambda_{\chi}} \right) \left[ {I}_{1}(\tilde{m}_\pi, \vec{p}+\vec{p}_a,\vec{p}_a) - (\vec{\sigma} \cdot \vec{p}_a) {I}_{2}(\tilde{m}_\pi, \vec{p} +\vec{ p}_a ) \right] \\
& + \left( \frac{6 g_A c_5 g_A^2 f_a}{f_\pi^4 \Lambda_{\chi}} \right) \tau^3 I_3(\tilde{m}_\pi,\vec{p}+\vec{p}_a)
\end{aligned}
\end{equation}

\begin{equation}
\begin{aligned}
(i_2) \rightarrow & \left(\frac{g_A \cm c_3 }{f_{\pi}^2 f_{a} \Lambda_{\chi}}\right) I_1 \left(m_\pi, \vec{p},\vec{ p}_a \right) - \left(\frac{g_A c_9 \cp}{2 f_{\pi}^2 f_{a} \Lambda_{\chi}}\right) \tau^3 I_1 \left(m_\pi, \vec{p},\vec{ p}_a \right) \\
& - \left( \frac{g_A \cm c_4}{f_{\pi}^2 f_{a} \Lambda_{\chi}} \right) \left[ I_{1}(m_\pi, \vec{p}, \vec{p}_a) - \vec{\sigma} \cdot \vec{p}_a I_{2}(m_\pi,\vec{p} ) \right] - \left( \frac{6 g_A c_5 g_A^2 f_a}{f_\pi^4 \Lambda_{\chi}} \right) \tau^3 {I}_3(m_\pi,\vec{p})
\end{aligned}
\end{equation}

\begin{figure}[htbp]
    \centering\includegraphics[width=.35\linewidth]{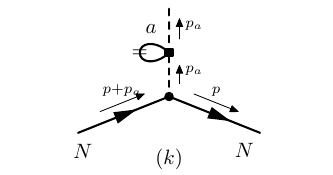}

 \caption{Additional 1-loop finite density diagram contributing at chiral order $\nu=3$ which is not $\Delta$ enhanced and therefore neglected. This can be matched to the effect considered in \cite{Balkin:2020dsr} coming from the medium-modified axion pion mixing angle.}
  \label{fig:vertex_diagrams_FD_nu3}
\end{figure}
For completeness, we also calculate the diagram shown in \Fig{fig:vertex_diagrams_FD_nu3}. This gives rise to a medium-modified axion pion mixing and yields
\begin{equation} \label{eq:c_5}
(k) = \left( \frac{4 k_f^3 g_A c_5 f_a g_A^2}{3 \pi^2 m_\pi^2 f_\pi^4  \Lambda_{\chi}} \right) \tau^3 ( \vec{\sigma} \cdot \vec{p}_a ).
\end{equation}
As it is not $\Delta$ enhanced, it is not included in our calculation. However, it has been included in the estimates given in \cite{Balkin:2020dsr}. Note also that this effect is important for a model-independent bound on the QCD axion \cite{Springmann:2024ret}.

Combining these results and the ones for symmetric matter above, we can assemble the density corrections for an arbitrary mixture of matter. In arbitrary matter, the propagator is given by \Eq{eq:HBChPT_propagator_isopin_Asymetric_T0}, \ie a combination of either the propagator with the identity matrix used in the section for symmetric matter and the one where the substitution \Eq{eq:sub_propagator} has been made which we used here.

\subsection{DFSZ axion additional results} \label{ap:DFSZresults}

\begin{figure}[h] 
  \centering
  \vspace{-0.5cm}
  \begin{subfigure}{.48\textwidth}
  \centering
  % include first image
\includegraphics[width=1.\textwidth]{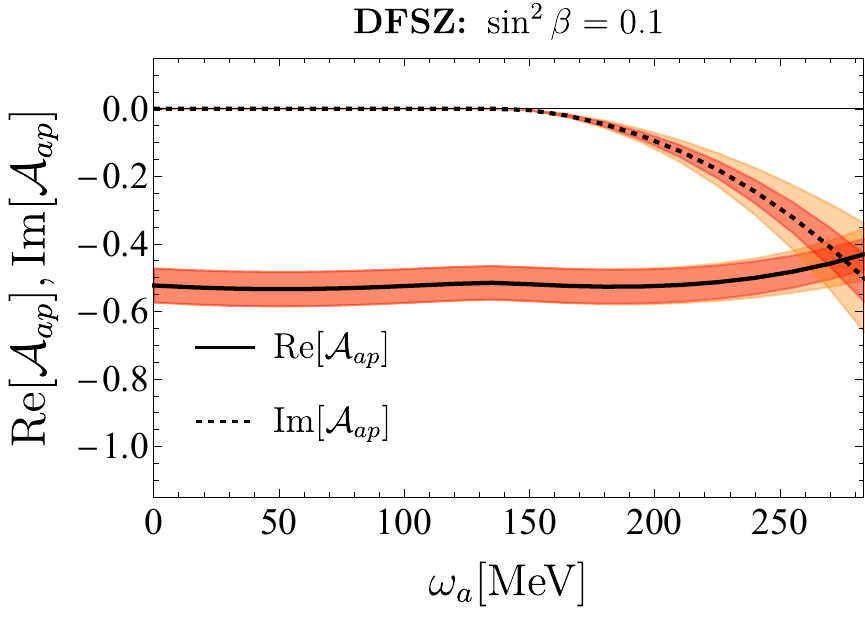}
\end{subfigure}
  \begin{subfigure}{.48\textwidth}
  \centering
  % include second image
\includegraphics[width=1.\textwidth]{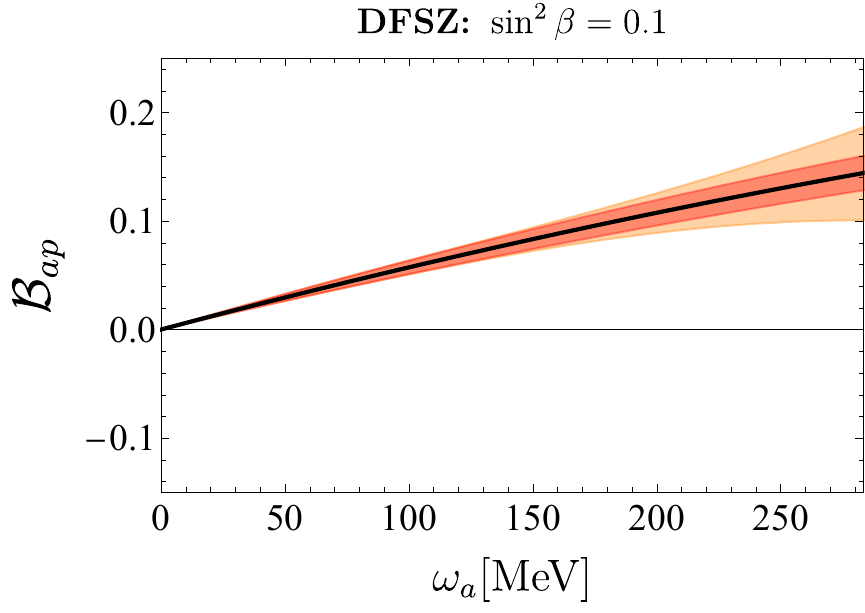}
\end{subfigure}
\begin{subfigure}{.48\textwidth}
  \centering
  % include first image
\includegraphics[width=1.\textwidth]{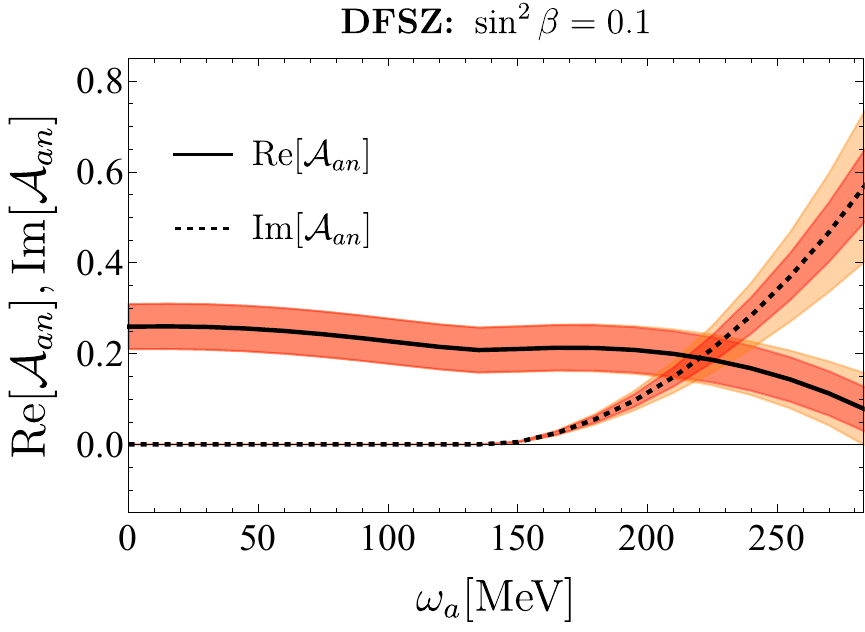}
\end{subfigure}
  \begin{subfigure}{.48\textwidth}
  \centering
  % include second image
\includegraphics[width=1.\textwidth]{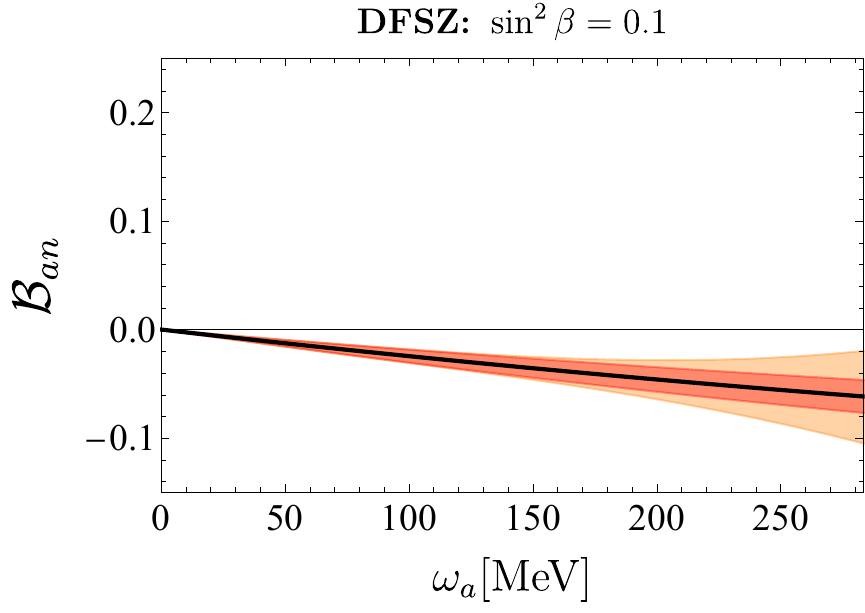}
\end{subfigure}
  \centering
  \begin{subfigure}{.48\textwidth}
  \centering
  % include first image
\includegraphics[width=1.\textwidth]{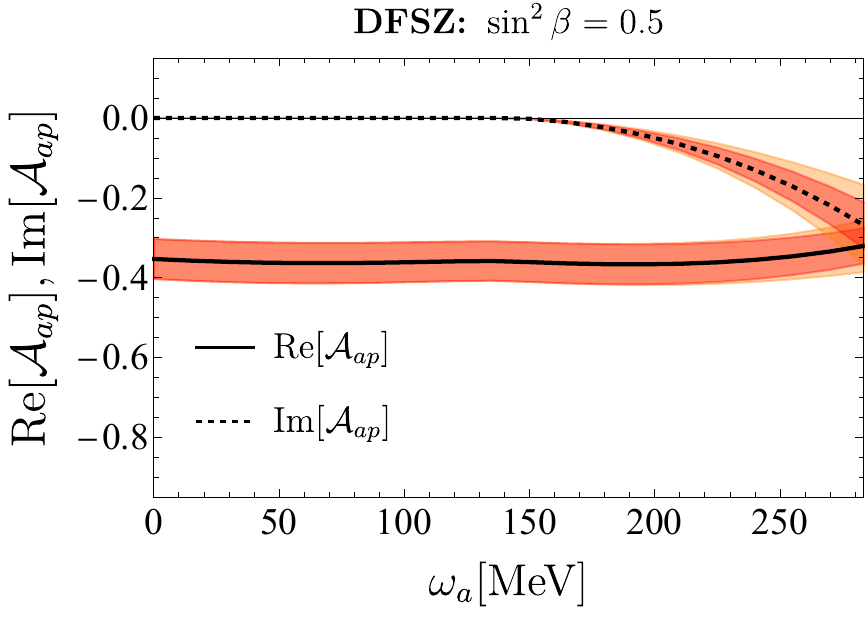}
\end{subfigure}
  \begin{subfigure}{.48\textwidth}
  \centering
  % include second image
\includegraphics[width=1.\textwidth]{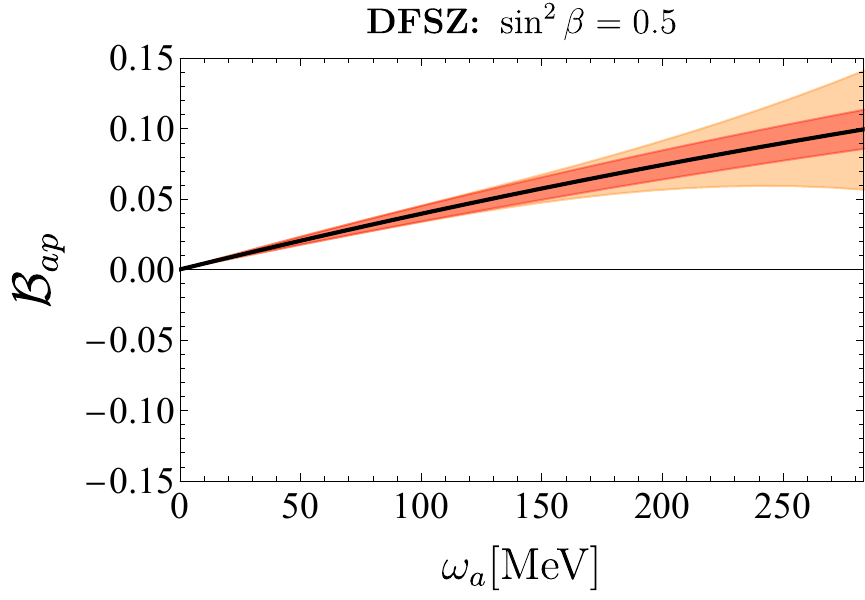}
\end{subfigure}
\begin{subfigure}{.48\textwidth}
  \centering
  % include first image
\includegraphics[width=1.\textwidth]{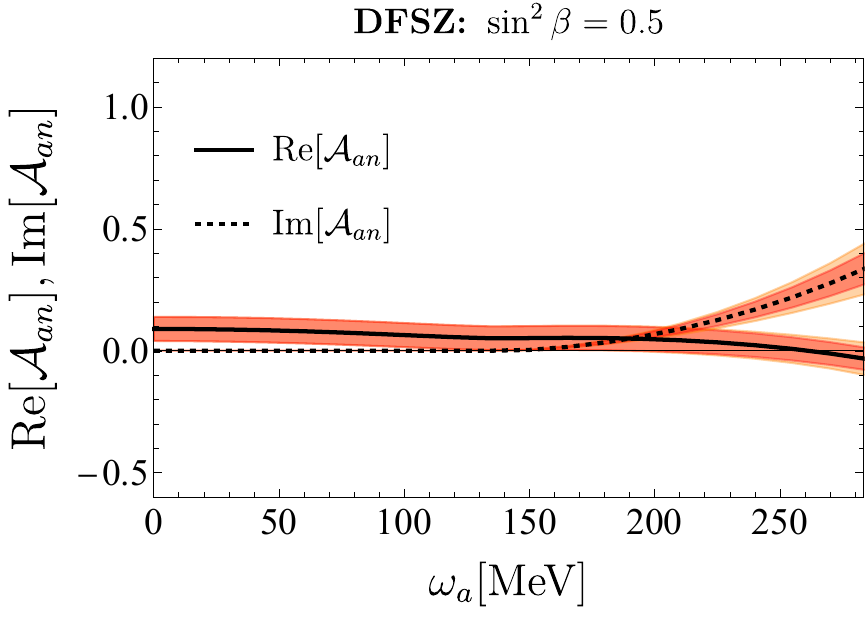}
\end{subfigure}
  \begin{subfigure}{.48\textwidth}
  \centering
  % include second image
\includegraphics[width=1.\textwidth]{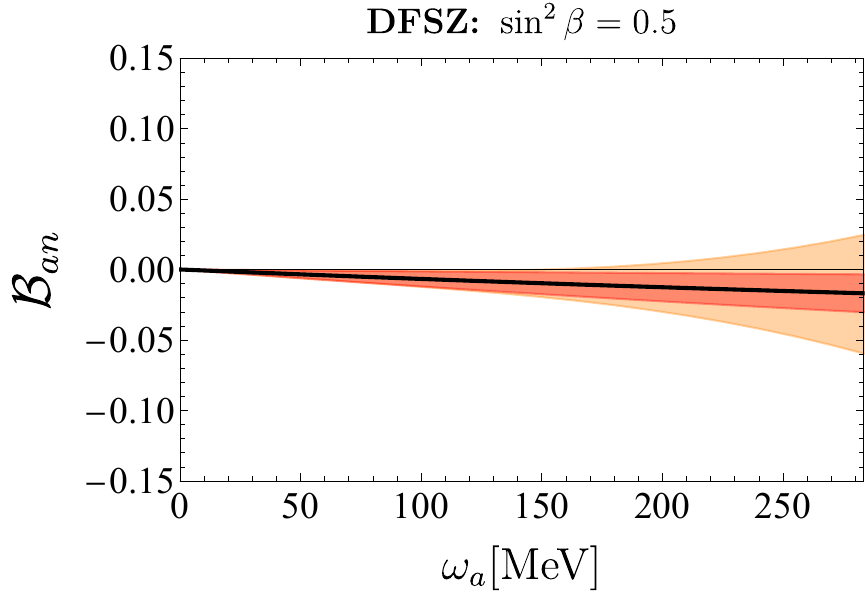}
\end{subfigure}

 \caption[]{Momentum dependence of the DFSZ for $\sin^2 \beta =0.1$ and $\sin^2 \beta =0.5$ axion-nucleon couplings. Error bars are shown as in \Fig{fig:axion_nucleon_coupling_KSVZ_vac}. At $\omega_a\sim m_\pi$ one can again see the pion production threshold, at which the coupling also gets an imaginary part.}\label{fig:axion_nucleon_coupling_DFSZ_vac}
\end{figure}

\begin{figure}[h] 
  \centering
  \begin{subfigure}{.49\textwidth}
  \centering
  % include first image
\includegraphics[width=1.\textwidth]{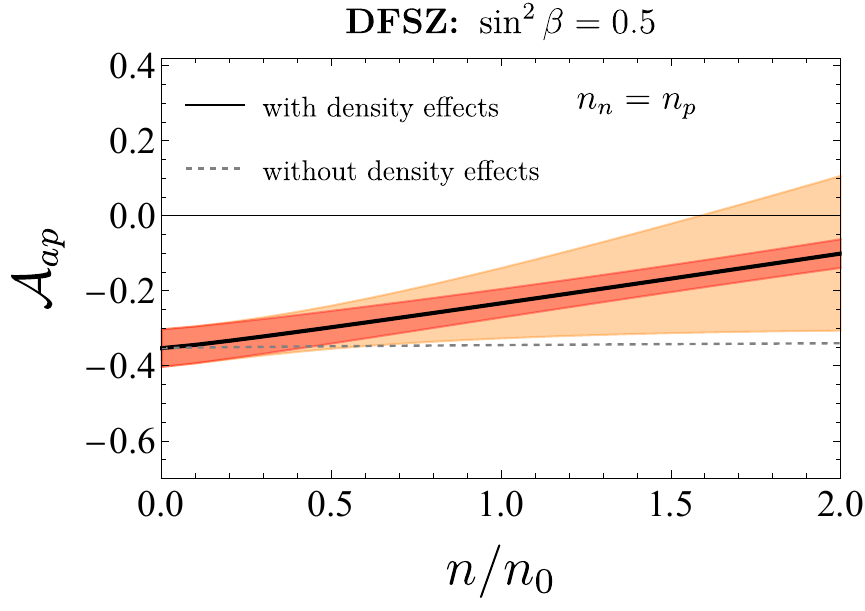}
\end{subfigure}
  \begin{subfigure}{.49\textwidth}
  \centering
  % include second image
\includegraphics[width=1.\textwidth]{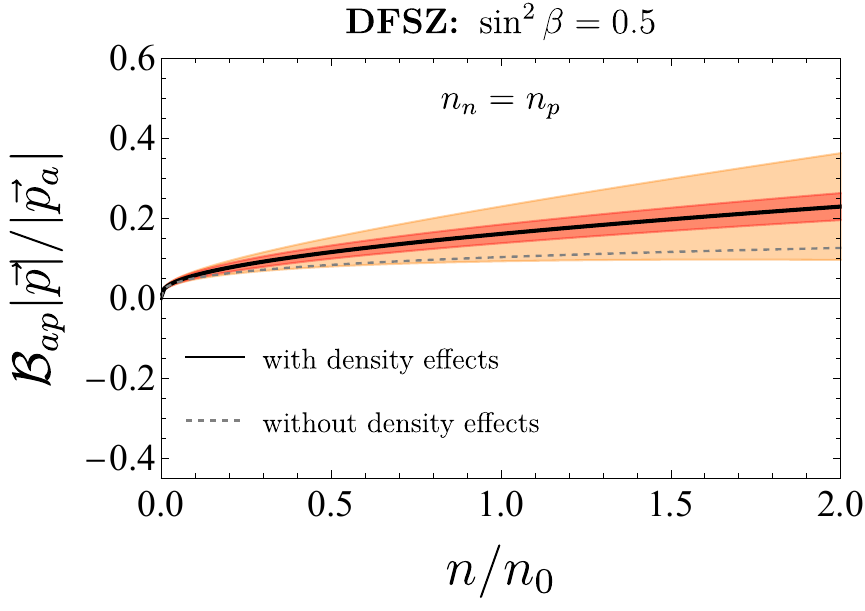}
\end{subfigure}
\begin{subfigure}{.49\textwidth}
  \centering
  % include first image
\includegraphics[width=1.\textwidth]{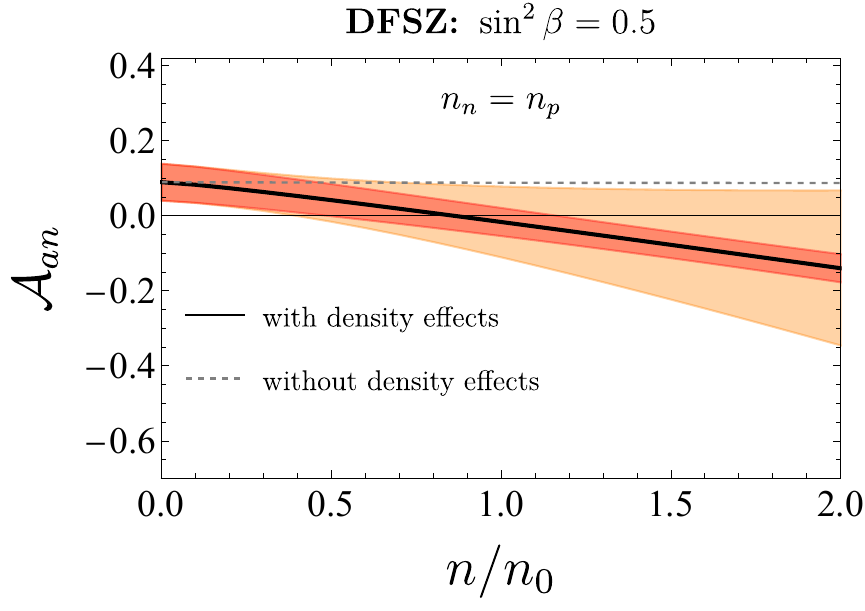}
\end{subfigure}
  \begin{subfigure}{.49\textwidth}
  \centering
  % include second image
\includegraphics[width=1.\textwidth]{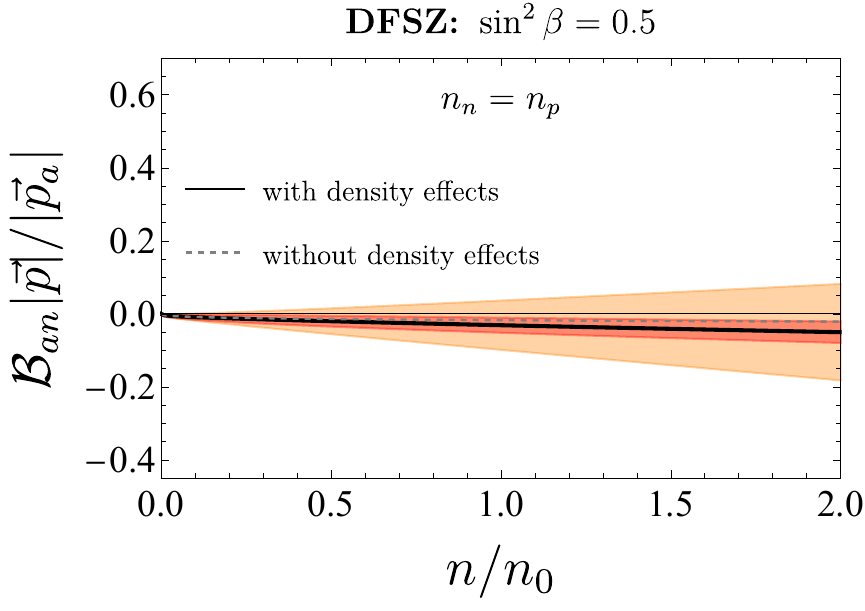}
\end{subfigure}
 \caption[]{Density dependence of the DFSZ axion-nucleon couplings with $\sin^2 \beta = 1/2$ in symmetric nuclear matter as a function of density in units of nuclear saturation density $n_0$. Error bars are shown as in \Fig{fig:axion_nucleon_coupling_KSVZ_vac}. The gray dashed line again shows the results of \Eq{eq:vertex_result}}
  \label{fig:axion_nucleon_coupling_DFSZ_0_5_sym_matter}
\end{figure}

\begin{figure}[h] 
  \centering
  \begin{subfigure}{.49\textwidth}
  \centering
  % include first image
\includegraphics[width=1.\textwidth]{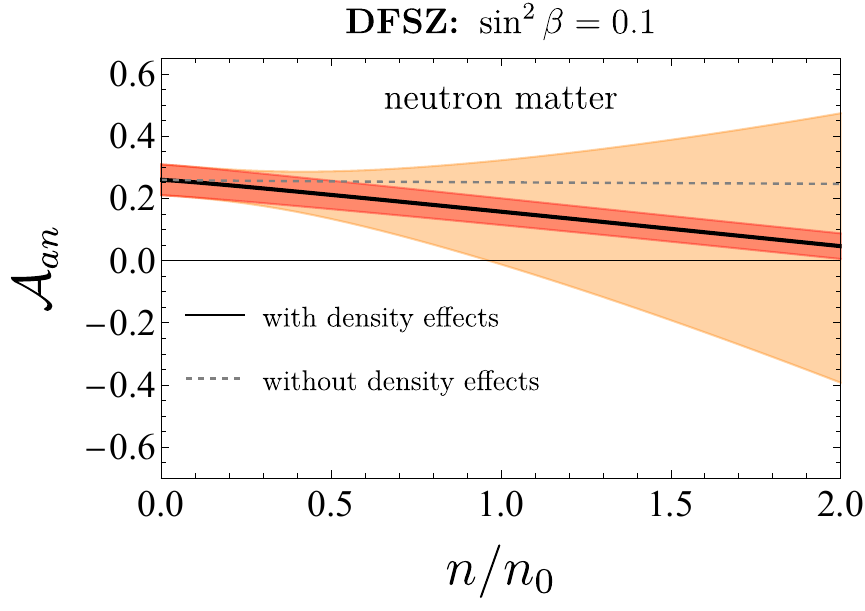}
\end{subfigure}
\begin{subfigure}{.49\textwidth}
  \centering
  % include second image
\includegraphics[width=1.\textwidth]{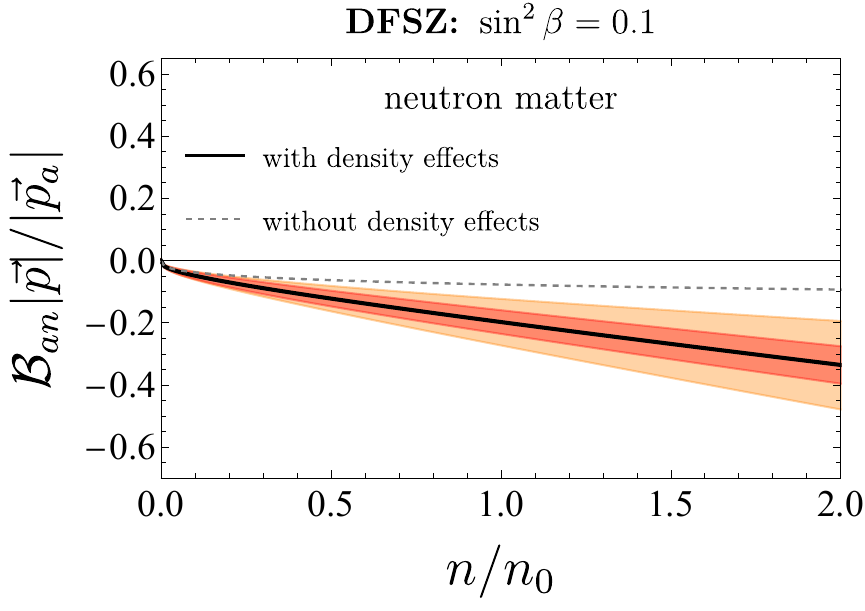}
\end{subfigure}
\begin{subfigure}{.49\textwidth}
  \centering
  % include first image
\includegraphics[width=1.\textwidth]{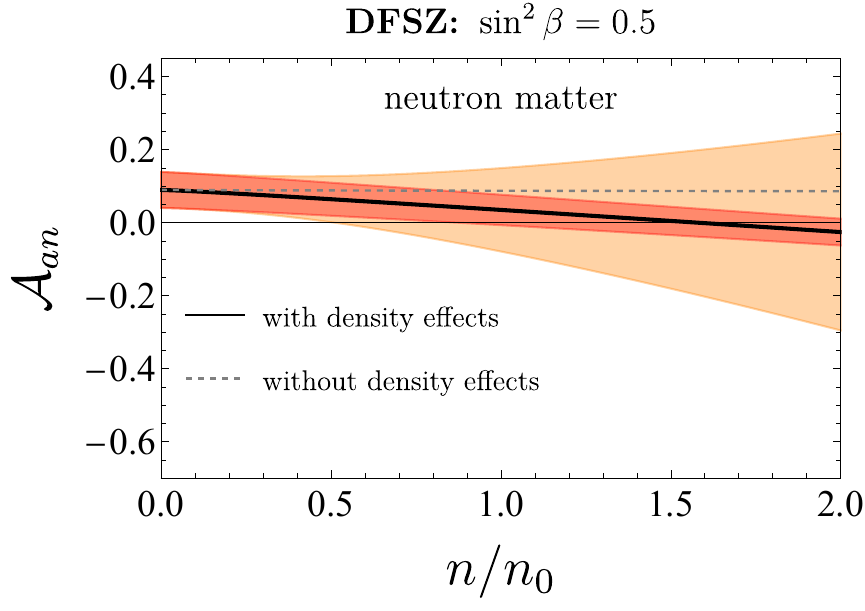}
\end{subfigure}
  \begin{subfigure}{.49\textwidth}
  \centering
  % include second image
\includegraphics[width=1.\textwidth]{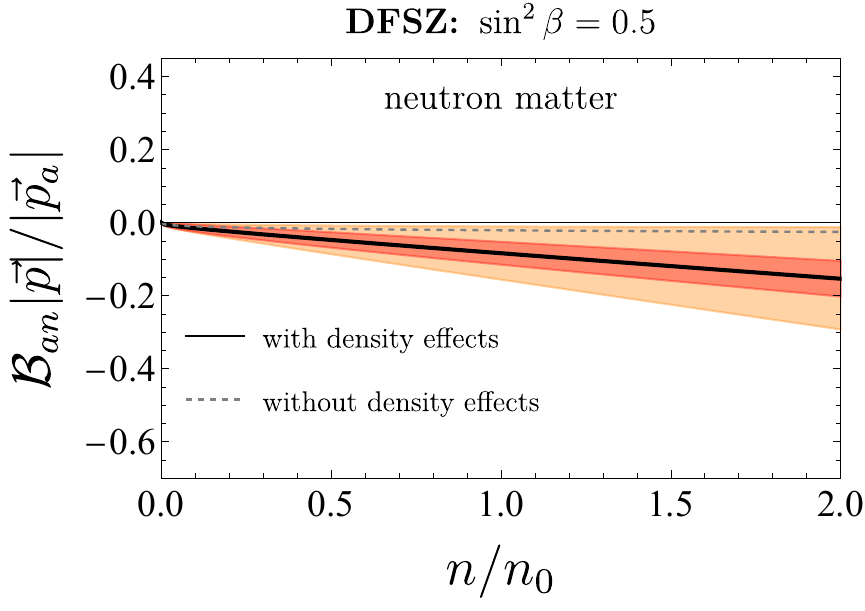}
\end{subfigure}
\begin{subfigure}{.49\textwidth}
  \centering
  % include first image
\includegraphics[width=1.\textwidth]{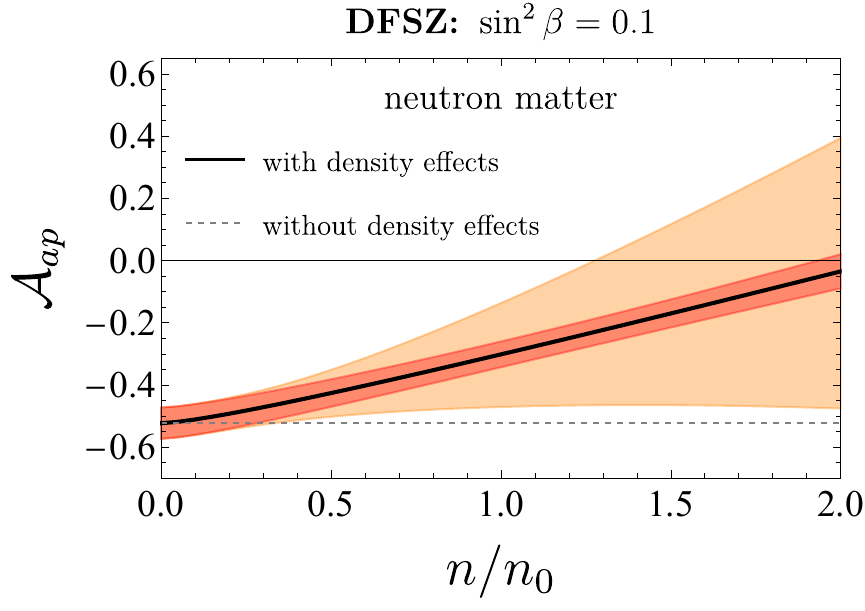}
\end{subfigure}
  \begin{subfigure}{.49\textwidth}
  \centering
  % include second image
\includegraphics[width=1.\textwidth]{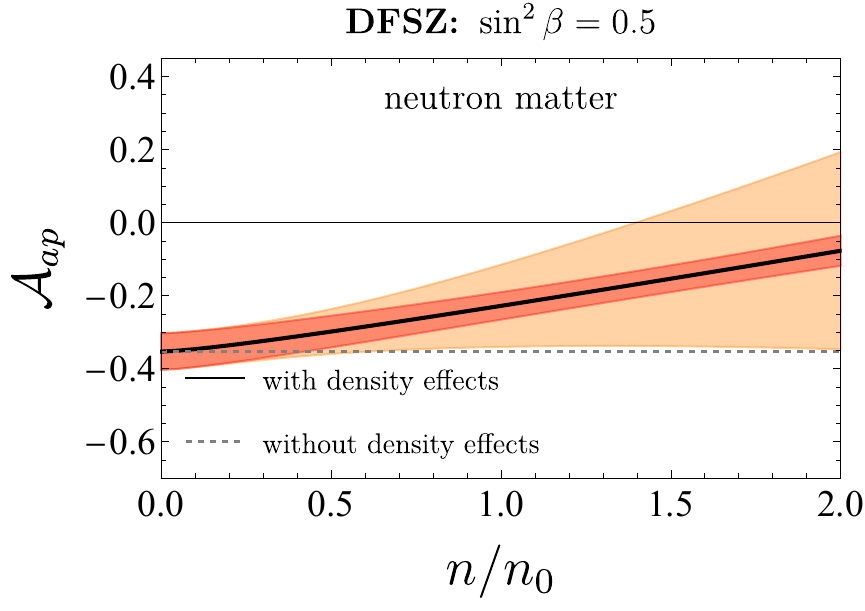}
\end{subfigure}
 \caption[]{Density dependence of the DFSZ axion-nucleon couplings with $\sin^2 \beta = 0.1$ (upper panel) as well as $\sin^2 \beta = 1/2$ (lower panel) in neutron matter as a function of density in units of nuclear saturation density $n_0$. Error bars are shown as in \Fig{fig:axion_nucleon_coupling_KSVZ_vac}. The gray dashed line again shows the results of \Eq{eq:vertex_result}.}\label{fig:DFSZneutronmatter}
\end{figure}

Here we show additional results for the DFSZ axion for different values of $\sin^2\beta$ as well as for pure neutron matter. For readability, these figures are not shown in the main text. In \Fig{fig:axion_nucleon_coupling_DFSZ_vac}, we show the momentum dependence of the DFSZ axion-nucleon coupling for $\sin^2 \beta =0.1$ and $\sin^2 \beta =1/2$ respectively. In \Fig{fig:axion_nucleon_coupling_DFSZ_0_5_sym_matter} we show the results for the axion neutron and axion proton coupling in mixed matter for the DFSZ axion with $\sin^2 \beta =1/2$, while in \Fig{fig:DFSZneutronmatter} we show the axion neutron coupling for $\sin^2 \beta = 0.1$ as well as $\sin^2 \beta =1/2$ pure neutron matter. 

\clearpage

\section{Temperature corrections} \label{app:temp}

So far, when talking about finite density loop corrections presented in \Sec{sec:finite_density_couplings}, we neglected finite temperature effects.
In the following we verify that in the environments we consider, neglecting the temperature dependence in the propagator is indeed a good approximation.
Note, however, that while temperature corrections in the propagator are not important, temperature changes the momentum scale at which the form factors have to be evaluated. We show below that this effect, which, of course, was taken into account in \Sec{sec:supernova}, is indeed non-negligible.

As explained in \App{app:nuclen_prop_FD}, when the temperature is not set to zero, the 11-component of the nucleon propagator in the non-relativistic limit is given by \Eq{eq:NRfiniteTnucprop}.
When evaluating the resulting diagrams at finite temperature, occurring standard integrals, defined in \Eq{eq:int_1}, can be numerically evaluated once the Fermi surface is smeared out.
\begin{figure}[h] 
  \centering
  \begin{subfigure}{.49\textwidth}
  \centering
  % include first image
\includegraphics[width=1.\textwidth]{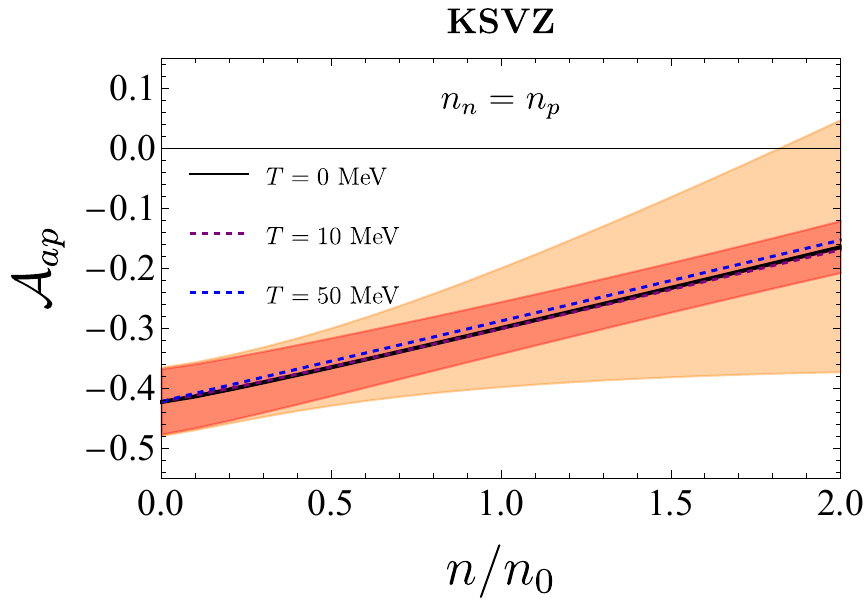}
\end{subfigure}
  \begin{subfigure}{.49\textwidth}
  \centering
  % include second image
\includegraphics[width=1.\textwidth]{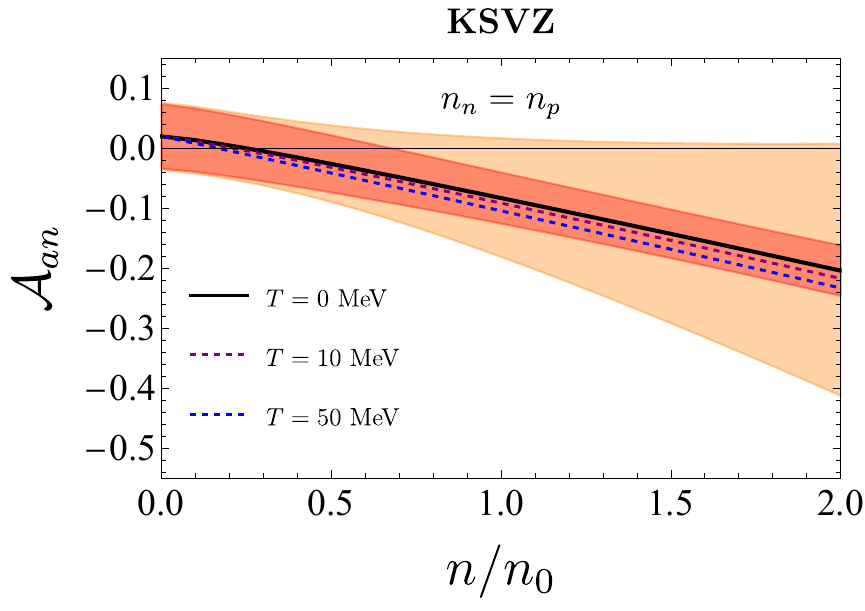}
\end{subfigure}
  \begin{subfigure}{.49\textwidth}
  \centering
  % include first image
\includegraphics[width=1.\textwidth]{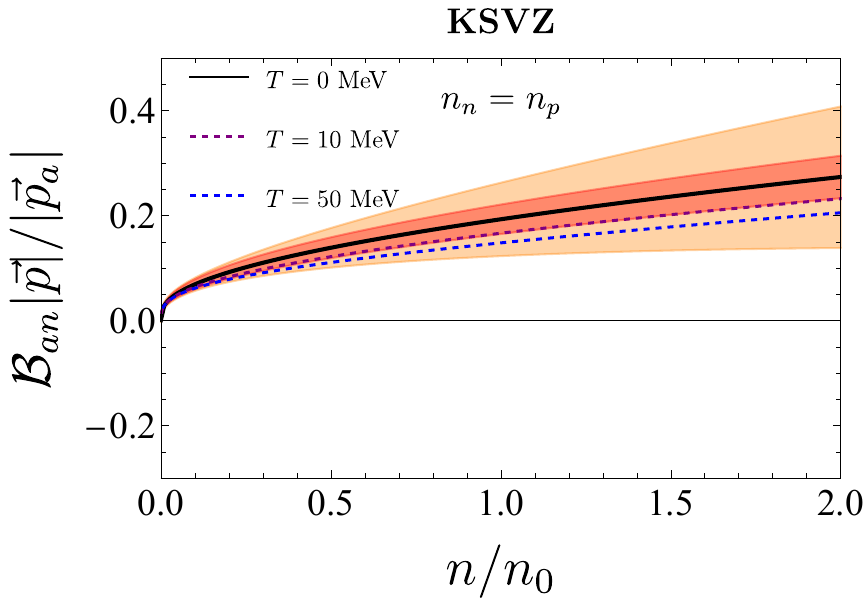}
\end{subfigure}
  \begin{subfigure}{.49\textwidth}
  \centering
  % include second image
\includegraphics[width=1.\textwidth]{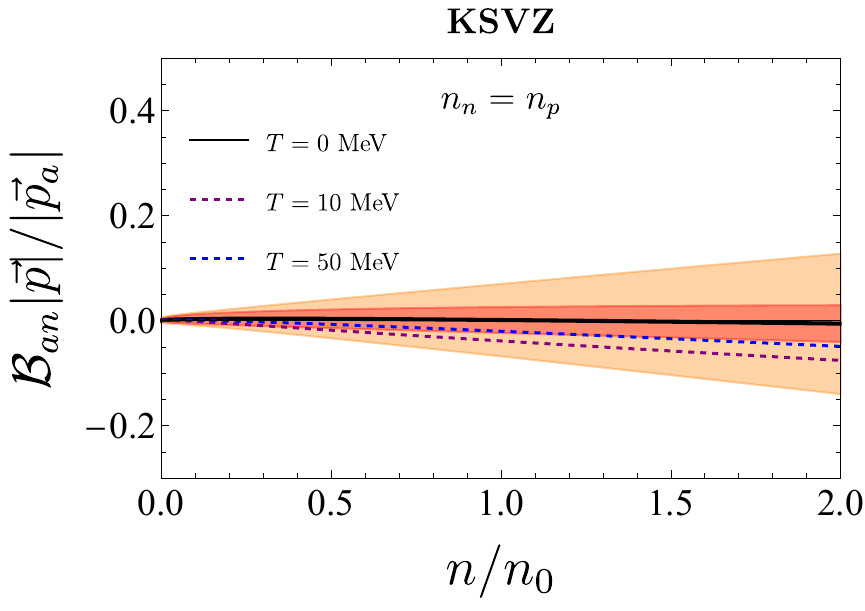}
\end{subfigure}
 \caption[]{Density dependence of the KSVZ axion-nucleon couplings in isospin symmetric nuclear matter as a function of density in units of nuclear saturation density $n_0$ for different temperatures $T$. Error bars are shown as in \Fig{fig:axion_nucleon_coupling_KSVZ_vac}. The black solid line is, as in \Fig{fig:axion_nucleon_coupling_KSVZ_sym_matter}, the mean value for $T=0$. The two dashed lines are the mean values for $T=10 ~\mathrm{MeV}$ (green line) and $T=50 ~\mathrm{MeV}$ (purple line).}  \label{fig:axion_nucleon_coupling_KSVZ_sym_matter_Temp}
\end{figure}
Doing so for typical SN temperatures $T \lesssim 50 ~\mathrm{MeV}$ one finds only small changes compared to the zero temperature case. 
These are within the uncertainties of the low energy constants summarized in \Tab{tab:constants}. 
As an example, we show in \Fig{fig:axion_nucleon_coupling_KSVZ_sym_matter_Temp} the effect of temperature corrections to axion-nucleon couplings for the KSVZ model in symmetric nuclear matter for $T=10\MeV$ and $T=50\MeV$. 
We conclude that, to a good approximation, we can neglect finite temperature corrections to the nucleon propagator at typical SN temperatures and densities. 

However, the momenta at which the form factors $\mathcal{A}_{aN}$ and $\mathcal{B}_{aN}$ are evaluated when calculating the axion emissivity, see \Eq{eq:emiss}, heavily depend on the temperature as the variance of the Fermi-Dirac distribution $f_i$ occurring in the emissivity roughly goes as $\sim \sqrt{3 m_N T}$. This temperature dependence is what we are studying in the following.

\begin{figure}[h!] 
  \centering
  \begin{subfigure}{.49\textwidth}
  \centering
  % include first image
\includegraphics[width=1.\textwidth]{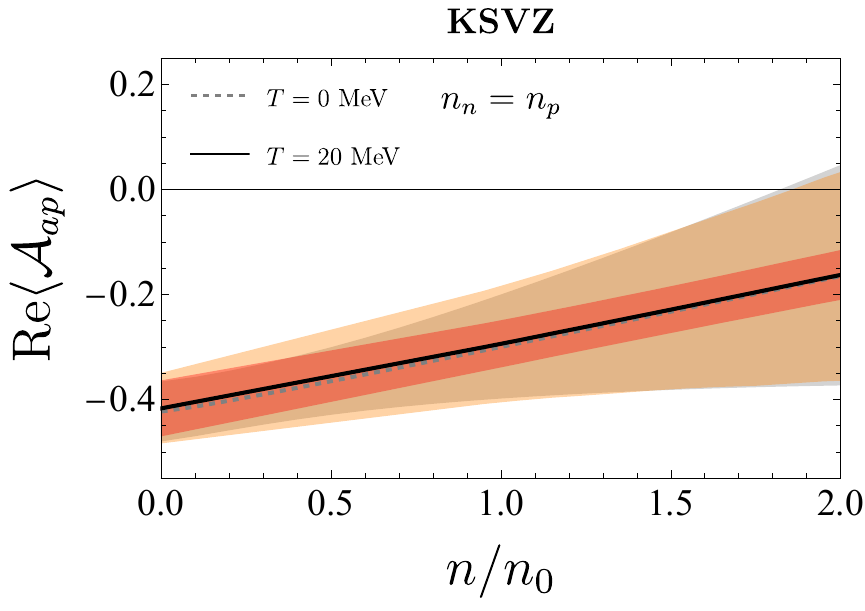}
\end{subfigure}
  \begin{subfigure}{.49\textwidth}
  \centering
  % include second image
\includegraphics[width=1.\textwidth]{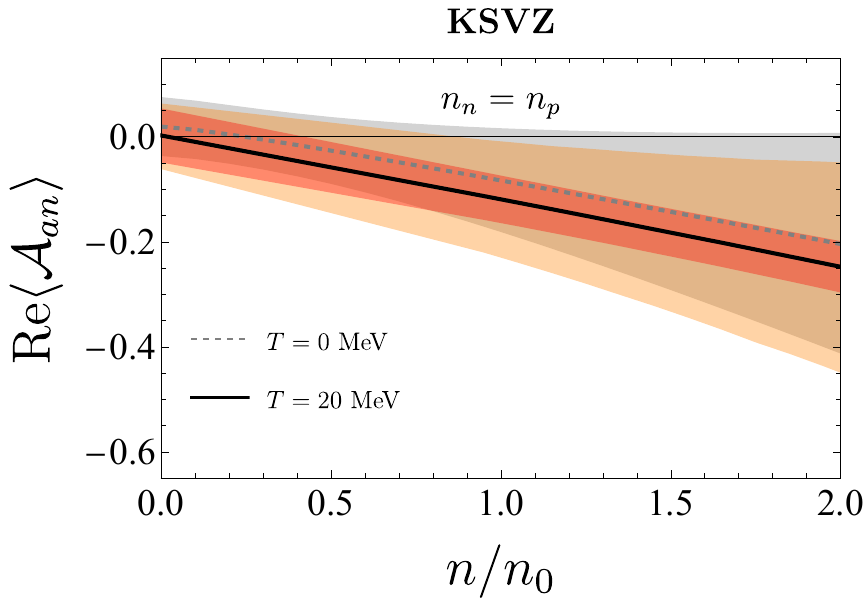}
\end{subfigure}
  \centering
  \begin{subfigure}{.49\textwidth}
  \centering
  % include first image
\includegraphics[width=1.\textwidth]{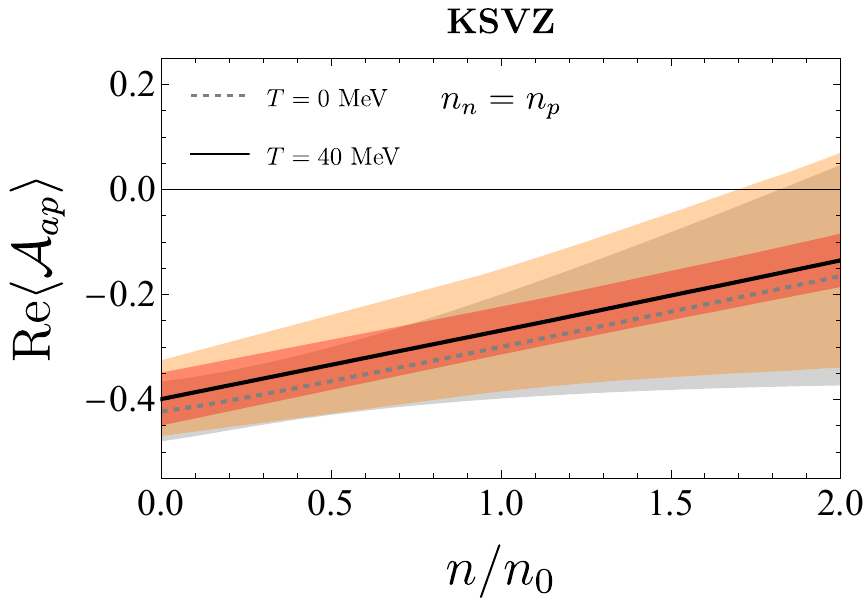}
\end{subfigure}
  \begin{subfigure}{.49\textwidth}
  \centering
  % include second image
\includegraphics[width=1.\textwidth]{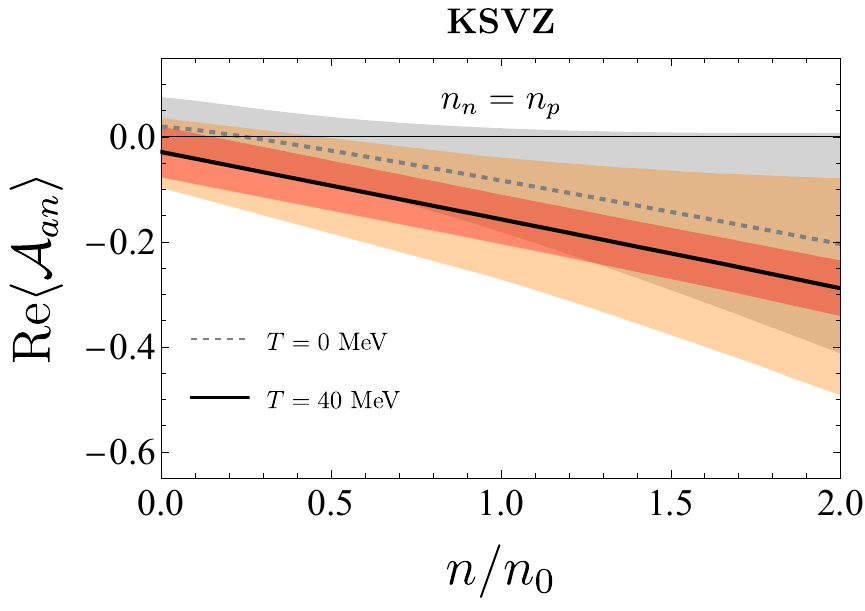}
\end{subfigure}
\begin{subfigure}{.49\textwidth}
  \centering
  % include first image
\includegraphics[width=1.\textwidth]{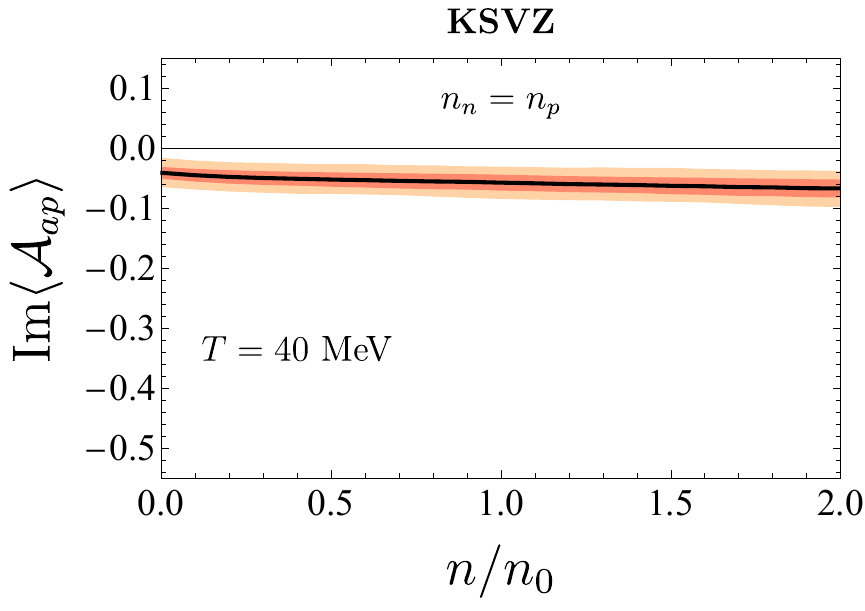}
\end{subfigure}
  \begin{subfigure}{.49\textwidth}
  \centering
  % include second image
\includegraphics[width=1.\textwidth]{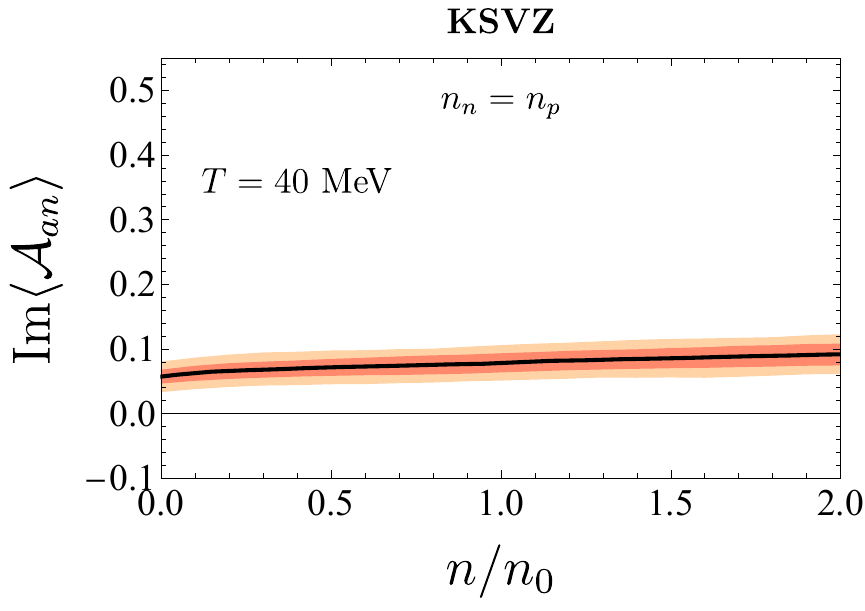}
\end{subfigure}
 \caption[]{Density dependence of the KSVZ axion-nucleon couplings for $T= 20~\mathrm{MeV}$ and $T=40~\mathrm{MeV}$. Density is in units of nuclear saturation density $n_0$. Error bars are shown as in \Fig{fig:axion_nucleon_coupling_KSVZ_vac}. The dashed gray line represents the $T \rightarrow 0~\mathrm{MeV}$ result, with the gray area representing the combined error. 
 For $T=20~\mathrm{MeV}$, we only show the real part of the form factor, as the imaginary part is negligibly small, as typical momenta are not above the pion threshold. On the other hand for $T=40~\mathrm{MeV}$, we show both real and imaginary parts.
}
  \label{fig:axion_nucleon_Temp}
\end{figure}

We calculate the expectation value of the form factors $\mathcal{A}(p, k_f^{p/n}, p_a)$ and $\mathcal{B}(p, k_f^{p/n}, p_a)$ which we define by
\begin{equation} \label{eq:exp_value}
\langle \mathcal{A} \rangle \equiv \mathcal{N} \int \big( \prod_i d \Pi_{i}  \big) d \Pi_{a} \mathcal{A}(p, k_f^{p/n}, p_a) \delta^{(4)}\left(\textstyle \sum_{i} p_i-p_{a}\right) f_{1} f_{2}\left(1-f_{3}\right)\left(1-f_{4}\right),
\end{equation}
using the same integration kernel and, therefore, also kinematics as in \Eq{eq:emiss}. Here $\mathcal{N}$ is a normalization constant defined such that $\langle \mathcal{A} \rangle = 1$ for $\mathcal{A}=1$. We choose a value for $T$ and scan over values for densities $n$ up to $2 n_0$, where the corresponding value of the chemical potential $\mu$ for each case is determined from the relation
\begin{equation}
    n(\mu,T) = 4 \int \frac{d^3 k}{(2 \pi)^3} f_{\rm FD}(k,\mu,T).
\end{equation}
We evaluate \Eq{eq:exp_value} for the momentum the form factor depends on $p = p_{1/2}$ being an incoming momentum (as this leads to the strongest effect) at the two temperatures $T=20~\mathrm{MeV}$ and $T=40~\mathrm{MeV}$, which we show in \Fig{fig:axion_nucleon_Temp}.
One can clearly see that this temperature dependence cannot be neglected and is included in the calculation of the axion emissivity from SN in \Sec{sec:supernova} as can also be seen from \Fig{fig:axion_SN_couplings}.

\begin{figure}[h!] 
  \centering
  \begin{subfigure}{.49\textwidth}
  \centering
  % include first image
\includegraphics[width=1.\textwidth]{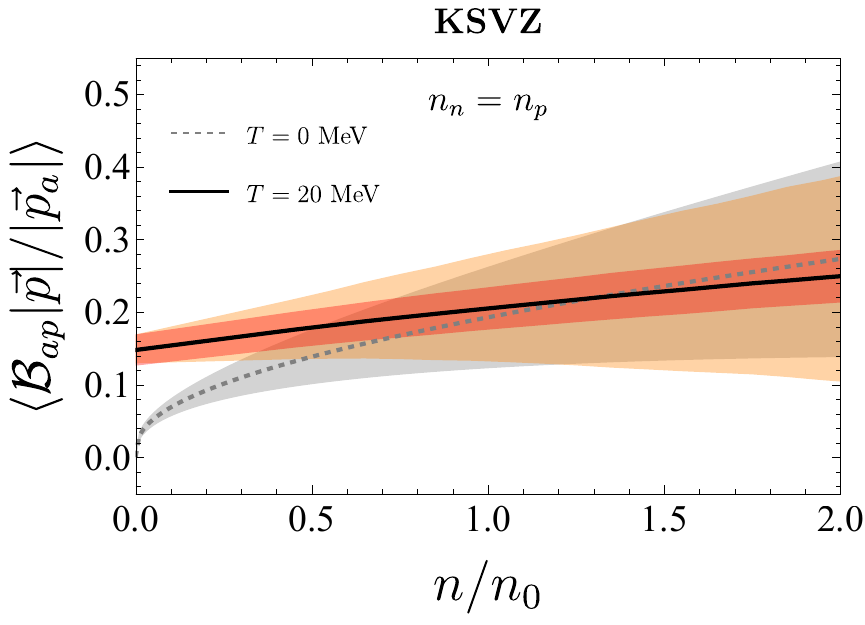}
\end{subfigure}
  \begin{subfigure}{.49\textwidth}
  \centering
  % include second image
\includegraphics[width=1.\textwidth]{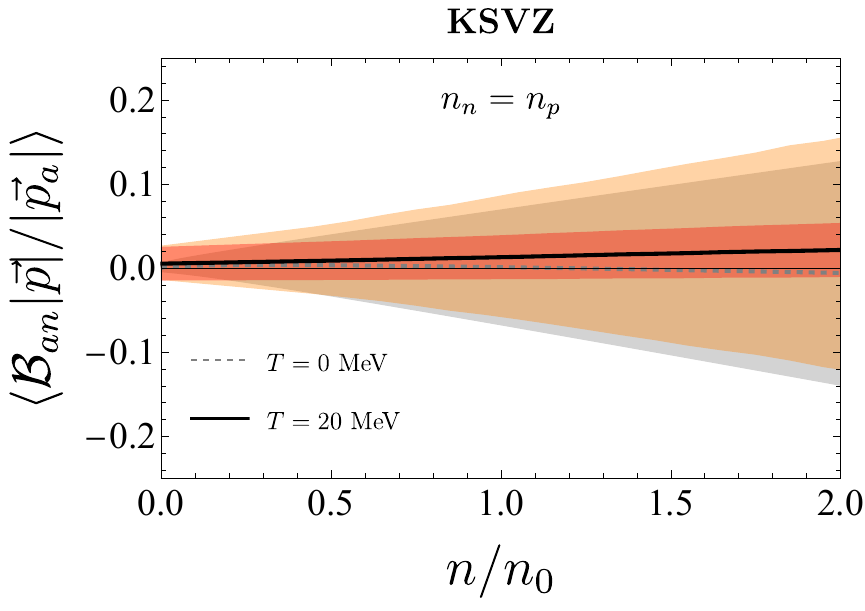}
\end{subfigure}
\begin{subfigure}{.49\textwidth}
  \centering
  % include first image
\includegraphics[width=1.\textwidth]{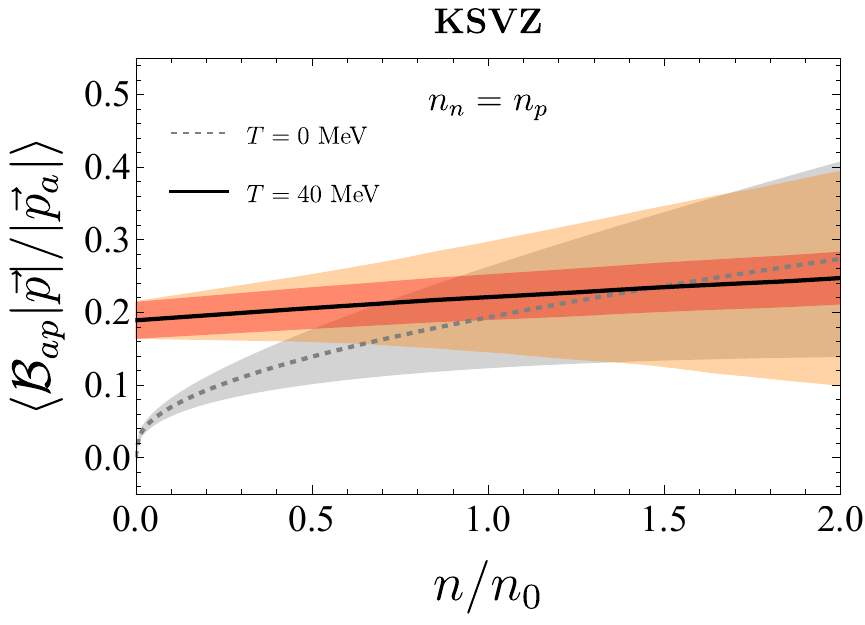}
\end{subfigure}
  \begin{subfigure}{.49\textwidth}
  \centering
  % include second image
\includegraphics[width=1.\textwidth]{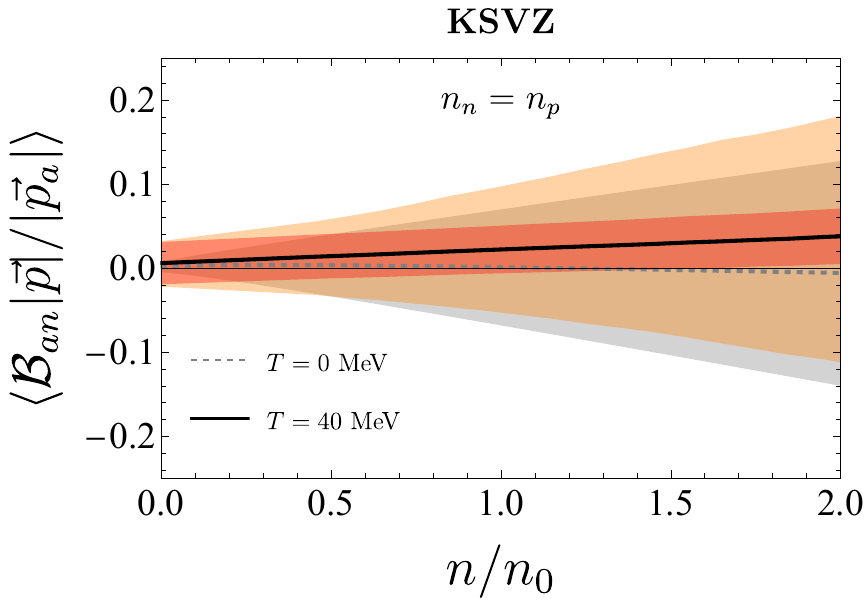}
\end{subfigure}
 \caption[]{Density dependence of the KSVZ axion-nucleon couplings for $T=20~\mathrm{MeV}$ and $T=40~\mathrm{MeV}$. Density is in units of nuclear saturation density $n_0$. Error bars are shown as in \Fig{fig:axion_nucleon_coupling_KSVZ_vac}. The dashed gray line represents the $T \rightarrow 0~\mathrm{MeV}$ result, while the gray area represents the combined error.}
  \label{fig:axion_nucleon_Temp_B}
\end{figure}

\FloatBarrier

\bibliographystyle{JHEP}
\addcontentsline{toc}{section}{Bibliography}
\bibliography{bibliography}

\end{document}